\newcounter{saveeqn}%
\newcommand{\alpheqn}{\setcounter{saveeqn}{\value{equation}}%
\stepcounter{saveeqn}\setcounter{equation}{0}%
\renewcommand{\theequation}
    {\mbox{\arabic{saveeqn}-\arabic{equation}}}}%
\newcommand{\reseteqn}{\setcounter{equation}{\value{saveeqn}}%
\renewcommand{\theequation}{\arabic{equation}}}%

\newcounter{savesec}%
\newcommand{\resec}{\setcounter{savesec}{\value{section}}%
\stepcounter{savesec}\setcounter{section}{0}%
\renewcommand{\thesection}
    {\mbox{\arabic{savesec}.\arabic{section}}}}%
\newcommand{\resetsec}{\setcounter{section}{\value{savesec}}%
\renewcommand{\thesection}{\arabic{section}}}%

\newcounter{savefig}%
\newcommand{\refig}{\setcounter{savefig}{\value{figure}}%
\stepcounter{savefig}\setcounter{figure}{0}%
\renewcommand{\thefigure}
    {\mbox{\arabic{savefig}.\arabic{figure}}}}%
\newcommand{\resetfig}{\setcounter{figure}{\value{savefig}}%
\renewcommand{\thefigure}{\arabic{section}}}%

\newcounter{savetab}%
\newcommand{\retab}{\setcounter{savetab}{\value{table}}%
\stepcounter{savetab}\setcounter{table}{0}%
\renewcommand{\thetable}
    {\mbox{\arabic{savetab}.\arabic{table}}}}%
\newcommand{\resettab}{\setcounter{table}{\value{savetab}}%
\renewcommand{\thetable}{\arabic{table}}}%

\def\ahat{{\mathaccent "7E {\cal A}}}
\def\bhat{{\mathaccent "7E {\cal B}}}
\def\chat{{\mathaccent "7E {\cal F}}}
\def\dhat{{\mathaccent "7E {\cal D}}}
\def\shat{{\mathaccent "7E {\cal S}}}
\def\proj{{\cal P}}
\def\ket{\vert \vert  \{ \emptyset \} \rangle}
\def\ket2{\vert \vert \otimes \{ R \} \rangle}
\def\sqr{$^{2}$}
\def\T{$T$}
\def\P{$P$}
\def\Q{$Q$}
\def\G{$G$}
\def\M{{\bf M}}
\def\H{{\bf H}}
\def\I{{\bf I}}
\def\Pr{{\bf P}}
\def\Tr{{\bf T}}
\def\eq{\enskip =\enskip}
\def\pls{\enskip+\enskip}
\def\mns{\enskip -\enskip}
\def\oper{O^{\bf k}_{R}}
\def\aug{\tilde{\cal H}}
\def\envfn{\chi^{\alpha}_{RL}}
\def\struc{S^{\alpha}_{RL,R^{\prime}L^{\prime}}}
\def\rR{r_{R}}
\def\aug{\vert R,L,\{\emptyset\}\rangle }
  \def\ket{\vert \vert	\{ \emptyset \} \rangle}
  \def\ket2{\vert \vert \otimes \{ R \} \rangle}
  \def\sqr{$^{2}$}
\def\dpr{\prime\prime}
\def\tpr{\prime\prime\prime}
\def\pr#1{ Phys.Rev. {\bf B#1}}
\def\pj#1{\proj_{{\cal #1}}}
\def\barr#1{{\overline{#1}}}
\def\jpc#1{J.Phys. Condensed Matter {\bf #1}}
\def\prl#1{ Phys. Rev. Lett. {\bf #1}}
\def\.#1{\mathaccent 95#1}
\def\^#1{\mathaccent 94 #1}
\def\~#1{\mathaccent "7E #1}
\def\Ir{{\mbox{I}}}
\def\Mr{{\mbox{M}}}
\def\Hr{{\mbox{H}}}
\def\sund{\mathaccent 22{\sigma}}
\def\equal{\enskip =\enskip}
\def\plus{\enskip +\enskip}
\def\minus{\enskip -\enskip}
\def\eq{\enskip =\enskip}
\def\pls{\enskip +\enskip}
\def\mns{\enskip -\enskip}
\def\Gund{\mathaccent 22 {G}}
\def\ul#1{\underline{#1}}
\def\ac#1{\mathaccent 95#1}
\def\td#1{\mathaccent "7E#1}
\def\un#1{\underline{#1}}
\def\nbox{\raisebox{.6ex}{\fbox{{\scriptsize{\phantom{$\sqrt{}$}}}}}$\:$}
\def\ybox{\raisebox{.6ex}{\fbox{{\scriptsize{$\sqrt{}$}}}}$\:$}
\def\pbox#1{\raisebox{.6ex}{\fbox{{#1}}}$\:$}
\def\c#1{\mbox{\bf #1}}
\def\und#1{$\underline{\mbox{\bf #1}}\:$}
\def\unit{{\cal I}}
\def\trans{{\cal T}}
\def\proj{{\cal P}}
\def\T{$T$}
\def\P{$P$}
\def\Q{$Q$}
\def\G{$G$}
\def\M{{\bf M}}
\def\H{{\bf H}}
\def\I{{\bf I}}
\def\Pr{{\bf P}}
\def\Tra{{\bf T}}
\def\diag{\varepsilon_{i}}
  \def\proj{{\cal P}}
  \def\trans{{\cal T}}
  \def\ket{\vert \vert	\{ \emptyset \} \rangle}
  \def\ket2{\vert \vert \otimes \{ R \} \rangle}
  \def\sqr{$^{2}$}
\def\k{{\bf k}}
  \def\ahat{{\mathaccent "7E  A}}
  \def\bhat{{\mathaccent "7E  B}}
  \def\chat{{\mathaccent "7E  C}}
  \def\fhat{{\mathaccent "7E  F}}
  \def\dhat{{\mathaccent "7E  D}}
  \def\shat{{\mathaccent "7E  S}}
  \def\phat{{\mathaccent "7E  P}}
  \def\jhat{{\mathaccent "7E  J}}
  \def\khat{{\mathaccent "7E  K}}
  \def\ohat{{\mathaccent "7E o}}
\def\ve{\varepsilon}
\def\car{\{{\cal C}\}}
\def\gt{\; > \;}
\def\lt{\: < \:}
\def\dpr{\prime\prime}
\def\tpr{\prime\prime\prime}
\def\pr#1{ Phys.Rev. {\bf B#1}}
\def\jpc#1{J.Phys. Condensed Matter {\bf #1}}
\def\prl#1{ Phys. Rev. Lett. {\bf #1}}
\def\.#1{\mathaccent 95#1}
\def\^#1{\mathaccent 94 #1}
\def\~#1{\mathaccent "7E #1}
\def\sund{\mathaccent 22{\sigma}}
\def\equal{\enskip =\enskip}
\def\plus{\enskip +\enskip}
\def\minus{\enskip -\enskip}
\def\eq{\enskip =\enskip}
\def\pls{\enskip +\enskip}
\def\mns{\enskip -\enskip}
\def\Gund{\mathaccent 22 {G}}
\def\ul#1{\underline{#1}}
\def\ac#1{\mathaccent 95#1}
\def\td#1{\mathaccent "7E#1}
\def\un#1{\underline{#1}}
\def\nbox{\raisebox{.6ex}{\fbox{{\scriptsize{\phantom{$\sqrt{}$}}}}}$\:$}
\def\ybox{\raisebox{.6ex}{\fbox{{\scriptsize{$\sqrt{}$}}}}$\:$}
\def\pbox#1{\raisebox{.6ex}{\fbox{{#1}}}$\:$}
\def\und#1{$\underline{\mbox{\bf #1}}\:$}
\def\unit{{\cal I}}
\def\trans{{\cal T}}
\def\proj{{\cal P}}
\def\T{$T$}
\def\P{$P$}
\def\Q{$Q$}
\def\G{$G$}
\def\M{{\bf M}}
\def\H{{\bf H}}
\def\I{{\bf I}}
\def\Pr{\tilde{P}}
\def\Tr{\tilde{T}}
\def\aug{\vert R,L,\{\emptyset\}\rangle }
  \def\ket{\vert \vert	\{ \emptyset \} \rangle}
  \def\ket2{\vert \vert \otimes \{ R \} \rangle}
  \def\sqr{$^{2}$}
\def\dpr{\prime\prime}
\def\tpr{\prime\prime\prime}
\def\pr#1{ Phys.Rev. {\bf B #1}}
\def\pj#1{\proj_{{\cal #1}}}
\def\barr#1{{\overline{#1}}}
\def\jpc#1{J.Phys. Condensed Matter {\bf #1}}
\def\prl#1{ Phys. Rev. Lett. {\bf #1}}
\def\.#1{\mathaccent 95#1}
\def\^#1{\mathaccent 94 #1}
\def\~#1{\mathaccent "7E #1}
\def\Ir{{\mbox{I}}}
\def\Mr{{\mbox{M}}}
\def\Hr{{\mbox{H}}}
\def\sund{\mathaccent 22{\sigma}}
\def\equal{\enskip =\enskip}
\def\plus{\enskip +\enskip}
\def\minus{\enskip -\enskip}
\def\eq{\enskip =\enskip}
\def\pls{\enskip +\enskip}
\def\mns{\enskip -\enskip}
\def\Gund{\mathaccent 22 {G}}
\def\ul#1{\underline{#1}}
\def\ac#1{\mathaccent 95#1}
\def\td#1{\mathaccent "7E#1}
\def\un#1{\underline{#1}}
\def\nbox{\raisebox{.6ex}{\fbox{{\scriptsize{\phantom{$\sqrt{}$}}}}}$\:$}
\def\ybox{\raisebox{.6ex}{\fbox{{\scriptsize{$\sqrt{}$}}}}$\:$}
\def\pbox#1{\raisebox{.6ex}{\fbox{{#1}}}$\:$}
\def\c#1{\mbox{\bf #1}}
\def\und#1{$\underline{\mbox{\bf #1}}\:$}
\def\unit{{\cal I}}
\def\trans{{\cal T}}
\def\proj{{\cal P}}
\def\T{$T$}
\def\P{$P$}
\def\Q{$Q$}
\def\G{$G$}
\def\M{{\bf M}}
\def\H{{\bf H}}
\def\I{{\bf I}}
\def\Pr{{\bf P}}
\def\Tra{{\bf T}}
\def\diag{\varepsilon_{i}}
  \def\proj{{\cal P}}
  \def\trans{{\cal T}}
  \def\ket{\vert \vert	\{ \emptyset \} \rangle}
  \def\ket2{\vert \vert \otimes \{ R \} \rangle}
  \def\sqr{$^{2}$}
\def\k{{ ( k}}
  \def\ahat{{\mathaccent "7E  A}}
  \def\bhat{{\mathaccent "7E  B}}
  \def\chat{{\mathaccent "7E  C}}
  \def\fhat{{\mathaccent "7E  F}}
  \def\dhat{{\mathaccent "7E  D}}
  \def\shat{{\mathaccent "7E  S}}
  \def\phat{{\mathaccent "7E  P}}
  \def\jhat{{\mathaccent "7E  J}}
  \def\khat{{\mathaccent "7E  K}}
  \def\ohat{{\mathaccent "7E o}}
\def\ve{\varepsilon}
\def\car{\{{\cal C}\}}
\def\gt{\; > \;}
\def\lt{\: < \:}

\def\be{\begin{equation}}
\def\ee{\end{equation}}

\def\d{{\rm d}}

 3

\def\plus{\enskip +\enskip}
\def\equal{\enskip =\enskip}
\def\k{{\bf k}}
\def\aug{{\tilde{\cal H}}}

\documentclass[12pt]{thesis}
\usepackage{epsfig,amssymb,amsmath}

\usepackage[round,comma]{authyear}
\def\lsim{\lower.5ex\hbox{$\; \buildrel < \over \sim \;$}}
\def\gsim{\lower.5ex\hbox{$\; \buildrel > \over \sim \;$}}
\def\etal{{\sl et al.}}
\begin{document}
\setcounter{section}{0}
\setcounter{figure}{0}
\setcounter{table}{0}
\vskip 3cm
\thispagestyle{empty}
%{\baselineskip 25pt

%~~~~~~~~~~~~~~~~~~~~~~~~~~~~~~~~~~~~~~~~~~~~~~~~~~~~~~~~~~~~~~~~~~~~~~~~~~~~~~
%                    THE TITLE
\centerline{\LARGE\bf NUMERICAL SIMULATION OF SPECTRAL}
\vskip 0.2cm
\centerline{\LARGE\bf AND TIMING PROPERTIES OF}
\vskip 0.2cm
\centerline{\LARGE\bf GALACTIC BLACK HOLES}
\vskip 3cm
\begin{center}
{\Large\bf  Thesis submitted for the degree of\\
            Doctor of Philosophy (Science)\\
            in \\
            Physics (Theoretical)\\
            by \\
            Sudip Kumar Garain\\ }
\end{center}
\vfill
\centerline{\LARGE\bf Department of Physics}
\vskip 0.2cm
\centerline{\LARGE\bf University of Calcutta}
\vskip 0.2cm
\centerline{\LARGE\bf 2013}
\vskip 0.2cm

\newpage
\pagestyle{newheadings}
\pagenumbering{roman}
\setcounter{page}{1}
\vskip 2cm
\centerline{\underline{\bf ABSTRACT}}

\vskip 0.4cm

%~~~~~~~~~~~~~~~~~~~~~~~~~~~~~~~~~~~~~~~~~~~~~~~~~~~~~~~~~~~~~~~~~~~~~~~~~~~~~~~
A black hole accretion may have both the Keplerian and the sub-Keplerian 
components. We consider the most general accretion flow configuration, 
namely, two-component advective flow (TCAF) in which the Keplerian disk 
is immersed inside a low angular momentum, accreting sub-Keplerian halo 
component around a black hole. The Keplerian component supplies low energy 
(soft) photons while the sub-Keplerian component supplies hot electrons 
which exchange their energy with the soft photons through Comptonization 
or inverse-Comptonization processes. In the sub-Keplerian component, a 
shock is generally formed due to the centrifugal force. The shock could be 
standing, oscillating and/or propagating. The post-shock 
region is known as the CENtrifugal pressure dominated BOundary Layer (CENBOL).  
In this thesis, we study the spectral and timing properties of such an accretion 
flow around a non-rotating, galactic black hole using a series of numerical simulations.

The spectral and the timing properties of TCAF have been extensively
studied since the model was proposed by Chakrabarti \& Titarchuk 
in 1995. However, the studies are mostly analytical. 
Some time dependent numerical simulation of the sub-Keplerian
flow including the dissipative effects (viscous and radiative cooling) 
have been performed. The 
findings are the key inputs of understanding several observed features
of black hole candidates.
In this thesis, using numerical simulation, we rigorously prove
some of the conjectures of the TCAF model. In the work presented in this
thesis, we have considered for the first time the presence of both the
Keplerian and the sub-Keplerian flow in a single simulation. The Keplerian
disk resides on the equatorial plane and is the standard disk from
which low energy photons having multi-color blackbody spectrum is 
emitted. The hydrodynamics as well as the thermal properties of the 
sub-Keplerian halo are simulated using a finite difference code which 
uses the principle of total variation diminishing (TVD). The Comptonization 
between the photons and the hot electrons is simulated using a Monte
Carlo code. These two codes are then coupled and the resulting localized 
heating and cooling are included in the coupled code. 

In Chapter 1, we give a general introduction about the accretion 
flow models and relevant radiative processes. We also discuss about the
developments of the numerical techniques to study the dynamics as
well as the radiative processes inside the accretion flow.

In Chapter 2, we describe the Monte Carlo simulation procedure for computing the
Comptonized spectrum from a two component advective flow in presence
of outflow. To reduce the time consumption, we parallelize this code.
The effects of the thermal and the bulk motion Comptonization on 
the soft photons emitted from a Keplerian disk by the CENBOL, the 
pre-shock sub-Keplerian disk and the outflowing jet are discussed here.
We study the emerging spectrum when a converging inflow and a
diverging outflow (generated from the CENBOL) are simultaneously present.

In Chapter 3, we describe the development of a time dependent
radiation hydrodynamic simulation code. Here, we couple the
Monte Carlo code with a time dependent hydrodynamic simulation code.
The details of the hydrodynamic code and the coupling procedure are
presented. Using this code, we study the spectral and timing
properties of the TCAF. The accreting halo is assumed to be of zero
angular momentum and spherically symmetric. 
We find that in presence of an axisymmetric disk, an originally
spherically symmetric accreting Compton cloud could become axisymmetric.
We also find the emitted spectrum to be direction dependent. We also 
explore the effects of the bulk motion of the halo on the emerging spectrum. 

In Chapter 4 and Chapter 5, we study the TCAF when the accreting halo
has some angular momentum with respect to the black hole.
Because of this, a shock is formed in the halo and outflows
are seen to form from the accretion disk. In Chapter 4, we study 
the effects of the Compton cooling on the outflow in a TCAF using 
the time dependent radiation hydrodynamic simulation code.
By simulating several cases for different inflow
parameters, we show that the temperature of the CENBOL region
is lowered and the outflow rate is reduced for higher Keplerian 
disk rate.  The spectrum is also found to become softer.
We thus find a direct correlation between the outflow rate 
and the spectral state of accreting black holes.
In Chapter 5, we study quasi-periodic oscillations (QPOs) in radiative
transonic accretion flows. We run several cases by varying the disk 
and the halo rates. Low Frequency QPOs are found for several combinations 
of disk and halo rates. We find that the QPO frequency increases and 
the spectrum becomes softer as we increase the Keplerian disk rate. 
An earlier prediction that QPOs occur when the infall 
time scale roughly matches with the cooling time scale, originally 
obtained using a power-law cooling, remains valid even for Compton 
cooling. We present these results. 

In Chapter 6, we draw the conclusions and discuss future plans. 
 % imp

%~~~~~~~~~~~~~~~~~~~~~~~~~~~~~~~~~~~~~~~~~~~~~~~~~~~~~~~~~~~~~~~~~~~~~~~~~~~~~~
\newpage
\vskip 3cm
\centerline{\Large\bf ACKNOWLEDGMENTS}  %imp
\vskip 2cm

It is my great pleasure to express my sincerest gratitude to my supervisor
Prof. Sandip K. Chakrabarti who has been a wonderful person with his inspiring guidance, 
enthusiastic and persistent support throughout the years of my Ph.D. period. I have not
seen so far any scientist as energetic and motivated as him so closely. I am 
indebted to him for introducing me to this fascinating topic of astrophysics 
and giving me the opportunity to work in his group. With his
clear insight in various problems (not necessarily only related to Astrophysics and space science), 
he has always motivated me to find my goals during this work. Working under 
his supervision has been a rich and rare experience and certainly will help me in the
future path of my life. 

I express my gratitude to all the academic and non-academic staff of S. N. Bose  
National Centre for Basic Sciences (SNBNCBS), Kolkata. I thank all the Professors
who taught me during my stay here since I joined the IPhD programme in 2006.
I thank Prof. Arup K. Raychaudhuri, Director, SNBNCBS, for giving me an
opportunity to work here and use the infrastructure of the S N Bose Centre.
I would like to acknowledge the South Asian Physics Foundation for sponsoring me
to attend the International Conference on Accretion and Outflow in Black Hole 
Systems in October, 2010, held in Kathmandu, Nepal. I also acknowledge
the the organizers of the Thirteenth Marcel Grossmann Meeting for providing
me with the partial financial support to attend the prestigious conference.
I am thankful to the organizers of the conference on Spectral and Timing 
Properties of Accreting Objects at European Space Agency, Madrid, for giving me 
an opportunity to attend it with full financial support. Attending all these
conferences gave me a broad overview of the current events in the field of 
astrophysics and enriched me with the essence of the subject.  

I would like to acknowledge the Indian Centre for Space Physics (ICSP), 
Kolkata where I had spent many sessions attending and giving seminars, taking
courses. I am thankful to ICSP for giving me an opportunity to take part 
actively on several experiments conducted by ICSP and get first hand
experience on real scientific experiments. I thank all the members of ICSP. 
In particular, I would like to acknowledge the members of the black hole 
astrophysics department of ICSP, specially Dr. Anuj Nandi (presently at 
ISRO, Bangalore), Dr. Dipak Debnath, Dr. Partha Sarathi Pal (presently at 
SNBNCBS), Dr. Chandra Bahadur Singh and Mr. Santanu Mondal, with whom I 
had many fruitful discussions.

My heartiest thanks go to my all friends and colleagues of SNBNCBS whose active
support and fruitful help drove my research up to the present mark. We shared very 
precious moments together and the joys are beyond to express in a few words. 
The unconditional support and love of my seniors, batchmates and juniors never 
let me feel lonely since the first day at SNBNCBS. I do not  want to mention
anyone's name in particular, as all the unforgettable moments I shared with them will 
always remain in my memory and I shall cherish them throughout my life.  

I should thank all my colleagues, past and present, in the astrophysics
department. I must mention the names of Dr. Himadri Ghosh (presently at ICSP)
and Dr. Kinsuk Giri, with whom I have collaborated. I must mention the other two scholars, namely,
Sujay Pal (presently at ICSP) and Tamal Basak. The presence of all of them 
made the working atmosphere so relaxed that disappointments could never grasp me.  
I must thank former students of our group, namely, 
Dr. Indranil Chattopadhyay (Indra-da, at ARIES, Nainital) and Dr. Santabrata Das (Santa-da,
at IIT/Guwahati) who visited SNBNCBS on various occasions and had fruitful 
discussions on several topics with us. I specially thank Indra-da with 
whom I have collaborated on several problems. I also thank some of the younger 
colleagues of mine, namely, Arnab Deb, Abhishek Roy and Arpita Nandi.   

The main encouragements behind this effort came from my family. I am happy 
to acknowledge the debts to my family members for their unconditional support 
and encouragement to maintain the interest and enthusiasm for my research.

%~~~~~~~~~~~~~~~~~~~~~~~~~~~~~~~~~~~~~~~~~~~~~~~~~~~~~~~~~~~~~~~~~~~~~~~~~~~~~~~
  %imp   

%~~~~~~~~~~~~~~~~~~~~~~~~~~~~~~~~~~~~~~~~~~~~~~~~~~~~~~~~~~~~~~~~~~~~~~~~~~~~~~
\newpage
\vskip 1cm
\centerline{\Large\bf PUBLICATIONS IN REFEREED JOURNALS}
\vskip 1cm

\begin{enumerate}
\item Quasi Periodic Oscillations in a Radiative Transonic Flow: Results of a Coupled Monte Carlo-TVD Simulation:
{\bf Sudip K. Garain}, Himadri Ghosh, Sandip K. Chakrabarti, to appear, {\bf Mon. Not. of R. Astron. Soc.}

\item Effects of Compton Cooling on Outflow in a Two Component Accretion Flow around
a Black Hole: Results of a Coupled Monte Carlo-TVD Simulation:
{\bf Sudip K. Garain}, Himadri Ghosh, Sandip K. Chakrabarti, 2012, {\bf Astrophysical Journal}, 758, 114.

\item VLF Signals in Summer and Winter in the Indian Sub-Continent using
Multi-Station Campaigns: Sandip K. Chakrabarti et al., 2012, {\bf Indian J Phys.}, 86, 323.

\item Effects of Compton Cooling on the Hydrodynamic and the Spectral Properties of a Two
Component Accretion Flow around a Black Hole:
Himadri Ghosh, {\bf Sudip K. Garain}, Kinsuk Giri, Sandip K. Chakrabarti, 2011, {\bf Mon. Not. of R. Astron. Soc.}, 416, 959.

\item Monte-Carlo Simulations of Thermal Comptonization Process in a Two Component Accretion Flow
Around a Black Hole in presence of an Outflow:
Himadri Ghosh, {\bf Sudip K. Garain}, Sandip K. Chakrabarti, Philippe Laurent, 2010, {\bf Int. Jour. of Mod. Phys. D}, 19, 607. 
\end{enumerate}

\clearpage

\vskip 1cm
\centerline{\Large\bf PUBLICATIONS IN PROCEEDINGS}
\vskip 1cm

\begin{enumerate}
\item Numerical Simulation of Spectral and Timing Properties of a Two 
Component Advective Flow around a Black Hole: {\bf Sudip K. Garain}, 
Himadri Ghosh, Sandip K. Chakrabarti, submitted,  Proc. of Recent 
Trends in the Study of Compact Objects: Theory and Observation -- 
2013, eds. S. Das, A. Nandi \& I. Chattopadhyay.

\item Effects of Compton Cooling on Outflows in a Two Component Accretion 
Flow around a Black Hole: {\bf Sudip K. Garain}, Himadri Ghosh, Sandip K. 
Chakrabarti, submitted,  Proc. of the Twelfth Marcel 
Grossmann Meeting on General Relativity (2012), eds. Kjell Rosquist, 
Robert T Jantzen, Remo Ruffini. 

\item How Plasma Composition Affects the Relativistic Flows and the Emergent 
Spectra: Indranil Chattopadhyay, {\bf Sudip K. Garain}, Himadri Ghosh, submitted, Proc.
of International Conference on Astrophysics and Cosmology (2012), Tribhuvan University, Nepal.

\item Effect of Equation of State and Composition on Relativistic Flows: 
I. Chattopadhyay, S. Mandal, H. Ghosh, {\bf S. Garain}, R. Kumar, 
D. Ryu, 2012, Proc. of Gamma Ray Bursts, Evolution of Massive Stars and 
Star Formation at High Redshift: A Bilateral Indo-Russian Workshop 
(ASI Conference Series, Vol. 5, 81-89), eds. S. B. Pandey, V. V. Sokolov \& Yu A. Schekinov. 

\item Monte-Carlo Simulations of Comptonization Process in a Two Component 
Accretion Flow around a Black Hole in Presence of an Outflow: Himadri Ghosh, 
{\bf Sudip K. Garain}, Kinsuk Giri, Sandip K. Chakrabarti, 2012, Proc. of 
the Twelfth Marcel Grossmann Meeting on General Relativity, eds. 
Thibault Damour, Robert T. Jantzen and Remo Ruffini, World Scientific: 
Singapore, p. 985. 
\end{enumerate}
   %imp
%
%~~~~~~~~~~~~~~~~~~~~~~~~~~~~~~~~~~~~~~~~~~~~~~~~~~~~~~~~~~~~~~~~~~~~~~~~~~~~~~
\newpage
\tableofcontents
\listoffigures   %*****
%\listoftables     %****

%~~~~~~~~~~~~~~~~~~~~~~~~~~~~~~~~~~~~~~~~~~~~~~~~~~~~~~~~~~~~~~~~~~~~~~~~~~~~~~
\newpage
\pagestyle{myheadings}
\pagenumbering{arabic}

%~~~~~~~~~~~~~~~~~~~~~~~~~~~~~~~~~~~~~~~~~~~~~~~~~~~~~~~~~~~~~~~~~~~~~~~~~~~~~~
\alpheqn
\resec
\refig
\retab

%%%%%%%%%%%%%%%%%%%%%%%%%%%%%%%%%%%%%%%%%%%%%%%%%%%%%%%%%%%
% Chapter 1 : Introduction
%%%%%%%%%%%%%%%%%%%%%%%%%%%%%%%%%%%%%%%%%%%%%%%%%%%%%%%%%%%

\def\k{{\bf k}}
\def\aug{{\tilde{\cal H}}}
\def\ahat{{\mathaccent "7E {\cal A}}}
\def\bhat{{\mathaccent "7E {\cal B}}}
\def\chat{{\mathaccent "7E {\cal F}}}
\def\dhat{{\mathaccent "7E {\cal D}}}
\def\shat{{\mathaccent "7E {\cal S}}}
\def\proj{{\cal P}}
\def\ket{\vert \vert  \{ \emptyset \} \rangle}
\def\ket2{\vert \vert \otimes \{ R \} \rangle}
\def\sqr{$^{2}$}
\def\T{$T$}
\def\P{$P$}
\def\Q{$Q$}
\def\G{$G$}
\def\M{{\bf M}}
\def\H{{\bf H}}
\def\I{{\bf I}}
\def\Pr{{\bf P}}
\def\Tr{{\bf T}}
\def\eq{\enskip =\enskip}
\def\pls{\enskip+\enskip}
\def\mns{\enskip -\enskip}
\def\oper{O^{\bf k}_{R}}
\def\aug{\tilde{\cal H}}
\def\envfn{\chi^{\alpha}_{RL}}
\def\struc{S^{\alpha}_{RL,R^{\prime}L^{\prime}}}
\def\rR{r_{R}}

\newpage
\markboth{\it Introduction}
{\it Introduction }
\chapter{INTRODUCTION}

%~~~~~~~~~~~~~~~~~~~~~~~~~~~~~~~~~~~~~~~~~~~~~~~~~~~~~~~~~~~~~~~~~~~~~~~~~~~~~~
Black holes can not be seen directly as no detectable radiation 
comes out from these objects. Their presence can only be 
{\em perceived} by observing the motion of other detectable objects 
around them and/or the radiation that comes out from the region close to them. 

Black holes are compact objects. They are broadly classified in two categories 
depending on their estimated mass range: stellar mass black holes and 
super-massive black holes. Generally, the stellar mass black holes have 
mass $\sim ~ 10~ M_\odot$ (where $M_\odot=1.99\times 10^{33}$g is the 
mass of the Sun). Many of such type have been found within our galaxy (e.g., Cygnus X-1, 
GRS 1915+105, GRO J1655-40, GX 339-4 etc.). On the other hand, the super-massive black 
holes have a mass generally $\geq 10^6 ~M_\odot$ and are found mostly at 
the center of galaxies (e.g., Sagittarius A* in our galaxy, M87 etc.). 
By the phrase `galactic black holes', we mean the stellar mass class of 
black holes whereas the `extragalactic black holes' are the super-massive 
black holes. Recently, it has been reported that another class of black 
holes having mass in the intermediate range ($\sim 10^2-10^4 ~M_\odot$)
may have been observed (Colbert \& Mushotzky 1999; Dewangan, Titarchuk \& 
Griffiths 2006; Patruno, Zwart, Dewi \& Hopman 2006). These are called
intermediate mass black holes.
% Colbert E. J. M., Mushotzky R. F., 1999, ApJ, 519, 89
% Dewangan G. C., Titarchuk L., Griffiths R. E., 2006, ApJ, 637, L21
% Patruno A., Zwart S. P., Dewi J., Hopman C., 2006, MNRAS, 370, L6

Most of the observed galactic black hole candidates are in binary systems
(Remillard \& McClintock 2006; McClintock, Narayan \& Steiner 2013).
% Remillard R. A., McClintock J. E., 2006, Ann. Rev. of Astron. Astrophys., 44, 49
% McClintock J. E., Narayan R., Steiner J. F., 2013, arxiv[astro-ph:13031583v2]
The black hole is the accretor and the companion star is the donor. A part of the matter flowing out of
the surface of the companion star is accreted by the black hole (Shakura \& Sunyaev 1973, hereafter SS73). 
% Shakura N. I. \& Sunyaev R. A., 1973, A \& A, 24, 337
High luminosity X-rays are emitted from these binary systems. Depending 
on the mass of the companion, these X-ray binary systems are divided 
in two major categories (Bradt \& McClintock 1983), namely, 
% Bradt, H. V. D., & McClintock, J. E. 1983, ARA&A, 21, 13
the high mass X-ray binary (HMXB) and the low mass X-ray binary (LXMB). 
When the companion has a higher mass compared to the black hole, it is called HMXB 
and when the case is opposite, it is called LMXB. When the matter is accreted by the black holes, an 
accretion disk forms around it because of the angular momentum of the 
accreted matter with respect to the black hole (Lynden-Bell 1969;
Shakura 1972; Pringle \& Rees 1972; SS73). Viscosity within the 
% Lynden-Bell D., 1969, Nature, 223, 690
% Shakura N. I., 1972, Astron. Zh., 49, 921
% Pringle J. P., Rees M. J., 1972, A\&A, 21, 1
disk transports the angular momentum outward and thus making the 
accretion possible (SS73). In black hole binaries, accretion may take place 
in two ways (see, Frank, King \& Raine 2002, hereafter FKR, for details): 
a) by Roche lobe overflow and b) from winds of the companion. 
% Frank J., King A., Raine D. J., 2002, Accretion Power in Astrophysics (Cambridge University Press, UK)

In binary systems, the accretion by Roche lobe overflow may be understood 
by drawing the equipotential surfaces of a binary system with component 
masses $M_1$ and $M_2$ having angular velocity $\Omega$ (see, Chakrabarti
1996, hereafter C96, for details). 
% Chakrabarti S. K., 1996, Physics Report, 266, 229
The effective potential of the corresponding Newtonian system is given by,
$$ 
\Phi_{eff}(r)=-\frac{M_1}{|r-r_1|}-\frac{M_2}{|r-r_2|}-\frac{1}{2}|\Omega\times r|.
$$
\begin{figure}[h!]
\begin{center}
\includegraphics[width=9cm]{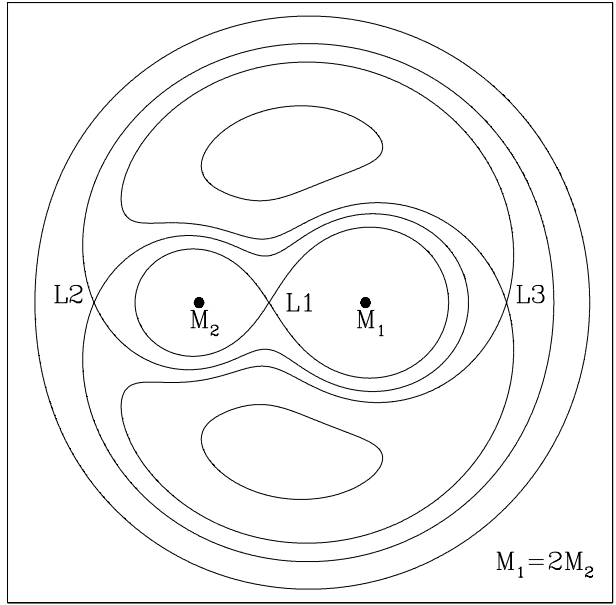}
\caption{Equipotential surfaces of a compact binary system with mass 
ratio $M_1/M_2 = 2$. Distances are in units of $GM_1/c^2$. $L_1, L_2 
{\rm~and~} L_3$ are called the Lagrange points where $\Phi_{eff}$ is locally 
or globally an extremum. Roche lobe overflow occurs when matter from 
$M_2$ fills its lobe (left section of the figure-of-eight formed by 
the innermost contour) and passes through $L_1$ to the star $M_1$ on 
the right (C96).}
\label{fig1.1}
\end{center}
\end{figure}
In Fig. \ref{fig1.1}, we show the effective potential for mass ratio
${M_1 \over M_2} = 2$. 
The innermost self-interacting contour marks the Roche lobe of two stars.
The lobes meet at $L_1$, the inner Lagrange point. Matter from the normal 
star $M_2$ overflows its Roche lobe and enters within the the Roche lobe 
of compact star $M_1$ through $L_1$, while remaining in the same plane 
as that of the binary orbit and eventually forming an accretion disk (C96).

In a galactic center, a black hole has no companion and the matter may come from the 
winds of star clusters or from the interstellar medium (Rees 1984; C96). 
% Rees M. J., 1984, Ann. Rev. Astron. Astrophys. 22, 471
This matter is expected to be of low angular momentum, quasi-spherical and mostly 
advecting. Matter tends to be almost freely falling till it `hits' the 
centrifugal barrier, which brakes the flow and causes the formation of 
a hot, puffed up region. The radial velocity of the matter of this 
region becomes very low and it starts spiraling into the black hole, 
thus forming a `thick' disk (C96).

\begin{figure}
\begin{center}
\includegraphics[width=10cm]{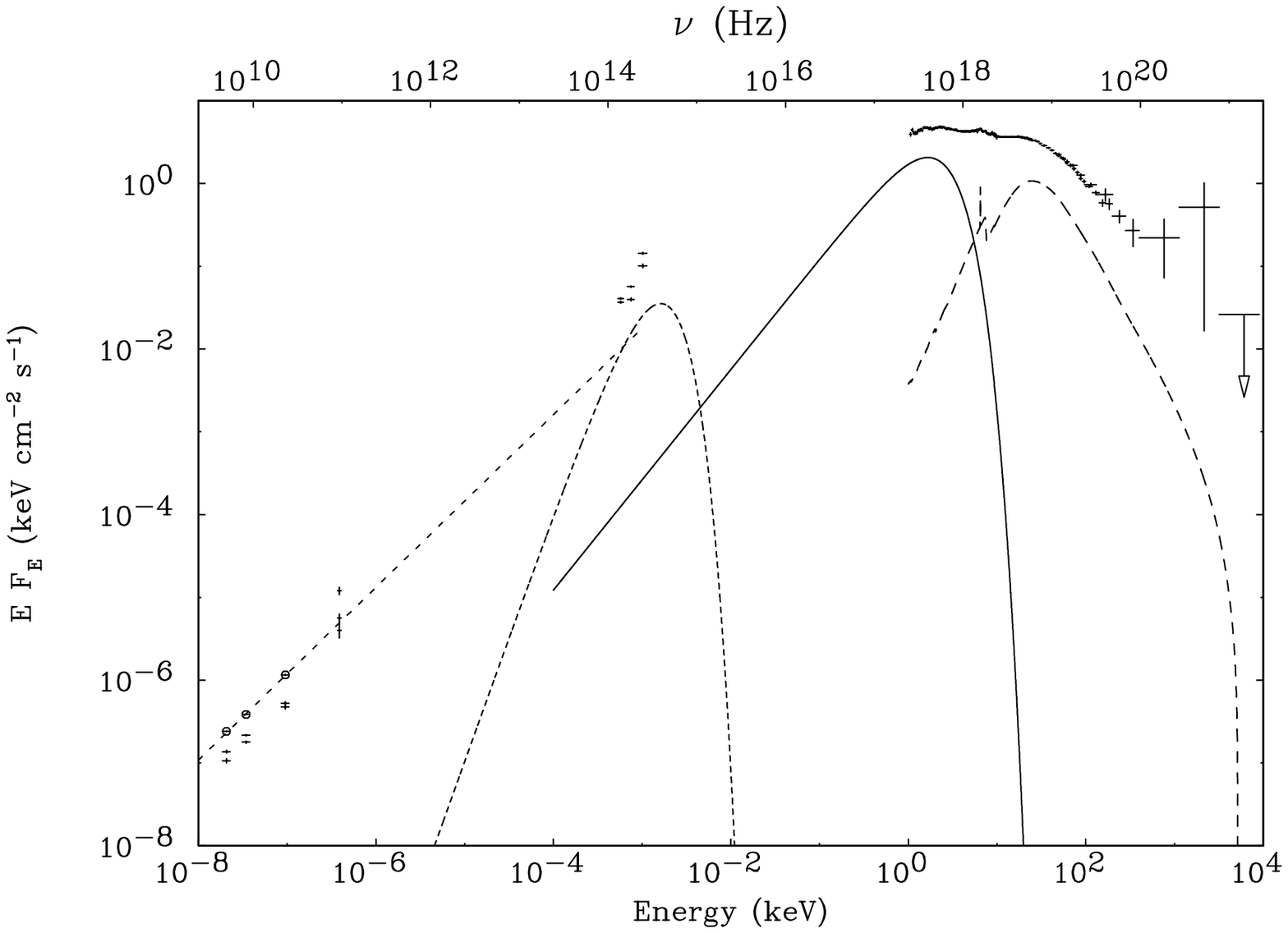}
\caption{Spectral energy distribution of GRS 1915+105. This Figure is taken from
Ueda et al. (2002). The details about the observation dates and the fitting can be
found in this reference.}
%Y. Ueda, K. Yamaoka, C. Sa´nchez-Ferna´ndez, V. Dhawan, S. Chaty, J. E. Grove, M. McCollough,
%A. J. Castro-Tirado, F. Mirabel, K. Kohno, M. Feroci, P. Casella, S. A. Trushkin, H. Castaneda,
%J. Rodrı´guez, P. Durouchoux, K. Ebisawa, T. Kotani, J. Swank, and H. Inoue, 2002, ApJ, 571, 918
\label{fig1.2}
\end{center}
\end{figure}

The radiations that are observed from the region around a black hole
are originated from the matter that is accreted onto it (FKR). The broadband 
spectrum shows the presence of photons from radio frequency through X-ray 
%SUDIP: SS433 IS NOT A GOOD CHOICE FOR A MULTIWAVELENGTH FIGURE SINCE IT IS NOT KNOWN 
%THAT IT IS A BLACK HOLE. Start replacing this.
to high frequency gamma rays (see, Fig. \ref{fig1.2}). 
However, different frequency bands are believed to be originated at 
different radial distances from the central black hole (see, SS73). 
These radiations carry the information about the properties of the 
central object as well as the matter that falls onto it. Therefore, the 
presence of a black hole can only be confirmed if it has a sufficient 
supply of matter to make it luminous enough for detection and further analysis.  

There are several models in the literature which explain the structure 
of the accretion disk and the processes in which these radiations are 
produced. In the following Sections, we discuss some of these theoretical 
models and the radiative processes relevant for such studies.

Before proceeding further, let us discuss about the units used in this 
thesis and the potential around the black holes. 
\paragraph{Units:}
The mass of the black hole ($M_{bh}$) is measured 
in the unit of Solar mass ($M_{\odot} = 1.99 \times 10^{33}$ g). 
The luminosity and the mass accretion rate are measured in the units of 
Eddington luminosity $L_{Edd} = 1.3~\times~10^{38}M_{bh} \rm{~erg~s^{-1}}$ 
and mass Eddington rate $\dot{m}_{Edd} = {L_{Edd}\over c^2} 
= 1.44~\times~10^{17}M_{bh} \rm{~gm~s^{-1}}$, respectively. The 
gravitational unit system i.e., $2G=1=M_{bh}=c$ has been used. Thus, the unit 
of velocity is $c$, the unit of distance is $2GM_{bh}/c^2$, 
unit of time is $2GM_{bh}/c^3$ and the unit of angular momentum 
is $2GM_{bh}/c$. But, if the unit of distance is $GM_{bh}/c^2$, 
unit of time will be $GM_{bh}/c^3$ and the unit of angular momentum will 
be $GM_{bh}/c$. However, the unit of specific energy is $c^2$ in both the cases.
In the following, we shall mention the unit of distance 
while writing any expression, the other units will be changed accordingly. 
Generally, the above units will be used unless stated otherwise.

\paragraph{Pseudo-Newtonian Potential:} In the case of astrophysical 
flows, it is not essential that one solves the problem using full general 
relativity. Paczy\'nski \& Wiita (1980, hereafter PW80) first suggested 
% Paczy\'{n}ski, B. and Wiita, P.J., 1980, A\&A,  88, 23
that for many practical purposes, one can use the pseudo-Newtonian potential, 
$$\phi = -\frac{GM_{bh}}{(r-r_g)},
$$ 
where, $r_g = 2GM_{bh}/c^2$, to capture the 
physical properties of Schwarzschild black hole.  As long as one is 
not interested in astrophysical processes `extremely' close 
(within 1-2 $r_g$), one can safely use this  potential and obtain 
satisfactory results. Paczy\'nski-Wiita potential accurately models 
general relativistic effects in the Newtonian theory that determine 
the motion of matter near a non-rotating black hole. The locations of the
marginally stable orbit $r_{ms}$, marginally bound orbit $r_{mb}$, and 
the form of the Keplerian angular momentum are exactly reproduced from 
this potential (PW80).

\section{Accretion disk models}
\subsection{Standard disk model}
Shakura \& Sunyaev (1973) proposed a thin disk model 
which assumed that the matter rotates in circular Keplerian orbits around 
the compact accretor. This is known as the `standard disk'. This disk is
thin in the sense that the half thickness at a radial distance $r$ is $H(r)<<r$. 
The calculations were done in a Newtonian geometry, which were redone 
including general relativistic effects by Novikov \& Thorne (1973). 
% Novikov \& Thorne, 1973, Black holes (Les astres occlus), p. 343 - 450
According to this model, matter loses 
its angular momentum because of viscosity and slowly spirals inward. 
In this process, matter also loses its gravitational energy, a part of 
which increases the kinetic energy of rotation and the other part
is converted into thermal energy and is radiated away from the surface. 
According to this model, the structure and radiation spectrum of the disk 
solely depend on the matter accretion rate ($\dot{m}$) and the viscosity parameter.

For the steady state disk, the radiation energy flux radiated from the disk surface 
at radius $r$ is given by (Shapiro \& Teukolsky 1983, hereafter ST83), 
%Shapiro,~S.~L. \& Teukolsky,~S.~A.~: {\emph{Black Holes, White Dwarfs and Neutron Stars}}, (John Wiley \& Sons, 1983)
$$
F(r) = 5\times 10^{26} M^{-2}_{bh} \dot{m}_{17} (2r)^{-3}(1-\sqrt\frac{3}{r}) 
\mathrm{~erg~cm^{-2}~s^{-1}}, 
$$ 
where, $M_{bh}$ is the mass of the black hole, $\dot{m}_{17}$ is the 
mass accretion rate in the units of $10^{17}$ g s$^{-1}$ and $r$ is in 
$2GM_{bh}/c^2$ unit.

Since the Shakura-Sunyaev disk is optically thick (optical depth 
$\tau~>1$), each element of the disk-face radiates as a blackbody spectrum 
with surface temperature $T_s(r)$ obtained by equating the dissipation 
rate to the blackbody flux, and hence, the local effective temperature 
is given by (ST83), 
\begin{equation}\label{eqno1.1}
T_s(r) = [\frac{F(r)}{\sigma}]^{1/4} = 
5\times 10^7 M_{bh}^{-1/2} \dot{m}_{17}^{1/4} (2r)^{-3/4} (1-\sqrt\frac{3}{r})^{1/4}\mathrm{~K}, 
\end{equation} 
where, $\sigma$ is the Stefan-Boltzmann constant. In case of accretion
around a stellar mass black hole, the effective temperature peaks around 
1keV, whereas for super-massive black holes, the radiation emitted from 
such a disk is in the ultra-violet region and is widely known as the 
big-blue bump (e.g., Malkan \& Sargent 1982; Malkan 1983; Sun \& Malkan 1989; Chakrabarti 2010).
% Malkan M. M., Sargent W. L. W., 1982, ApJ, 254, 22
% Malkan M. M., 1983, ApJ, 268, 582
% Sun W., Malkan M. M., 1989, ApJ, 364, 68
%Chakrabarti,~S.~K.~: 2010, \emph{AIP Conf. Proc of 4th Gamow Int. Conf. 
%on Astrophys. \& Cosm. After Gamow}, \textbf{1206},~244-262
For $\dot{m} \sim 10^{17} - 10^{18}$ g s$^{-1}$, the radiation
from the standard thin disk around a 10$M_{\odot}$ black hole generally 
extends up to 1-10 keV.\\

A typical spectrum from a black hole candidate shows that, sometimes, 
the radiation energy extends till MeV. Such a spectrum consists of a low 
energy component (E $\leq$ 10keV) and a high energy power-law tail (E $\geq$ 10keV).
% Thorne \& Price, ApJ, 1975, 195, L101;
% Eardley, Lightman \& Shapiro, ApJ, 1975, 199, L153
The Shakura-Sunyaev model cannot explain the origin of such high energy 
power-law component. To explain the high energy power-law tail part, it 
was proposed that apart from the cold standard disk, there must exist an 
optically thin ($\tau \leq $1) region where high energy 
X-ray photons get their energy from inverse-Compton scattering with the 
high energy thermal electrons (Thorne \& Price 1975; 
Eardley, Lightman \& Shapiro 1975; Katz 1976).
% Katz J. I., 1976, ApJ, 206, 910
Theories were developed for  an optically thin, geometrically thick 
accretion disk. Several other theoretical accretion disk models were 
developed: thick disk (PW80), ion-tori model (Rees, Begelman, 
Blandford \& Phinney 1982), slim disk (Abramowicz, Czerny, Lasota \& 
Szuszkiewicz 1988), transonic hybrid model (Chakrabarti 1989a, 1989b, 1990),
advection dominated accretion flow model (Narayan \& Yi 1994, 1995) etc.
% Paczy\'nski B. \& Witta P., A\&A, 1980, 88, 23
% Rees M., Begelman M., Blandford R. \& Phinney E., Nature, 1982, 295, 17
% Abramowicz M., Czerny B., Lasota J. P. \& Szuszkiewicz E., 1988, 332, 646
% Chakrabarti S. K., ApJ, 1989a, 337, 89L
% Chakrabarti S. K., ApJ, 1989b, 347, 365
% Chakrabarti S. K., 1990, Theory of Transonic Astrophysical Flows, World Scientific, Singapore
% Narayan R. \& Yi I., 1994, ApJ, 428, L13
% Narayan R. \& Yi I., 1995, ApJ, 444, 231

In the following subsections, we describe the above models very briefly.

\subsection{Thick disk}
Contrary to thin disk model of Shakura-Sunyaev (SS73), this type of disk is
geometrically thick. Thus, $H(r) \sim r$. The disk becomes thick when
pressure effects are significant such that the sound speed $a \sim (GM_{bh}/r)^{1/2}$ (Rees 1984; C96).
In these disks, the pressure gradient term in Euler equation cannot be neglected and thus,
the angular momentum does not remain Keplerian (Rees 1984; C96).
Pressure can become significant either because radiation force becomes
comparable to gravity (PW80) or because matter is unable to radiate the
energy efficiently and pressure is close to the internal energy (Rees et al. 1982).
In the former case, the thick disk is called radiation pressure dominated thick disk (PW80),
whereas in the later case, it is called ion pressure dominated thick disk i.e.,
ion-tori model (Rees et al. 1982).

Radiation pressure dominated thick disks are formed when accretion rates
onto the central object is super-Eddington i.e., when $\dot{m} > \dot{m}_{Edd}$.
For such accretion rates, luminosity of accretion disk
becomes comparable to or exceeds the Eddington luminosity. Radiation pressure
causes the flow to be non-Keplerian and puffs up the disk geometrically.
Structure of such disk becomes that of a torus and a funnel wall is produced
around the rotation axis of the matter (Rees 1984; C96; Chakrabarti 2010).
Radiation escapes through this funnel (Rees 1984; C96).

Ion pressure dominated thick disk are formed when accretion rate is very small
and the flow can not cool because of inefficient radiation process (Rees 1984; C96).
In such cases, internal energy is stored inside the disk
which increases its temperature comparable to the virial temperature.
The disk puffs up because of this (Rees 1984; C96).

However, in these thick disks, the advection terms has not been taken into account
and thus, these are not transonic. Close to a real solution which resemble a thick disk
is the CENBOL which is the post-shock region of a transonic flow.

\subsection{Slim disk}

Advective term, along with the pressure gradient term, were taken into account by
Abramowicz et al. (1988) and a new solution for the accretion flow was constructed.
The accretion rate of the disk is assumed to be moderate super-Eddington i.e.,
$\dot{m} \approx \dot{m}_{Edd}$ and hence, named slim accretion disk (Abramowicz et al. 1988).
The structure of the disk is not supposed to be neither thick nor thin ($H(r)< r$) for such accretion
rates. However, this type of disk model is not fully self-consisten (Chakrabarti 1998a, 2004). For instance, the
Abramowicz et al (1988) gave examples of a disk with 50 times critical rate (800 times the Eddington rate)
and the angular momentum sometimes decreases outward. Such a distribution would be unstable.
% Chakrabarti S. K., 1998, in Observational Evidence for the Black Holes in the Universe, p. 19 (arXiv:astro-ph/9807104)
% Chakrabarti S. K., 2004, in FRONTIERS IN ASTROPHYSICS, Eds S.K. Chakrabarti, p. 14 (arXiv:astro-ph/0402562)
 
Next, we discuss one of the most successful models which is based on the earlier theoretical 
solutions of Chakrabarti (1990), namely, two-component 
advective flow (TCAF) model, proposed by Chakrabarti \& Titarchuk (1995, 
hereafter CT95), which explains the spectral and timing properties of the 
accretion disk quite satisfactorily (e.g., Chakrabarti \& Manickam 2000, 
hereafter CM00; Nandi, Manickam, Rao \& Chakrabarti  2001; 
Smith, Heindl, Swank 2002; Chakrabarti \& Mandal 2006; 
Mandal \& Chakrabarti 2008; Dutta \& Chakrabarti 2010; Cambier \& Smith 2013).
% Chakrabarti S. K. \& Manickam S. G., ApJ, 2000, 53, 41L 
% Nandi A., Manickam, S. G., Rao, A. R. \& Chakrabarti, MNRAS, 2001, 324, 267
% Smith, D., Heindl, W. \& Swank, J., ApJ, 2002, 569, 362
% Chakrabarti, S. K. \& Mandal, S., ApJ, 2006, 49L
% Mandal, S. \& Chakrabarti, S. K., ApJ, 2008, 689, 17L
% Dutta, B. \& Chakrabarti, S. K., MNRAS, 404, 2136
% Hal J. Cambier and David M. Smith 2013 ApJ 767 46

\subsection{Advective accretion disk: Two Component Advective Flow (TCAF)}

TCAF model (CT95) is based on the physics of shock formation in a sub-Keplerian 
(low angular momentum) flow (transonic hybrid model). Unlike the 
Keplerian disk, this flow has a higher radial velocity which can reach 
up to the velocity of light $c$ at the horizon of the black hole and 
becomes supersonic (Mach number $M={v_r\over a} > 1$, where, $v_r$ 
and $a$ are the radial velocity and the sound speed, respectively). 
However, far away from the black holes, the matter is subsonic since 
its radial velocity $v_r\sim~0$ while $a>0$. Thus, a black hole accretion
is always transonic in nature (Chakrabarti 1990, hereafter C90). 
%Chakrabarti S. K., 1990, Theory of Transonic Astrophysical Flows, World Scientific, Singapore
It can be easily shown that a transonic flow is necessarily sub-Keplerian at the sonic point(s) (C90).
The accreted matter advects the mass, entropy, 
energy etc. along with it. As the sub-Keplerian flow approaches the black 
hole, at a certain radius $r=r_s$, the angular momentum becomes comparable 
to the local Keplerian angular momentum and the matter slows down due 
to centrifugal barrier. As a result, a shock is formed (C96). However, the 
formation of the shock depends on parameters like the specific energy and 
specific angular momentum of the flow, the heating and cooling mechanisms, 
and the viscosity present in the flow etc. Because of this shock, the kinetic 
energy of the incoming flow is converted into thermal energy and 
the matter is heated up. Therefore a CENtrifugal pressure dominated BOundary 
Layer (CENBOL) forms around the black hole (CT95). Inside the CENBOL, 
the density of matter also increases. Subsequently, the flow continues 
its journey to the black holes and accretes onto the black hole supersonically. 

\begin{figure}[h!]
\begin{center}
\includegraphics[width=12cm]{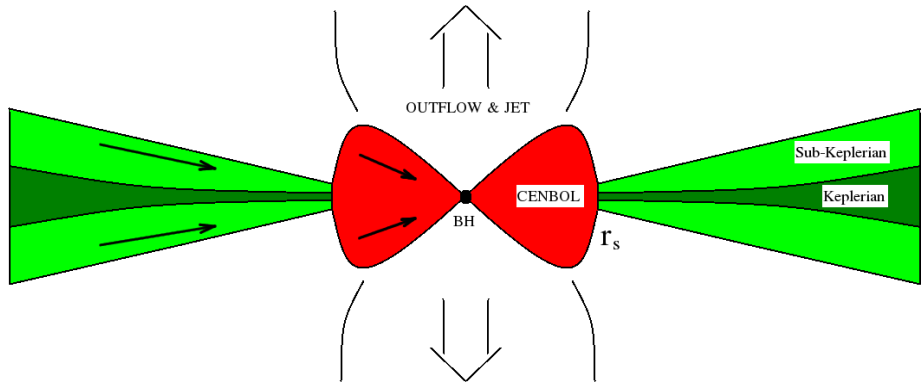}
\caption{Cartoon diagram of two-component advective flow (TCAF) model. 
The Keplerian disk is flanked by the sub-Keplerian flow. In the CENBOL 
region, both the components mix, and its density as well as temperature
increases. The outflow and the jet are produced from the CENBOL region. 
}
\label{fig1.3}
\end{center}
\end{figure}

TCAF consists of two major disk components, namely, a high viscosity, standard Keplerian 
disk component and a low viscosity sub-Keplerian halo component (CT95).
The Keplerian disk resides on the equatorial plane and it is flanked by the 
sub-Keplerian flow (Fig. \ref{fig1.3}). 

In the region $r<r_s$, the Keplerian disk is assumed to be evaporated 
because of heating. The Keplerian matter mixes up with the sub-Keplerian 
halo inside the CENBOL and form a single component (CT95). The outflows 
and jets are believed to be produced from the CENBOL region 
(CT95; Chakrabarti 1999, hereafter C99).   

A part of the low energy, blackbody photons (soft photons) that are emitted from 
the Keplerian disk, is intercepted by the hot electrons in the CENBOL. 
These photons are energized by the inverse-Compton scattering with these 
electrons and emerge as hard radiations. Thus, the
radiated spectra are produced from both the components and are a 
function of accretion rates (CT95; Chakrabarti 1997). The relative importance of 
the accretion rates of these two components determine whether the spectrum is going 
to become hard or soft. The transition from the hard to soft state is 
found to be smoothly initiated by the mass accretion rates of the disk 
(CT95). The fast variability of the photon counts are explained by 
the time variation of the flow dynamics. If the shock moves back and 
forth then the hard radiations are expected to be modulated by the 
frequency of this oscillations since they are mainly produced in the 
post-shock region (CM00; Chakrabarti 2005).
% Chakrabarti S. K., 2005, Astrophysics and Space Science, Volume 297, Issue 1-4, pp. 131 
This way an important observed feature of several black hole candidates, 
namely, the quasi-periodic oscillation (QPO) is explained. According 
to the TCAF solution, different length scales of the flow are responsible 
for the different QPO frequencies (Chakrabarti 2005).

\section{Radiative processes}
When the accretion takes place, the gravitational energy of the matter is 
released partly in the form of radiation. This radiation covers the  
entire electromagnetic spectrum that ranges from radio to $\gamma$-rays, 
as discussed previously. The production of radiation in different 
frequency-bands, basically depends on the nature of the medium and the physical 
processes associated with the system. In black hole astrophysics, the 
radiation is mostly dominated in the X-ray frequency. Below we discuss
some of the relevant radiative mechanisms. The emission 
mechanism of X-rays can be subdivided into two categories: thermal 
emission and non-thermal emission (Rybicki \& Lightman 1979, hereafter RL79;
Longair 2011).
% Rybicki, G. \& Lightman, A. P., 1979, Radiative Processes in Astrophysics (New York: Wiley Interscience)
% Longair M. S., 2011, High Energy Astrophysics (Cambridge University Press: New York)
\subsection{Thermal emission}
Thermal radiation is the radiation emitted by the matter in thermal equilibrium. 
The emitted radiation carries the information about the thermal nature of the body from where 
it is emitting. This emission mechanism can be subdivided in the following categories:
\paragraph{a) Blackbody radiation:}
Blackbody radiation comes out from a system which is in thermodynamic 
equilibrium. Once the radiation enters into the system, it does not emit from it 
unless an equilibrium is established. In astrophysics, the blackbody radiation 
is emitted mainly by optically thick medium (e.g., standard thin disk).
In such a medium, a single photon suffers from 
several scatterings. The intensity of the blackbody photons emitted from 
a system characterized by the temperature $T$ is given by the Planck's Law, 
$$
B_{\nu}=\frac{2h\nu ^3/c^2}{exp(h\nu/kT)-1}.
$$
\paragraph{b) Bremsstrahlung radiation:}
The acceleration or deceleration of a charged particle causes it to emit 
a photon. This is called bremsstrahlung or free-free emission. When a 
charged particle moves in the Coulomb field of another charged particle, 
the electric field causes the moving particle to emit bremsstrahlung. 
The radiation from a highly ionized medium which is in local thermal 
equilibrium (particles have Maxwell-Boltzmann velocity distribution) 
and optically thin (so that the radiation field is not in equilibrium), 
has a characteristic shape of continuous spectrum that is determined 
only by the temperature. This particular type of bremsstrahlung process 
is called the thermal bremsstrahlung.

The thermal bremsstrahlung spectrum falls off exponentially at higher energies 
and is characterized by the temperature $T$. The intensity $I$ of the 
radiation is given by, 
$$
I(\nu,T)=6.8\times 10^{-38}Z^2 n_e n_i T^{-1/2}e^{-n\nu/kT}g(\nu,T) \rm{~erg ~s^{-1} ~cm^{-3} ~Hz^{-1}},
$$ 
where, $Z$ is the atomic number, $n_e,n_i$ are the electron and ion number 
densities respectively and $g(\nu, T)$ is called the `Gaunt factor'. It 
is a slowly varying function of energy($h\nu$). A detail discussion is 
given in RL79.
\paragraph{c) Thermal Comptonization:}
Comptonization (i.e., Compton scattering) occurs when a photon is scattered 
by an electron. A significant part of the energy is transferred from one 
to the another. When electron gains energy from the photon, it is called 
the Compton scattering, whereas, when the reverse process occurs, i.e., photon gains 
energy from the electrons, it is called the inverse-Compton scattering. In 
astrophysics, photons are energized by the second process, i.e., by 
inverse-Comptonization, and by the term Comptonization, we generally mean this process.  

When the photon has a long wavelength (i.e., photon energy $h\nu<<m_ec^2=511\rm{keV}$), 
the scattering is closely elastic. It is called Thomson scattering. 
Electrons oscillate in the electric field of the wave, radiating 
the scattered wave as it does so. The scattering cross-section is 
$\sigma_T=8\pi r_e^2/3$, where, $r_e$ is the classical radius of the 
electron. But, when quantum effects enter, one considers Comptonization process.

Let us assume that a photon of energy $E=h\nu$ and momentum 
$\frac{E}{c}\boldsymbol{\Omega}$ is scattered by an electron of energy 
$\gamma m_ec^{2}$ and momentum ${\boldsymbol{p}} = \gamma m_e\boldsymbol{v}$,
with $\gamma = \left( 1 - \frac{v^2}{c^2}\right)^{-1/2}$. After 
scattering the photon has energy $E^\prime=h\nu^\prime$ and momentum 
$\frac{E^\prime}{c}\boldsymbol{\Omega^\prime}$. Defining 
$\mu = \boldsymbol{\Omega.v}$, $\mu^\prime = \boldsymbol{\Omega^\prime.v}$
and scattering angle $\alpha=\cos^{-1}\boldsymbol{\Omega.\Omega^\prime}$,
we find that, 
$$
\frac{\nu^\prime}{\nu}=\frac{1- \mu v/c}
{1- \mu^\prime v/c + \frac{h\nu}{\gamma m_ec^2}(1-\cos\alpha)}.
$$
If the electron is at rest ($v=0$), then,
$$
\frac{\nu^\prime}{\nu}=\frac{1}{1 + \frac{h\nu}{m_ec^2}(1-\cos\alpha)}.
$$

The scattering cross-section is given by the Klein-Nishina formula 
(Poznyakov, Sobol \& Sunyaev 1983, hereafter PSS83), 
% Poznyakov L., Sobol I. \& Sunyaev R., 1983, A \& Sp Physics Reviews, 2, 189
$$
\sigma = \frac{2\pi r_{e}^{2}}{x}\left[ \left( 1 - \frac{4}{x} - 
\frac{8}{x^2} \right) ln\left( 1 + x \right) + \frac{1}{2} + \frac{8}{x} - 
\frac{1}{2\left( 1 + x \right)^2} \right],
$$
where, $x$ is given by,
$$
x = \frac{2E}{m_e c^2} \gamma \left(1 - \mu \frac{v}{c} \right).
$$
Here, $r_{e} = e^2/m_e c^2$ is the classical electron radius and $m_e$ is the 
mass of the electron.

As mentioned earlier, in Comptonization, $\delta\nu<0$, i.e., a
photon losses its energy. However, 
when we consider the scattering of a photon by a moving electron and 
the electron has a sufficient kinetic energy compared to the photon energy, 
an inverse-Comptonization occurs. Also, in astrophysics, we have to consider 
the scattering of isotropic distribution of photons with isotropic 
distribution of electrons. For non-relativistic electrons in thermal 
equilibrium at temperature $T$, the expression for the energy transfer 
per scattering is given by (RL79),
$$
\frac{(\Delta h\nu)_{NR}}{h\nu}=\frac{(4kT-h\nu)}{m_ec^2}.
$$ 
If the electrons have a temperature high enough so that $4kT > h\nu$, the 
photons gain energy, while at a low electron temperature it is the other way around. 
In any case, the fractional energy gain is very small, so that many 
scatterings are required for a significant effect, leading to diffusion 
of the energy in phase space. Hence, the emitted spectrum depends upon 
the factor $y = \tau_{es}(\Delta h\nu)_{NR}/h\nu,$ where, $\tau_{es}$ 
is the electron scattering optical depth.

\subsection{Non-thermal emission}
Non-thermal radiations are emitted when the emitter particles 
are not in thermal equilibrium, i.e., not Maxwellian. Photons do not 
interact with the electrons completely since the matter falls very rapidly.
Non-thermal emission is very important in any environment where there are high energy particles.
\paragraph{a) Cyclotron radiation:}
This is basically the bremsstrahlung process due to the presence of a magnetic field. If an electron 
gas permeated by a magnetic field, the electrons will be 
forced to gyrate about the field lines, and the radiation that is 
emitted as a result of this acceleration is known as the Cyclotron 
radiation, provided the electrons are moving at non-relativistic speeds. 
The radiation is emitted at the gyro-frequency, which is proportional to 
the magnetic field strength, $B$, and is given by, $\omega_c=eB/m_e$. 
The radiation emitted is linearly polarized when viewed perpendicular 
to the direction of the field lines, and circularly polarized when viewed 
end-on. In this particular type of emission process, unless the field 
strength ($B$) is large, the acceleration is not particularly large, 
nor is the intensity which depends on the square of the acceleration.
\paragraph{b) Synchrotron radiation:}
When the velocity of the electrons gyrating in the magnetic field is 
relativistic ($v\sim c$), the radiation emitted is called synchrotron 
radiation. The frequency spectrum for this radiation is much more complex 
than the cyclotron radiation and can extend to many times the gyration 
frequency. The frequency of rotation, or gyro-frequency, in case of 
synchrotron radiation is given by $\omega_s=eB/\gamma m_e c$. As a result, 
the radiation produced is tightly beamed in a narrow angle about the 
forward direction of motion, by an amount determined by the Lorentz factor, 
$\gamma = 1/\sqrt{(1-v^2/c^2)}$. Hence in each rotation, a flash of light 
is observed with a duration $\delta t\approx (\gamma^3\omega_s\sin\alpha)^{-1}$. 
If the $\gamma$ is large, the width of the observed pulse ($\delta t$) can be 
much smaller than the gyration period $T_s = {2\pi \over \omega_s}$ (RL79).
%If the $\gamma$ is large, the radiation frequency can be far above the gyro-frequency.
%SUDIP CHECK THIS: dt and frequency cannot be compared. So sentence has to be changed.

In this process, the power radiated by mono-energetic electrons is given by,
$$
P=\frac{4}{3}\sigma_Tc\beta^2\gamma^2U_B,
$$
where, $\sigma_T$ is the Thomson scattering cross section, $\beta(=v/c)$ 
is the velocity of the electron, and $U_B(=B^2/8\pi)$ is the magnetic 
energy density. The energy spectrum of synchrotron radiation results from 
the superposition of the individual electron spectra, and the energy 
spectrum can be approximated as power law distribution, i.e., 
$N(E)\propto E^{-p}$. Therefore, the resulting synchrotron emission 
spectrum will also be a power law type with a spectral index $s={1\over2}(p-1).$
\paragraph{c) Non-thermal Comptonization:}
We considered the scattering of photons with electrons in thermal 
equilibrium. However, electrons may be energized at the shock front 
by the shock acceleration process (Blandford \& Eichler 1987; C96) 
% Blandford R., Eichler D., 1987, Physics Report, 174, 2
very close to the black hole, where their 
kinetic energy become very high. These highly energetic electrons are 
called non-thermal electrons and with the presence of these, the process 
of Comptonization will be modified with respect to the thermal case.

The effect of non-thermal electrons on Comptonization will produce a 
high energy tail in the spectrum that is above the thermal cut-off. 
This high-energy tail is simply the characteristic of the superposition 
of the individual electron spectra of non-thermal electrons which 
have optical depth $(\tau) << 1$. Therefore, the spectral shape depends 
on the energy index $p$ of the power-law like, with an spectral index 
$s={1\over2}(p-1)$.

\section{Numerical simulations}

The accretion disk models that have been mentioned above, give the 
steady state behavior of the accretion flow. Also, restrictions are 
imposed in the form of assumptions in order to tackle analytically the 
non-linear processes involved in the accretion and associated dissipative 
processes. This way we loose the `details' that are present in the system 
and their dynamics. Numerical simulations help us to relax some 
restrictions by considering less number of assumptions. The use of a 
numerical code to solve the equations of fluid dynamics including
the associated dissipative processes allow us to be much less restrictive
in finding the detail features and their dynamics.  

In the literature, results of numerical simulation of hydrodynamics and
the radiative processes are present. In the following subsections, we 
give a brief overview of their developments and discuss the results 
that were obtained in these simulations.

\subsection{Hydrodynamic simulation}

The hydrodynamic simulation of the `thick' accretion disk (including shock)
was done much earlier than the models of thick accretion disk were made.
Wilson (1972) developed a finite difference, 
% Wilson, J. R., 1972, ApJ, 173, 431
fully general relativistic code which uses the method of first order backward
space-differencing technique to study the behavior of inviscid 
rotating accreting matter onto a black hole. It was shown that the accretion 
flow having a significant angular momentum is accompanied by a generally moving shock.
This code was later improved and a series of important simulations were 
performed (Hawley, Smarr \& Wilson 1984a, 1984b; Hawley \& Smarr 1986).
% Hawley, J., Smarr, L. \& Wilson, J., ApJ, 1984, 277, 296
% Hawley, J., Smarr, L. \& Wilson, J., ApJSS, 1984, 55, 211
% Hawley, J. F., & Smarr, L. L. 1986, in AIP Conf. Proc. 144, Magnetospheric
%Phenomena in Astrophysics, ed. R. Epstein & W. Feldman (Melville, NY:
%AIP), 263
In these simulations also, the shocks were found in the accretion of 
rotating matter. They found that the centrifugal barrier, or funnel wall, 
plays a key role in shock heating the inflowing supersonic fluid, 
producing a stationary thick disk with bipolar outflow. Subsequently, 
Eggum, Coroniti \& Katz (1987) studied the evolution and stability of 
existing, sub-Eddington standard Keplerian disk including the  
% Eggum, G. E., Coroniti, F. V., Katz, J. I., ApJ, 1987, 323, 634
% Eggum, G. E., Coroniti, F. V., Katz, J. I., ApJ, 1988, 330, 142
effect of viscosity and radiation transport in their numerical simulations.
In another paper, Eggum, Coroniti \& Katz (1988) studied super-Eddington
accretion disk using the same numerical code. They found that a thick disk 
really forms and found significant radiation driven winds and jet from 
the accretion disk. 

A systematic study of the thick accretion disk including shock using numerical 
simulation was started after the extensive theoretical works on transonic accretion flows were
done by Chakrabarti (1989b; C90). In the earlier works of Wilson, Hawley and others as mentioned
above, the shocks were found in the accretion flow, but they were not 
stable and traveled outward. No standing shocks were found. The extensive 
works on transonic flows provided a very good theoretical basis of the 
subject and provided an understanding of the 
parameter space spanned by the specific energy ($\epsilon$) and the 
specific angular momentum ($\lambda$) of the sub-Keplerian inflow at 
the outer boundary. One good reason why previous simulations did not find any standing 
shock may be the wrong choice of the flow parameters or presence of a significant amount
of numerical viscosity (C96). However, the theoretical 
works predicted that there are two types of solutions, with and without a shock,
for a particular set of parameters $(\epsilon, \lambda)$.

Chakrabarti \& Molteni (1993) showed by numerical 
simulation of inviscid, thin accretion flow that the shock could be
% Chakrabarti S. K., Molteni D., 1993, ApJ, 417, 671
common in the accretion flow onto a black hole. They found that the flow 
chooses the shock free solution in a truly unperturbed accretion (in one dimension).
A shock is generated when some perturbation is introduced in 
the flow. In a realistic accretion flow which is three dimensional, 
the turbulence is always expected to be present and this acts as the 
seed of the required perturbation. Therefore, it is almost certain that 
realistic accretion disks have centrifugal pressure supported shocks in them 
(Chakrabarti \& Molteni 1993; C96). Moreover, a shock solution 
has a higher entropy (Chakrabarti 1989b; C90) which makes it favorable choice as compared to 
the shock free solution for a given set of $(\epsilon, \lambda)$.

That the shock actually forms in a realistic accretion flow, was shown by
two dimensional numerical simulations of accretion flows 
(Molteni, Lanzafame \& Chakrabarti 1994, hereafter MLC94; 
Molteni, Ryu \& Chakrabarti 1996, hereafter MRC96;
Ryu, Chakrabarti \& Molteni 1997, hereafter RCM97). 
% Molteni D., Lanzafame G., Chakrabarti S. K., 1994, ApJ, 425, 161
% Molteni D., Ryu D., Chakrabarti S. K., 1996, ApJ, 470, 460
% Ryu D., Chakrabarti S. K., Molteni D., 1997, ApJ, 474, 378
They assumed axisymmetry and reduced the three dimensional problem to 
two dimensions. In these works, not only standing shocks but also oscillatory 
shocks were found to be formed. The consequent oscillations in the hydrodynamic 
and thermal properties of the accreting matter is believed to be
the origin of the quasi-periodic variabilites observed in black hole candidates. 
Outflows and jets are seen to be produced from the post-shock region. 
More importantly, two different numerical simulation codes written on 
totally different principles produce similar results and those 
match with the theoretically obtained results (MRC96). One is a 
Lagrangian code based on smooth particle hydrodynamics (SPH) scheme
(Lucy 1977; Gingold \& Monaghan 1977; Monaghan 1983, 1992) and 
% Lucy L., 1977, AJ, 82, 1013
% Gingold R. A., Monaghan J. J., 1977, MNRAS, 181, 375
% Monaghan J. J., 1983, Comput. Phys. Rep., 3, 71
% Monaghan J. J., 1992, ARA\&A, 30. 543 
the other is an Eulerian finite difference code based on total variation 
diminishing (TVD) scheme (Harten 1983; Ryu, Ostriker, Kang \& Cen 1993). 
% Harten A., 1983, Jour. Comput. Phys., 49, 357
% Ryu D., Ostriker J., Kang H., Cen R., 1993, ApJ, 414, 1

However, the occurrence and stability of the shocks depend on the transport 
phenomena e.g., viscous or radiative transport that are present in the flow. 
The simulations of thin, isothermal, viscous accretion disk using 
SPH code (Chakrabarti \& Molteni 1995) show that the 
% Chakrabarti S. K., Molteni D., 1995, MNRAS, 272, 80 
shock becomes weaker, wider and forms farther out in presence of weak 
viscosity. Shakura-Sunyaev (SS73) viscosity parameter $\alpha$ is used 
to tune the effect of viscosity. When the viscosity is increased, the shock 
starts traveling outward making the post-shock disk subsonic and the 
angular momentum distribution of this disk becomes similar to the 
Keplerian distribution. The flow becomes supersonic only close to the 
black hole, before falling onto it. The simulations of thick disk 
including viscous effects (Lanzafame, Molteni \& Chakrabarti 1998)
% Lanzafame G., Molteni D., Chakrabarti S. K., 1998, MNRAS, 299, 799
generalize the above results and show that even in thick disk, the 
centrifugally driven standing shock drifts away and eventually disappears.
Similar conclusions are found more recently by Giri \& Chakrabarti (2012)
% Giri K., Chakrabarti S. K., 2012, MNRAS, 421, 666 
where the simulations of viscous, thick accretion disk were done using TVD
code. Actually, it was found that above a certain value of 
$\alpha$, the critical viscosity $\alpha_c$, as predicted before by C90 and C96, the shock disappears. The 
critical viscosity parameter depends upon the injected flow parameters.
Another important finding of the above simulations is the oscillation
of the shock when the viscosity is low. 

The oscillation of the shock was found in an inviscid flow also. In RMC97, the 
authors reported the presence of unstable shocks for the accretion of thin, 
cold, adiabatic gas in an axisymmetric, two dimensional simulation. The 
angular momentum of the incoming gas was around the marginally stable value.
The reason of the formation of such an unstable shock is that the
Rankine-Hugoniot conditions are not satisfied. In another significant 
simulation of inviscid accretion flows which include radiative transfer 
(Molteni, Sponholz \& Chakrabarti 1996, hereafter MSC96)
%Molteni, D., Sponholz, H., \& Chakrabarti, S. K. 1996, ApJ, 457, 805
the shock oscillation was found when the cooling time scale of the
post-shock region {\em roughly} matches with the infall time scale of the 
matter onto the black hole. The simulations were performed in both one 
and two dimensions. They used the power-law cooling 
$\Lambda \propto \rho^2 T^\zeta$, where, $\rho$ and $T$ are the density 
and temperature, respectively, of the accreting matter.
$\zeta=0.5$ represents the bremsstrahlung cooling. In Chakrabarti, Acharyya
\& Molteni (2004, hereafter CAM04), effect of Comptonization is included
%Chakrabarti, S. K., Acharyya, K., Molteni, D. 2004A&A...421....1C
and the reflection symmetry about the equatorial plane is removed in an
axisymmetric, two dimensional simulation of thick disk using SPH code. 
Compton cooling per unit mass from the post-shock matter with accretion 
rate $\dot{m}_c$ is mimicked by multiplying an efficiency factor $\eta$ 
to bremsstrahlung cooling with accretion rate $\dot{m}_b={\dot{m}_c \over \eta}$. 
Similar conclusions as MSC96 are obtained along with an extra effect of 
vertical shock oscillation apart from the radial oscillation.      

The above mentioned shock oscillations in the thick accretion disk result
in an significant observational effect, namely, the QPOs. As mentioned 
earlier, in TCAF model the high energy photons of the observed spectra 
from the black hole candidates are produced in the post-shock
region. Therefore the hydrodynamic and thermal variability present in this
region will be reflected in the observed variations of the high energy photon 
count rate. This is actually observed in the realistic data (CM00; 
Rao, Naik, Vadawale \& Chakrabarti 2000). 
% Chakrabarti S. K., Manickam S. G., 2000, ApJ, 53, 41L   
% Rao A. R., Naik S., Vadawale S. V., Chakrabarti S. K, 2000, A\&, 360, 25L

\subsection{Radiative transfer}
One of the most important radiative processes in the context of galactic 
black holes is the Compton scattering between the electrons and the radiation 
field. In order to explain the power-law part in the hard state spectrum 
from the black hole candidates, multiple inverse-Compton scattering of 
the low energy photons by the weakly relativistic or relativistic, hot 
electrons is assumed to be mainly responsible.
Several theoretical models have been developed to compute the spectral
shape from the corona (Katz 1976; Sunyaev \& Trumper 1979; Sunyaev \& Titarchuk 1980,1985).
% Katz, J. I., ApJ, 1976, 206, 910
% Sunyaev, R., Trumper, J., Nature, 1979, 279, 506
% Sunyaev, R. A., Titarchuk, L. G., A\&A, 1980, 86, 121 
% Sunyaev R. A., Titarchuk L. G., 1985, A\&A, 143, 374
In these references, calculations have been done using the Kompaneets equation
(Kompaneets 1956; RL79).
% Kompaneets, A. S. 1956, Zh. Eksp. Teor. Fiz., 31, 876 [English transl. Soviet Phys. - JEPT, 4, 730 (1957)]
% Rybicki, G. \& Lightman, A. P., 1979, Radiative Processes in Astrophysics (New York: Wiley Interscience)
However, theoretical calculations are restricted by several constraints.
For example, the spectra can be described adequately by analytic methods 
given in the above references for non-relativistic ($kT_e<<m_ec^2$) or 
ultra-relativistic ($kT_e>>m_ec^2$) cases, but in mildly relativistic plasma 
($kT_e \sim m_ec^2$), it is hard to treat the problem analytically (PSS83). 
This restriction was removed later by Titarchuk and his collaborator 
(Titarchuk 1994; Hua \& Titarchuk 1995; Titarchuk \& Lyubarskij 1995)
% Titarchuk L., 1994, ApJ, 434, 570
% Hua X., Titarchuk L., 1995, ApJ, 449, 188
% Titarchuk L., Lyubarskij Y., 1995, ApJ, 450, 876
where they derived the spectral indices for wide ranges of the optical depth
and electron temperatures of the cloud. However, the computations
are done for a particular geometry such as a slab or spherical type and the variation
of the optical depth or the temperature inside a cloud is not taken into account,
rather some average values are considered. However, it is seen from numerical
simulations that the geometry of the electron cloud in the accretion disk 
is much more complicated and different regions have different temperatures
as well as different optical depths. Therefore by taking simplified assumptions
such as a particular geometry or averaged temperature or optical depth while
computing the spectral shape, we actually lose the details of an accretion
flow configuration. In such conditions, the correct spectra would be obtained by 
Monte Carlo methods. 

A Monte Carlo method is a numerical method of solving mathematical 
problems by random sampling (Sobol 1994). This method can be applied 
% Sobol I. M., 1994, A Primer to the Monte Calro Method, CRC Press, US
to solve any mathematical problem, not just those of probabilistic character (PSS83). 
The Monte Carlo simulation technique has been applied to study the Compton
scattering problem. Poznyakov, Sobol \& Sunyaev (1976, 1977, 1983) 
% Poznyakov, L., Sobol, I. \& Sunyaev, R., Soviet Astronomy Letters, 1976, 2, 55 (Translation)
% Poznyakov, L., Sobol, I. \& Sunyaev, R., Soviet Astronomy, 1977, 21, 708 (Translation)
% Poznyakov, L., Sobol, I. \& Sunyaev, R., A \& Sp Physics Reviews 1983, 2, 189
computed the X- and $\gamma$ ray radiations for spherical cloud and slab 
disks of thermal relativistic plasma of a given optical depth with a given 
temperature. The low frequency photon source was either a point source 
(for a spherical cloud) at the center or a flat disk (slab geometry) or 
distributed within an electron cloud. These simulations show that the 
power-law part of the spectrum is produced by the inverse-Comptonization 
of the low frequency photons, firmly establishing early theoretical 
work on Comptonization 
(Katz 1976; Sunyaev \& Trumper 1972; Sunyaev \& Titarchuk 1980).
In the simulation, the energy of the injected Planck photons,
the momentum of the Maxwell electrons inside the electron cloud, mean 
free path of the photon and the Compton scattering -- all these events
have been modelled by Monte Carlo techniques (PSS83). However, in these
calculations, the Comptonization by the thermal Maxwellian electrons alone
has been considered. In subsequent works, Laurent \& Titarchuk 
(1999, 2001, 2007) incorporated the effects of bulk velocity of electrons 
% Laurent, P. \& Titarchuk, L., ApJ, 1999, 511, 289
% Laurent, P. \& Titarchuk, L., ApJ, 2001, 562, 67L
% Laurent, P. \& Titarchuk, L., ApJ, 2007, 656, 1056
in addition to its thermal motions in the Monte Carlo code and computed
the effects of up-scattering of converging inflow on the output spectrum. 
They assumed the geometry of the Compton cloud to be spherical and the 
electrons to approach 
the central black hole with free fall velocity. The simulation of the
Compton scattering is similar to the method described in PSS83 with 
the difference that it takes into account the exact motion of the 
electron, which is the composition of its free fall motion with its
Brownian thermal motion. Simulations in both flat as well as curved
space were done. They showed that in both the soft and the hard states, the 
spectra are formed by the up-scattering of the disk soft photons by the
converging electrons. The effect of momentum transfer is found to be so 
strong that even a cooler Compton cloud can produce a power-law 
component extending to energies comparable with the kinetic energy of 
electrons in the converging flow.

The first computation of spectral properties of the TCAF using Monte Carlo 
method was done by Ghosh, Chakrabarti \& Laurent (2009). 
% Ghosh H., Chakrabarti S. K., Laurent P., 2009, Int. J. Mod. Phys. D, 18, 1693    
In this study, they computed the effects of thermal Comptonization of 
soft photons coming from an extended Keplerian disk by hot electrons
of the CENBOL. The effect of bulk motions was not taken into account, however.
The simulations show that the state transition of the black
hole candidate can be explained either by varying the size of the 
CENBOL or by changing the central density of the CENBOL, which is
governed by the sub-Keplerian accretion rate. 

\clearpage
%\end{document}
%@@@@@@@@@@@@@@@@@@@@@@@@@@@@@@@@@@@@@@@@@@@@@@@@@@@@@

	\reseteqn
	\resetsec
	\resetfig
	\resettab

%~~~~~~~~~~~~~~~~~~~~~~~~~~~~~~~~~~~~~~~~~~~~~~~~~~~~~~~~~~~~~~~~~~~~~~~~~~~~~~
\alpheqn
\resec
\refig
\retab

%%%%%%%%%%%%%%%%%%%%%%%%%%%%%%%%%%%%%%%%%%%%%%%%%%%%%%%%%%%
% Chapter 2 : Computational Procedure of Spectral Properties of TCAF Model
%%%%%%%%%%%%%%%%%%%%%%%%%%%%%%%%%%%%%%%%%%%%%%%%%%%%%%%%%%%

\def\k{{\bf k}}
\def\aug{{\tilde{\cal H}}}
\def\ahat{{\mathaccent "7E {\cal A}}}
\def\bhat{{\mathaccent "7E {\cal B}}}
\def\chat{{\mathaccent "7E {\cal F}}}
\def\dhat{{\mathaccent "7E {\cal D}}}
\def\shat{{\mathaccent "7E {\cal S}}}
\def\proj{{\cal P}}
\def\ket{\vert \vert  \{ \emptyset \} \rangle}
\def\ket2{\vert \vert \otimes \{ R \} \rangle}
\def\sqr{$^{2}$}
\def\T{$T$}
\def\P{$P$}
\def\Q{$Q$}
\def\G{$G$}
\def\M{{\bf M}}
\def\H{{\bf H}}
\def\I{{\bf I}}
\def\Pr{{\bf P}}
\def\Tr{{\bf T}}
\def\eq{\enskip =\enskip}
\def\pls{\enskip+\enskip}
\def\mns{\enskip -\enskip}
\def\oper{O^{\bf k}_{R}}
\def\aug{\tilde{\cal H}}
\def\envfn{\chi^{\alpha}_{RL}}
\def\struc{S^{\alpha}_{RL,R^{\prime}L^{\prime}}}
\def\rR{r_{R}}

\newpage
\markboth{\it Computational Procedure of Spectral Properties }
{\it Computational Procedure of Spectral Properties }
\chapter{\uppercase {Computational Procedure of Spectral Properties of the  TCAF Model}}

%We study the spectral properties of the accretion disk using a
%Monte Carlo code (PSS83; Ghosh et al. 2009). 

In this Chapter, we shall discuss the algorithm of the serial Monte Carlo code and its parallelization 
process. Next, we shall discuss how this code is used to study the spectral 
properties of TCAF in presence of outflows.  

%~~~~~~~~~~~~~~~~~~~~~~~~~~~~~~~~~~~~~~~~~~~~~~~~~~~~~~~~~~~~~~~~~~~~~~~~~~~~~~~
\section{Monte Carlo code for Comptonization and its parallelization}

Here we briefly discuss the algorithm of computation of Comptonized 
spectrum using a Monte Carlo code. The algorithm is similar to the
one used by Laurent \& Titarchuk (1999) and Ghosh et al. (2009). 
Then, the implementation of parallelization will be described in detail.
% Ghosh H., Chakrabarti S. K., Laurent P., 2009, IJMPD, 18, 1693

In our simulation, each photon is tracked beginning from its origin till 
its detection. The photon source may be a point source or any disk 
(e.g., a standard SS73 disk) or distributed within the electron cloud 
(e.g., bremsstrahlung, synchrotron etc.). Its initial energy may be 
drawn from the Planck's distribution function (for blackbody type source) 
or from any other given distribution e.g., power-law etc. It can be given 
any initial preferential direction or it may be emitted isotropically. 
So, once a photon is generated within the accretion disk, we know its 
location (co-ordinate), initial energy and initial direction of motion.
Certain scattering condition is associated with each photon. For the 
present case, we set a target optical depth $\tau_c$ at which a particular 
photon will suffer scattering. This value is computed from the exponential 
law $P(\tau) = exp(-\tau)$ (Laurent \& Titarchuk 1999). Next, the optical thickness 
$\tau = n_e \sigma l$ along the photon path is integrated taking into 
account the variation in the electron number density $n_e$ and the scattering 
cross-section $\sigma$. If integrated $\tau$ for that particular photon becomes 
$\tau_c$ before leaving the electron cloud, a Compton scattering is 
simulated following PSS83. After the scattering, the direction as well as 
energy of the photon change and it is again tracked in the same way as 
mentioned above. On the other hand, if $\tau < \tau_c$ before leaving the Compton cloud, the photon 
remains unscattered. This way, each photon is tracked till its escape 
from the electron cloud or it is absorbed by the black hole. Once a 
photon is out, it is captured and used for spectrum determination.

In the Monte Carlo code, the initial properties of a photon (e.g., its 
location, energy etc.), scattering criteria (e.g., target optical depth, 
mean free path of photons etc.) and the scattering event (e.g., selection 
of scattering electron, its momentum and energy etc.) are modelled using
Monte Carlo technique (PSS83). The pseudo-random number generator that 
has been used here is taken from Wichmann \& Hill (1982).
% Wichmann, B. A. \& Hill, I. D., {\textit{Applied Statistics}}, \textbf{31}, 2, 1982, 188.

\begin{figure}
\centering{
\includegraphics[height=9.5truecm]{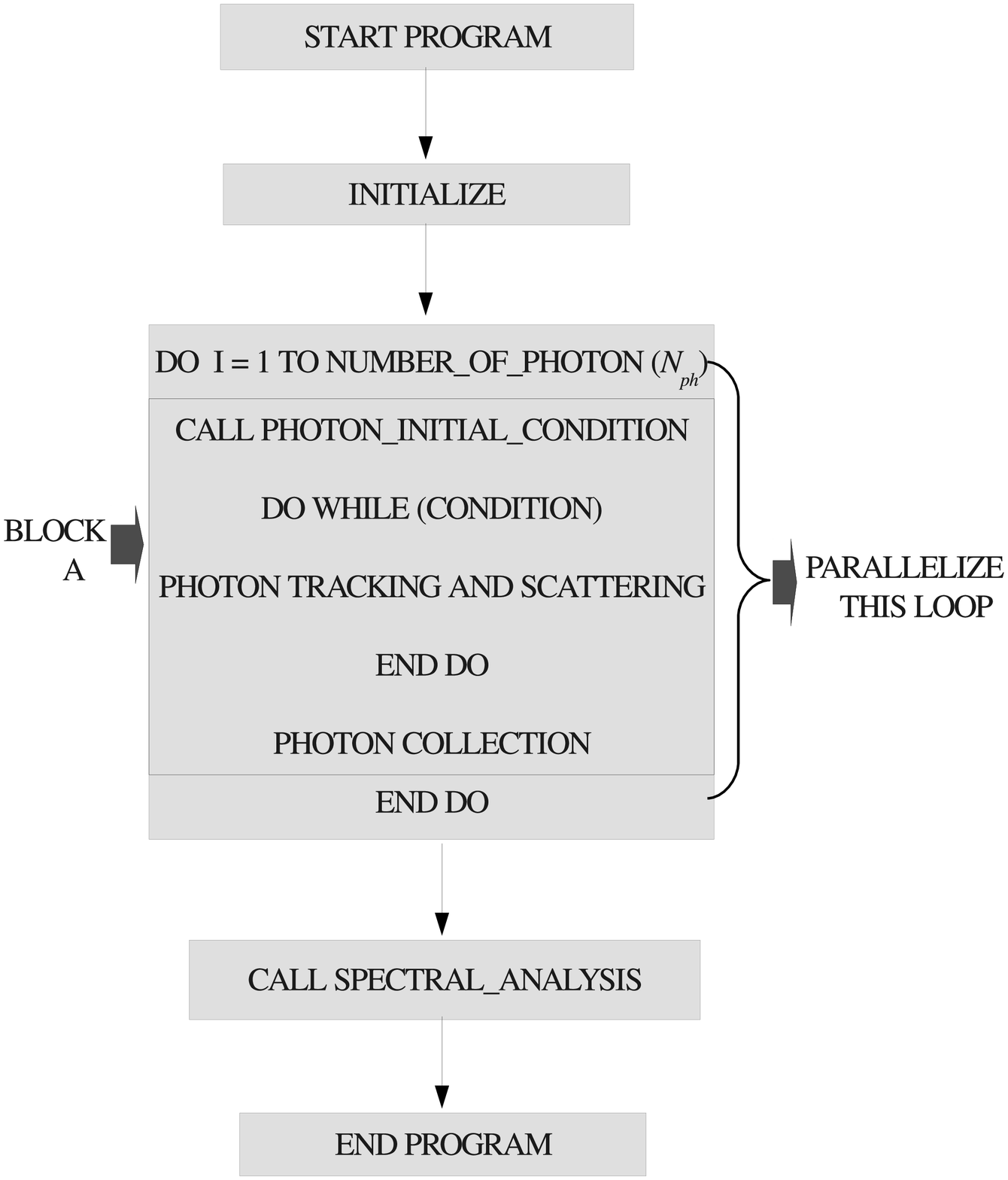}
\hskip -3mm
\includegraphics[height=9.5truecm]{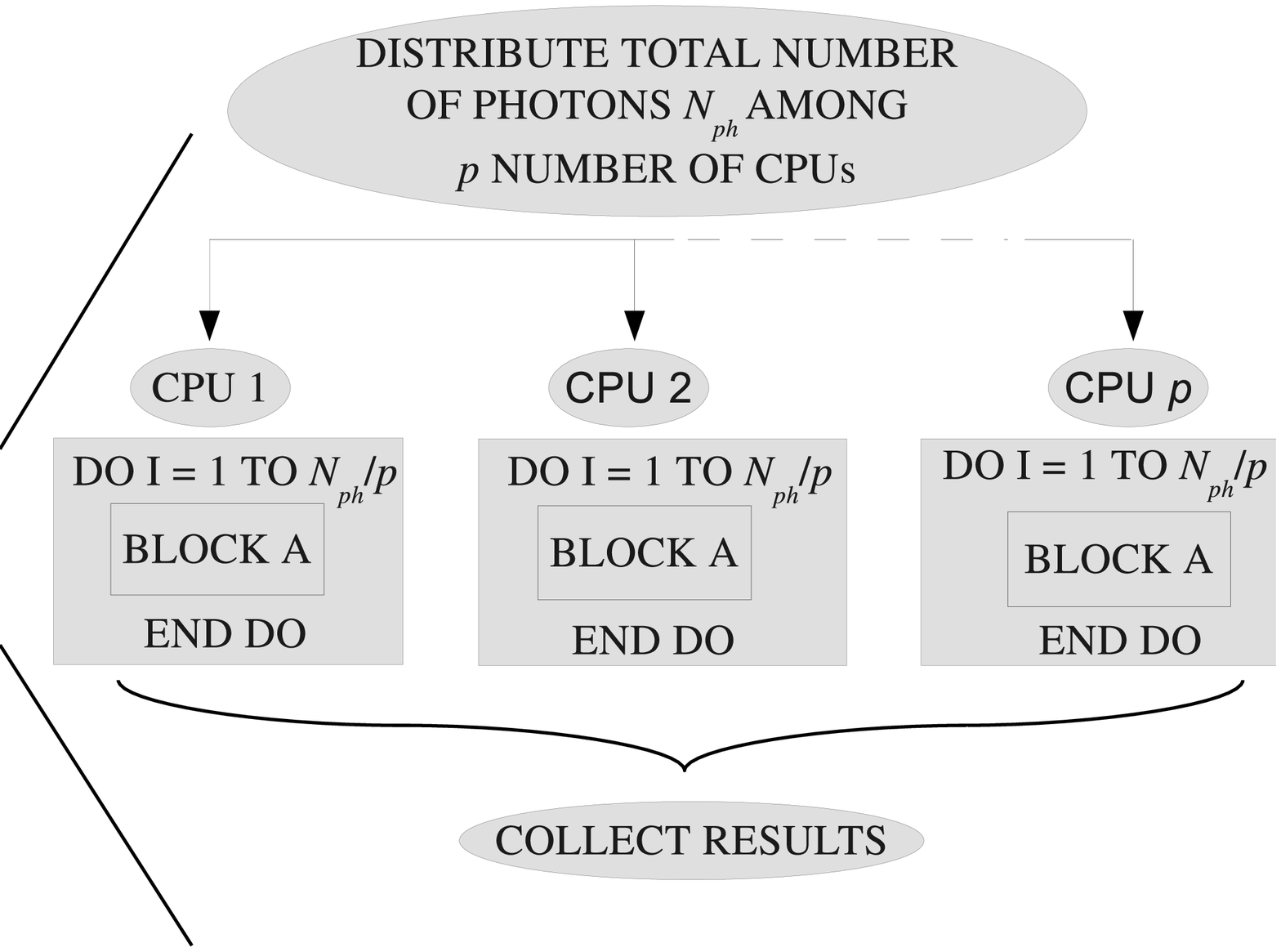}}
\caption{Flowchart of the serial code (left) and its parallelization (right).
}
\label{fig2.1}
\end{figure}

In a serial run, a single processor processes photons one by one and at 
the end, we get the full spectrum. For a steady state system, the behavior 
of one photon inside the accretion disk is independent of another one,
i.e., processing of photons are mutually exclusive. So, it does not 
matter if we process more than one photon simultaneously. In a 
prallelized code, that is what is done. Depending on the number of 
available processors, the number of simultaneously injected photons is 
decided. So, if we wish to inject $N_{ph}$ number of photons and we have $p$ 
number of available processors, then each processor processes $N_{ph}/p$ 
number of photons simultaneously in the same electron cloud. 
In Fig. \ref{fig2.1}, we show the flowcharts of the serial code and
after its parallelization. In the serial code, `Block A' is the main 
computational block which is called $N_{ph}$ times under a `Do' loop. 
In the parallel code, we have broken this single `Do' 
loop in $N_{ph}/p$ loops and run the same block A in $p$ number of 
processors. At the end, we collect the results from each processors 
and sum them to get the final spectrum.

One important practical problem of parallelization in this case is the 
choice of set of seed values for the pseudo-random number generator 
used here. In a serial run, the initial seed values are chosen such 
that the cycle length is very long, and the cycle length depends on the 
initial seed values. In parallelized case, since the same program with the same 
input files runs on different processors, same seed values will produce 
exactly same results. For this, we  have to be careful about the initial 
seed values for individual processors. We cannot choose any arbitrary 
seed values for each processors since that may hamper the cycle length. 
To circumvent this problem, we pick up some numbers from the series that 
is generated when the pseudo-random number generator is run. We use 
these numbers as the initial seed values for different processors. 
We have to make sure that these selected numbers are far away from each 
other in the series so that no repetition occurs and we do not get same 
results from different processors. 
%******************************
%SUDIP: Could you not choose different seed values for different components of 
%velocity vector, locations etc. Then each of those would have been independent and repeatations
%would not have occurred.
%**********************8

\subsection {Parallelization technique}
The parallelization has been done using Message-Passing Interface (MPI). 
As the name suggests, the communication between the multiple processors 
is done by message-passing. MPI is a library of functions and macros that 
can be used in C, FORTRAN, and C++ programs. Here, we have used MPI FORTRAN 
functions for parallelizing our Monte Carlo code written in FORTRAN.

There are many functions in MPI, out of which only a handful have been 
utilized here. Here, in brief, we describe the functions that we have used. 
More details can be found in Pacheco \& Ming (1997).
% Pacheco, P. S. \& Ming, W. C., {\textit{MPI Users' Guide in FORTRAN}}, (1997).

\textbf{MPI\_init and MPI\_finalize} -- The first function must be called 
before any other MPI function is called. After a program has finished using MPI 
library, the second function must be called. The syntaxes are as follows: 

\indent MPI\_init(Ierr) \\
\indent MPI\_finalize(Ierr) \\

The argument contains an error code. This argument is generally used in 
every FORTRAN MPI routines. 

\textbf{MPI\_COMM\_Rank} -- This function gives rank to the processors 
being used. Its syntax is as follows: 

\indent MPI\_COMM\_Rank(Comm, Myrank, Ierr) \\

The first argument is a communicator. A communicator is a collection of 
processors that communicate among themselves and take part in message 
passing when the program begins execution. In our program, we have used 
the communicator `MPI\_COMM\_WORLD'. It is pre-defined in MPI and 
consists of all the processors running when program execution begins.

\textbf{MPI\_COMM\_SIZE} -- This function is used to know the number 
of processors that is executing the program. Its syntax is: 

\indent MPI\_COMM\_SIZE(Comm, P, Ierr) \\

`P' gives the number of processors. This number may be used for various 
purposes. In our program, we have explicitly used this number while message-passing. 

\textbf{MPI\_Send and MPI\_Recv} -- These are the two functions we 
find mostly used in message-passing between different processors. 
The syntaxes are as follows:

\indent MPI\_Send(Message, Count, Datatype, Dest, Tag, Comm, Ierr) \\
\indent MPI\_Recv(Message, Count, Datatype, Source, Tag, Comm, Status, Ierr) \\

The first one sends a message to a designated processor, whereas the second
one receives from a processor. The contents of `Message' is stored in a 
block of memory referenced  by the argument message. `Message' may be a 
single number, an array of numbers or characters. `Count' and `Datatype' 
are the count values and the \textit{MPI} type datatype of `Message', respectively.
This type is not the Fortran type. In the following list, some of the MPI 
types and their corresponding Fortran types are listed.

\hspace{1cm}
\begin{center}
\begin{tabular}[h]{|c|c|}
\hline
MPI datatype & Fortran datatype \\
\hline
MPI\_integer & Integer \\
MPI\_real    & Real    \\
MPI\_double\_precision & Double precision \\
MPI\_complex & Complex \\
\hline
\end{tabular}
\end{center}
\hspace{1cm}

`Dest' and `Source' are two integer variables to mark the rank of the 
destination and the source processors of `Message', respectively. The 
`Tag' is also an integer that is used to distinguish messages received 
from a single processor.

\textbf{MPI\_Barrier} -- This function provides a mechanism for synchronizing 
all the processors in the communicator. Each processor pauses until every 
processors in communicator have called this function. It has the following syntax:

\indent MPI\_Barrier(Comm, Ierr) \\

\textbf{MPI\_Bcast} -- This is a \emph{collective communication} function, 
meaning the communication where usually all the processors are involved. 
Using this command a single processor can send the same data to every 
processors in a single call. The `Send-Recv' commands usually involve two 
processors - one sender and other receiver, whereas in collective 
communications like \emph{broadcast}, number of sender is one but 
receivers are all the other processors in the communicator. 
The syntax is as follows: 

\indent MPI\_Bcast(Message, Count, Datatype, Root, Comm, Ierr) \\

All the processors in a communicator have to call this function with the 
same argument `Root'. The contents of `Message' in processor `Root' is 
broadcasted to all the processors. 

\textbf{MPI\_Reduce} -- This is another collective communication
function. This is a global reduction operation in which all the processors
contribute data which is combined using a binary operation. The typical
binary operations are addition, max, min, product etc. The syntax is as
follows:
 
\indent MPI\_Reduce(Operand, Result, Count, Datatype, Operation, Root, Comm, Ierr)\\

MPI\_Reduce combines `Operand' stored in different processors to `Results'
in `Root' using operation `Operation'.
In the following table, we present some predefined operations.

\hspace{1cm}
\begin{center}
\begin{tabular}[h]{|c|c|}
\hline
Operation Name & Meaning \\
\hline
MPI\_sum & Addition \\
MPI\_max & Maximum \\
MPI\_min & Minimum \\
MPI\_prod & Product \\
\hline
\end{tabular}
\end{center}
\hspace{1cm}

\textbf{MPI\_Gather} -- This collective communication function is used to
gather data in one processor from all other processors. The syntax is as
follows:

MPI\_Gather(Send\_Val, Send\_Count, Send\_Type, Recv\_Val, Recv\_Count,
Recv\_Type, Root, Comm, Ierr)\\

Each processor sends the contents of `Send\_Val' to processor `Root' and 
the `Root' concatenates the received data in processor rank order, i.e.,
data from processor 0 is followed by data from processor 1, which is 
followed by processor 2 and so on.

\section{Spectral properties of TCAF in presence of a jet}

We use the above mentioned Monte Carlo code to study the
spectral properties of a TCAF in presence of a jet around a galactic black
hole (Ghosh, Garain, Chakrabarti \& Laurent 2010, hereafter GGCL10) .
% Ghosh H., Garain S. K., Chakrabarti S. K., Laurent P., 2010, IJMPD, 19, 607
Computation of the spectral characteristics have so far concentrated
only on the advective accretion flows (CT95; Chakrabarti \& Mandal 2006;
Dutta \& Chakrabarti 2010) 
% Chakrabarti S. K., Mandal S., 2006, ApJ, 642, L49
% Dutta B. G., Chakrabarti S. K., 2010, MNRAS, 404, 2136
and the jet was not included. In the Monte Carlo simulations of 
Laurent \& Titarchuk (2007) outflows in isolation were considered
and  not in conjunction with inflows.
In the following work, we obtain the outgoing spectrum in presence of
both inflows and outflows (GGCL10). We also include a Keplerian disk inside an
advective flow which is the source of the soft photons. We show how the spectrum
depends on the flow parameters of the inflow, such as the accretion rates 
and the  shock strength. These results, as such, were anticipated earlier
(C99; Das, Chattopadhyay, Nandi \& Chakrabarti 2001). 
% Chakrabarti S. K., 1999, A\&A, 351, 185
% Das, S., Chattopadhyay, I., Nandi, A., Chakrabarti, S. K., 2001, A\&A, 379, 683.
The post-shock region being denser and hotter, 
behaves as the so-called `Compton cloud' in the classical model of 
Sunyaev and Titarchuk (1980) and is known as the CENtrifugal
pressure supported BOundary Layer or CENBOL. The variation of the size of 
the Compton cloud, and therefore the basic Comptonized component of the 
spectrum is thus a function of the basic parameters of the flow, 
such as the energy, accretion rate and the angular momentum. Since the 
intensity of soft photons determines the Compton cloud temperature, the 
result depends on the accretion rate of the Keplerian component also.
In our result, we see the effects of the bulk motion Comptonization 
(CT95) because of which even a cooler CENBOL produces a harder spectrum. 
At the same time, the effects of down-scattering due the outflowing 
electrons are also seen, because of which even a hotter CENBOL causes 
the disk-jet system to emit lesser energetic photons. 

\subsection{Simulation set up}
\begin{figure}
\centering{
\includegraphics[height=5.truecm,angle=0]{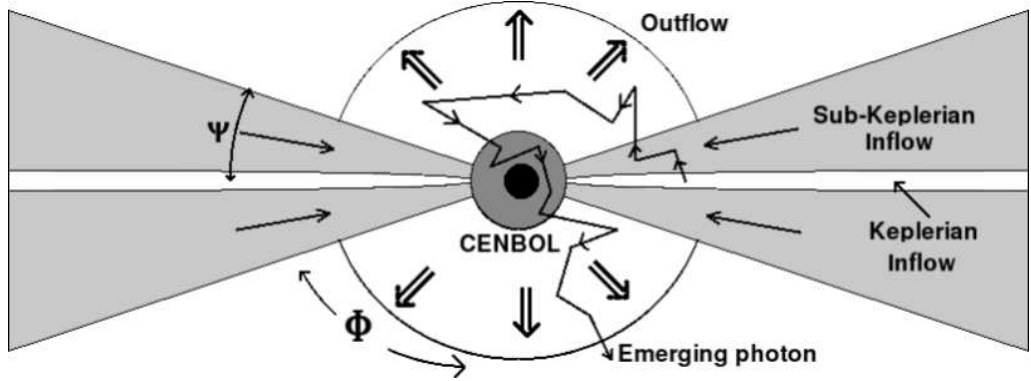}}
\caption{A schematic diagram of the geometry of our Monte Carlo Simulations
presented in this Chapter. The spherical inflowing post-shock region
surrounds the black hole and it is surrounded by the Keplerian disk on the
equatorial plane. A diverging conical outflow is present on the top of the
post-shock region. A tenuous sub-Keplerian flow above and below the Keplerian
disk is also present (CT95). Typical photon scatterings are shown by 
zig-zag paths (GGCL10).}
\label{fig2.2}
\end{figure}

In Fig. \ref{fig2.2}, we present a schematic diagram of our simulation set up
(GGCL10). The components of the hot electron clouds, namely, the CENBOL, the jet 
and the sub-Keplerian flow, intercept the soft photons emerging out of the 
Keplerian disk and reprocess them via inverse-Compton scattering. A photon 
may undergo a single, multiple or no scattering at all with the hot 
electrons in between its emergence from the Keplerian disk and its 
detection by the telescope at a large distance. The photons which enter 
the black holes are absorbed. The CENBOL, though toroidal in nature (CT95),
is chosen to be of spherical shape for simplicity. The sub-Keplerian 
inflow in the pre-shock region is assumed to be of wedge shape of a 
constant angle. The outflow, which emerges from the CENBOL in this picture 
is also assumed to be of constant conical angle.

\subsection{Temperature, velocity and density profiles inside the Compton cloud}

We assume the black hole to be non-rotating and we use the pseudo-Newtonian
potential (PW80) to describe the geometry around a black hole (Chapter 1). This 
potential is $-\frac{1}{2(r-1)}$ ($r$ is in the unit of Schwarzschild 
radius $r_g=2GM_{bh}/c^2$). Velocities and angular momenta are measured accordingly. 

As a simple example, we use the Bondi accretion and wind solutions to compute the density, 
velocity and temperature in the inflowing (inside sub-Keplerian inflow 
and CENBOL) and outflowing regions of the CENBOL, respectively (GGCL10). Bondi solution 
(Bondi 1952) was originally done to describe the accretion of 
% Bondi, H., 1952, MNRAS 112, 195.
matter which is at rest at infinity onto a star at rest. The motion of matter is steady and spherically symmetric. 
The equation of the motion of this matter around the black hole in the steady state is given by (C90),
$$
u\frac{du}{dr}+\frac{1}{\rho}\frac{dP}{dr} +\frac{1}{2(r-1)^2}=0.
$$
Here, $u$ is the velocity, $\rho$ is the density and $P$ is the thermal pressure.
This is the Eulerian equation written in the spherical polar coordinate
system $[r, \theta, \phi]$. $\theta$ and $\phi$ derivatives have been
removed because of spherical symmetry and the time derivative has been
removed since we consider the steady state.
Integrating this equation, we get the expression of the conserved 
specific energy as (C90),
\begin{equation}\label{eqno2.1}
\epsilon=\frac{u^2}{2}+na^2-\frac{1}{2(r-1)}.
\end{equation}
Here, $a$ is the adiabatic sound speed, given by $a=\sqrt{\gamma P/\rho}$, 
$\gamma$ being the adiabatic index and is equal to $\frac{4}{3}$ in our 
case. The conserved mass flux equation, as obtained from the continuity 
equation, is given by (C90),
\begin{equation}\label{eqno2.2}
\dot{M}=\Omega\rho u r^2,
\end{equation}
where, $\Omega$ is the solid angle subtended by the flow. 
For an inflowing matter, $\Omega$ is given by,
$$
\Omega_{in}=4\pi \sin\Psi,
$$
where, $\Psi$ is the half-angle of the conical inflow (see, Fig. \ref{fig2.2}). 
For the outgoing matter, the solid angle is given by,
$$
\Omega_{out}=4\pi(1-\cos\Phi),
$$
where, $\Phi$ is the half-angle of the conical outflow (see, Fig. \ref{fig2.2}). 
From Eq. (\ref{eqno2.2}), we get
\begin{equation}\label{eqno2.3}
\dot{\mu}=a^{2n}ur^2.
\end{equation}
The quantity $\dot{\mu}=\frac{\dot{M}\gamma ^n K^n}{\Omega}$ is called the 
entropy accretion rate (Chakrabarti 1989b; C90), 
$K$ being the constant measuring the entropy of the flow,
and $n=\frac{1}{\gamma -1}$ is called the polytropic index. We take 
derivative of Eq. (\ref{eqno2.1}) and (\ref{eqno2.3}) with respect 
to $r$ and eliminating $\frac{da}{dr}$ from both the equations (C90), we get 
the gradient of the velocity as,
\begin{equation}\label{eqno2.4}
\frac{du}{dr}=\frac{\frac{1}{2(r-1)^2}-\frac{2a^2}{r}}{\frac{a^2}{u}-u}.
\end{equation}
This equation is solved numerically using 4$^{th}$ order Runge Kutta 
method. Solving these equations, we obtain the radial variations of $u$, 
$a$ and finally the temperature profile using 
$T(r)=\frac{\mu a^2(r) m_p}{\gamma k_B}$,
where $\mu=0.5$ is the mean molecular weight, $m_p$ is the proton mass 
and $k_B$ is the Boltzmann constant. Using Eq. (\ref{eqno2.2}), we calculate 
the mass density $\rho$, and hence, the number density variation of 
electrons inside the Compton cloud. We ignore the electron-positron 
pair production inside the cloud.

The flow is supersonic in the pre-shock region and sub-sonic in the  
post-shock (CENBOL) region. The shock forms at the location of the 
CENBOL surface (CT95). We chose this surface at a location where the Mach 
number $M=2$. This location depends on the specific energy $\epsilon$.
In our simulation, we have chosen $\epsilon = 0.015$ so that we get 
$R_{s} = 10$ (GGCL10). We simulated a total of six cases: for Cases 1(a-c), 
we chose the halo accretion rate $\dot{m}_h = 1$ and the disk accretion 
rate $\dot{m}_d = 0.01$, and for Cases 2(a-c), 
the values are listed in Table 2 (GGCL10). The velocity variation of the 
sub-Keplerian flow is the inflowing Bondi solution (pre-shock point). 
The density and the temperature of this flow have been calculated according 
to the above mentioned formulas. Inside the CENBOL, both the Keplerian 
and the sub-Keplerian components are merged. The velocity variation of the 
matter inside the CENBOL is assumed to be the same as the Bondi accretion 
flow solution reduced by the compression ratio $R$ due to the shock. 
The compression ratio (i.e., the ratio between the post-shock and 
pre-shock densities) $R$ is also used to compute the density and the 
temperature profile. 

\begin{figure}
\centering{
\includegraphics[height=9.0truecm,angle=270]{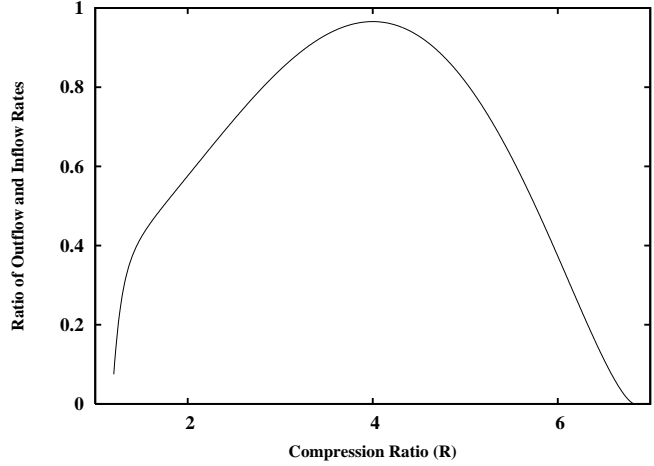}}
\caption{Ratio of the outflow and the inflow rates as a function of the 
compression ratio of the inflow when the outflow is adiabatic.
The jet angle $\Phi\sim60^\circ$ was used (GGCL10).}
\label{fig2.3}
\end{figure}

When the outflow is adiabatic, the ratio of the outflow to the inflow 
rate is given by (Das et al., 2001), 
\begin{equation}\label{eqno2.5}
R_{\dot{m}} = \frac{\Omega_{out}}{\Omega_{in}} \left( \frac{f_0}{4 \gamma}\right)^3 
\frac{R}{2} \left[ \frac{4}{3} \left( \frac{8(R-1)}{R^2} -1 \right) \right]^{3/2},
\end{equation}
where, $f_0 = {R^2 \over R-1}$ and we have used $n=3$ for a relativistic flow. 
Using this and the velocity variation obtained from the wind branch of 
the Bondi solution, we compute the density variation inside the jet. 
In our simulation, we have used $\Phi = 58^\circ$ and $\Psi = 32^\circ$ (GGCL10). 
Figure \ref{fig2.3} shows the variation of the percentage of matter 
in the outflow for these particular parameters.

\subsection{Keplerian disk}
The soft photons are produced from a Keplerian disk whose inner edge 
coincides with the CENBOL surface, while the outer edge is located at 
$500 r_g$. The source of soft photons has a multi-color blackbody 
spectrum coming from a standard disk (SS73). We assume the disk to 
be optically thick and the opacity due to free-free absorption is more 
important than the opacity due to scattering. The emission is blackbody 
type with the local surface temperature (Eq. \ref{eqno1.1}):
$$
T(r)= 5 \times 10^7 (M_{bh})^{-1/2}(\dot{m_d}_{17})^{1/4} (2r)^{-3/4}
 \left[1- \sqrt{\frac{3}{r}}\right]^{1/4} \mathrm{~K.} 
$$

Photons are emitted from both the top and bottom surfaces of the disk at
each radius. Total number of photons emitted from the disk 
surface at a radius $r$ is obtained by integrating over all frequencies 
($\nu$) and is given by,
\begin{equation}\label{eqno2.6}
n_\gamma(r) = \frac{4\pi}{c^2} \left(\frac{kT}{h}\right)^3 \times 1.202
\mathrm{~cm^{-2}~s^{-1}.}
\end{equation}
Thus, the disk between radius $r$ to $r+\delta r$ produces $dN(r)$ number
of soft photons:
\begin{equation}\label{eqno2.7}
dN(r) =  4\pi r\delta r n_{\gamma}(r) \mathrm{~s^{-1}}.
\end{equation}

The soft photons are generated isotropically between the inner and 
the outer edges of the Keplerian disk. Their positions are randomized using 
the above distribution function (Eq. \ref{eqno2.7}). All the results 
of the simulations presented here have used the number of injected 
photons as $6.4\times10^8$. We chose $M_{bh} = 10 M_\odot$ and 
$\delta r = 0.5 r_g$ (GGCL10).

\subsection{Simulation procedure}
The simulation procedure is the same as used in Ghosh at al. (2009) and GGCL10.
To begin a Monte Carlo simulation, we generate photons from the Keplerian 
disk with randomized locations as mentioned in the earlier Section. 
The energy of the soft photons at radiation temperature $T(r)$ is 
calculated using the Planck's distribution formula, where the number 
density of the photons [$n_\gamma(E)$] having an energy $E$ is expressed 
by (PSS83),
$$
n_\gamma(E) = \frac{1}{2 \zeta(3)} b^{3} E^{2}(e^{bE} -1)^{-1}, 
$$
where, $b = 1/k_BT(r)$ and $\zeta(3) = \sum^\infty_1{l}^{-3} = 1.202$, the
Riemann zeta function. Using another set of random numbers, we obtain the
direction of the injected photon and with yet another random number, we
obtain a target optical depth $\tau_c$ at which the scattering takes place.
The photon is followed within the sub-Keplerian matter till the total optical 
depth ($\tau$) reached $\tau_c$. The increase in optical depth ($d\tau$) 
during its traveling of a path of length $dl$ inside the sub-Keplerian 
matter is given by: $d\tau = \rho_n \sigma dl$, where $\rho_n$ is the 
electron number density.

The total scattering cross section $\sigma$ is given by Klein-Nishina formula:
$$
\sigma = \frac{2\pi r_{e}^{2}}{x}\left[ \left( 1 - \frac{4}{x} - 
\frac{8}{x^2} \right) ln\left( 1 + x \right) + \frac{1}{2} + \frac{8}{x} - 
\frac{1}{2\left( 1 + x \right)^2} \right],
$$
where, $x$ is given by,
$$
x = \frac{2E}{m_e c^2} \gamma \left(1 - \mu \frac{v}{c} \right).
$$
Here, $r_{e} = e^2/m_ec^2$ is the classical electron radius and $m_e$ is the mass 
of the electron.

We have assumed here that a photon of energy $E$ and momentum 
$\frac{E}{c}\boldsymbol{\Omega}$ is scattered by an electron of energy 
$\gamma m_ec^{2}$ and momentum ${\boldsymbol{p}} = \gamma m_e \boldsymbol{v}$,
with $\gamma = \left( 1 - \frac{v^2}{c^2}\right)^{-1/2}$ 
and $\mu = \boldsymbol{\Omega.v}$.
At the point where a scattering is allowed to take place, the photon 
selects an electron and the energy exchange is computed using the 
Compton or inverse-Compton scattering formula. The electrons are 
assumed to obey relativistic Maxwell distribution inside the sub-Keplerian 
matter. The number $dN(\boldsymbol{p})$ of Maxwellian electrons having 
momentum between $\boldsymbol{p}$ to $\boldsymbol{p} + d\boldsymbol{p}$
is expressed by,
$$
dN(\boldsymbol{p}) \propto exp[-(p^2c^2 + m_e^2c^4)^{1/2}/kT_e]d\boldsymbol{p}.
$$

\subsection{Results and discussions}

In a given simulation, we assume one Keplerian disk rate ($\dot{m}_d$) 
and one sub-Keplerian halo rate ($\dot{m}_h$) (GGCL10). The specific energy of the 
halo provides hydrodynamic (such as number density of the electrons 
and the velocity variation) and the thermal properties of matter. The 
shock location of the CENBOL is chosen where the Mach number $M=2$ 
for simplicity and the compression ratio $R$ (i.e., jump in density) 
at the shock is assumed to be a free parameter.

\begin{figure}
\centering{
\vskip 2.0cm
\includegraphics[height=15truecm,angle=270]{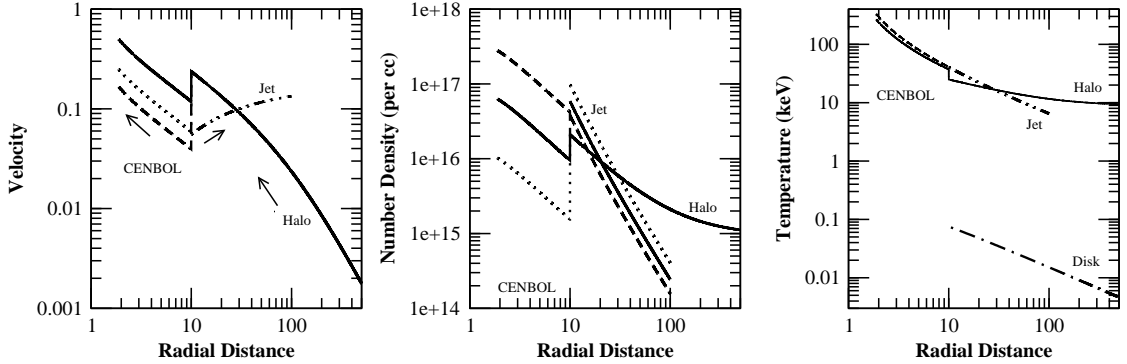}}
%\vskip -6.0cm
\caption{(a-c): Velocity (left), density (middle) and the temperature 
(right) profiles of the Cases 1(a-c) as described in Table 1
are shown with solid ($R=2$), dotted ($R=4$) and dashed ($R=6$) curves.
$\dot{m_d} = 0.01$ and $\dot{m_h} = 1$ were used (GGCL10).}
\label{fig2.4}
\end{figure}

In Fig. \ref{fig2.4}(a-c), we present the velocity, electron number 
density and temperature variations as a function of the radial distance 
from the black hole for specific energy $\epsilon=0.015$. 
$\dot{m_d} = 0.01$ and $\dot{m_h} = 1$ were chosen. Three cases were 
run by varying the compression ratio $R$. These are given in Column 2 
of Table 1. The corresponding percentage of matter going in the
outflow is also given in Column 2. In the left panel, the bulk velocity 
variation is shown. Since we chose the pseudo-Newtonian potential, 
the radial velocity is not exactly unity at $r=1$, the horizon,
but it becomes unity just outside. In order not to over estimate the 
effects of bulk motion Comptonization which is due to the momentum 
transfer of the moving electrons to the photons, we shift the horizon 
just outside $r=1$ where the velocity is unity. The solid, dotted and 
dashed curves are the velocity for $R = 2$ (Case 1a), $4$ (Case 1b) 
and $6$ (Case 1c) respectively. The same line style is used in other 
panels. The velocity variation within the jet does not change with $R$, 
but the density (in the unit of $cm^{-3}$) does (middle panel).
The double dot-dashed line gives the velocity variation of the matter 
within the jet for all the above cases. The arrows show the direction 
of the bulk velocity (radial direction in accretion, vertical direction in jets). 
The last panel gives the temperature (in keV) of the electron cloud in 
the CENBOL, jet, sub-Keplerian and Keplerian disk. Big dash-dotted line 
gives the temperature profile inside the Keplerian disk.

\begin{figure}
\centering{
\vskip 2.0cm
\includegraphics[height=15truecm,angle=270]{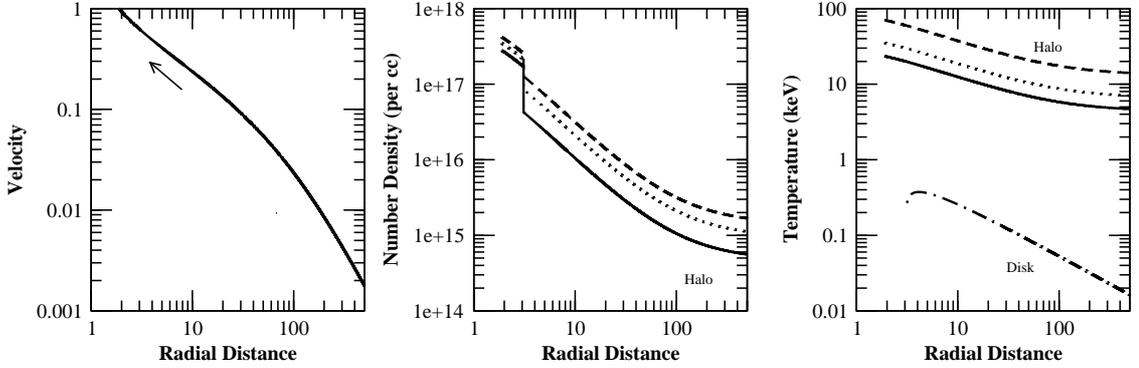}}
%\vskip -6.0cm
\caption{(a-c):
Velocity (left), density (middle) and the temperature (right) profiles 
of Cases 2(a-c) as described in Table 2 are shown with solid 
($\dot{m_h} = 0.5$), dotted ($1$) and dashed ($1.5$) curves. 
$\dot{m_d} = 1.5$ was used throughout. Velocities are the same for all 
the disk accretion rates (GGCL10).}
\label{fig2.5}
\end{figure}

In Fig. \ref{fig2.5}(a-c), we show the velocity (left), number density 
of electrons (middle) and temperature (right) profiles of Cases 2(a-c) 
as described in Table 2. Here we have fixed $\dot{m_d}=1.5$ and
$\dot{m_h}$ is varied: ${\dot {m_h}}=\ 0.5$ (solid), $1$ (dotted) and 
$1.5$ (dashed). No jet is present in this case ($R=1$). To study the 
effects of bulk motion Comptonization, the temperature of the electron 
cloud has been kept low for these cases. The temperature profile of the 
Keplerian disk for the above cases has been marked as `Disk' .

{\footnotesize
\begin{center}
\begin {tabular}[h]{|c|c|c|c|c|c|c|c|c|c|}
\hline
\multicolumn{10}{|c|}{Table 1 (GGCL10)}\\
\hline Case & R, $P_m$ & $N_{int}$ & $N_{cs}$ & $N_{cenbol}$ & $N_{jet}$ & $N_{subkep}$ & $N_{cap}$ & $p$ & $\alpha$\\
\hline 1a & 2, 58 & 2.7e+8 & 4.03e+8 & 1.35e+7 & 7.48e+7 & 8.39e+8 & 3.34e+5 & 63 & 0.43 \\
\hline 1b & 4, 97 & 2.7e+8 & 4.14e+8 & 2.38e+6 & 1.28e+8 & 8.58e+8 & 3.27e+5 & 65 & 1.05  \\
\hline 1c & 6, 37 & 2.7e+8 & 3.98e+8 & 5.35e+7 & 4.75e+7 & 8.26e+8 & 3.07e+5 &  62 & -0.4 \\
\hline
\end{tabular}
\end{center}
}
In Table 1, we summarize the results of all the cases in 
Fig. \ref{fig2.4}(a-c). In Column 1, various cases are marked. 
In Column 2, the compression ratio (R) and percentage $P_m$ of the total 
matter that is going out as outflow (see, Fig. \ref{fig2.4}) are 
listed. In Column 3, we show the total number of photons (out of the 
total injection of $6.4 \times 10^8$) intercepted by the CENBOL and 
jet ($N_{int}$) combined. Column 4 gives the number of photons 
($N_{cs}$) that have suffered Compton scattering inside the flow. 
Columns 5, 6 and 7 show the number of scatterings which took place 
in the CENBOL ($N_{cenbol}$), in the jet ($N_{jet}$) and in
the pre-shock sub-Keplerian halo ($N_{subkep}$) respectively. 
A comparison of them will give the relative importance of these 
three sub-components of the sub-Keplerian disk. The number
of photons captured ($N_{cap}$) by the black hole is given
in Column 8. In Column 9, we give the percentage $p$ of the total
injected photons that have suffered scattering through CENBOL and the jet.
In Column 10, we present the energy spectral index
$\alpha$ [$I(E) \sim E^{-\alpha}$] obtained from our simulations.

{\footnotesize
\begin{center}
\begin {tabular}[h]{|c|c|c|c|c|c|c|c|c|}
\hline
\multicolumn{9}{|c|}{Table 2 (GGCL10)}\\
\hline Case & $\dot{m_h}$, $\dot{m_d}$ & $N_{int}$ & $N_{cs}$ & $N_{ms}$ &
$N_{subkep}$ & $N_{cap}$ & $p$ & $\alpha_1, \alpha_2$\\
\hline 2a & 0.5, 1.5 & 1.08e+6 & 2.13e+8 & 7.41e+5 & 3.13e+8 & 1.66e+5 & 33.34 & -0.09, 0.4\\
\hline 2b & 1.0, 1.5 & 1.22e+6 & 3.37e+8 & 1.01e+6 & 6.82e+8 & 2.03e+5 & 52.72  & -0.13, 0.75\\
\hline 2c & 1.5, 1.5 & 1.34e+6 & 4.15e+8 & 1.26e+6 & 1.11e+9 & 2.29e+5 & 64.87  & -0.13, 1.3\\
\hline
\end{tabular}
\end{center}
}

In Table 2, we summarize the results of simulations where we have 
varied $\dot{m}_h$, for a fixed value of $\dot{m}_d$. In all of these 
cases no jet comes out of the CENBOL (i.e., $R=1$). In the last column, 
we list two spectral slopes $\alpha_1$ (from $10$ to $100$keV) 
and $\alpha_2$ (due to the bulk motion Comptonization). Here, $N_{ms}$ 
represents the photons that have suffered scattering between $R_g=3$ 
and the horizon of the black hole.

\begin{figure}
\centering{
\includegraphics[width=10truecm]{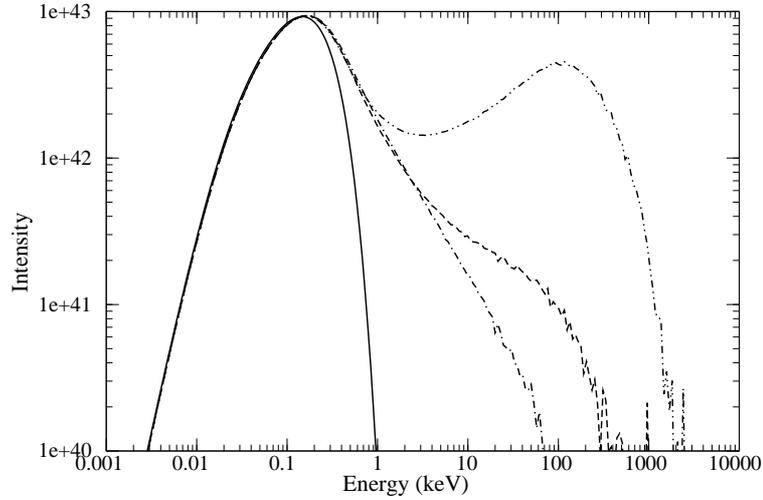}}
\caption{Variation of the emerging spectrum for different compression 
ratios. The solid curve is the injected spectrum from the Keplerian disk. 
The dashed, dash-dotted and double dot-dashed lines are for $R=2$ (Case 1a), 
$R=4$ (Case 1b) and $R=6$ (Case 1c), respectively. The disk and halo 
accretion rates used for these cases are $\dot{m}_d = 0.01$ and 
$\dot{m}_h = 1$ (GGCL10). See, text for details.}
\label{fig2.6}
\end{figure}

In Fig. \ref{fig2.6}, we show the variation of the spectrum in the 
three simulations presented in Fig. \ref{fig2.4}(a-c).
The dashed, dash-dotted and double dot-dashed lines are for 
$R=2$ (Case 1a), $R=4$ (Case 1b) and $R=6$ (Case 1c), respectively. 
The solid curve gives the spectrum of the injected photons. Since 
the density, velocity and temperature profiles of the pre-shock, 
sub-Keplerian region and the Keplerian flow are the same in all 
these cases, we find that the difference in the spectrum is mainly 
due to the CENBOL and the jet. In the case of the strongest shock
(compression ratio $R=6$), only $37\%$ of the total injected
matter goes out as the jet. At the same time, due
to the shock, the density of the post-shock region increases
by a factor of $6$. Out of the three cases, the effective density of 
the matter inside CENBOL is the highest and that inside the jet is
the lowest in this case. Due to the shock, the temperature
increases inside the CENBOL and hence the spectrum is the hardest.
Similar effects are seen for moderate shock ($R=4$) and
to a lesser extent, the low strength shock also ($R=2$).
When $R=4$, the density of the post-shock region
increases by the factor of $4$ while almost $97\%$ of total injected
matter (Fig. \ref{fig2.4}) goes out as the jet reducing the matter 
density of the CENBOL significantly. From
Table 1, we find that the $N_{cenbol}$ is the lowest and
$N_{jet}$ is the highest in this case (Case 1b). This
decreases the up-scattering and increases the down-scattering
of the photons. This explains why the spectrum is
the softest in this case (Chakrabarti 1998b). 
% Chakrabarti, S. K. 1998, Indian Journal of Physics, V. 72B, p. 565-569 (astro-ph/9810412)
In the case of low strength shock
($R=2$), $57\%$ of the inflowing matter goes out as jet, but due
to the shock the density increases by factor of $2$ in the
post-shock region. This makes the density similar to a case
as though the shock did not happen except that the temperature
of CENBOL is higher due to the shock. So the spectrum with
the shock would be harder than when the shock is not present.
The disk and the halo accretion rates used for these cases
are $\dot{m_d} = 0.01$ and $\dot{m_h} = 1$.

\begin{figure}
\centering{
\vskip 2.0cm
\includegraphics[height=5truecm]{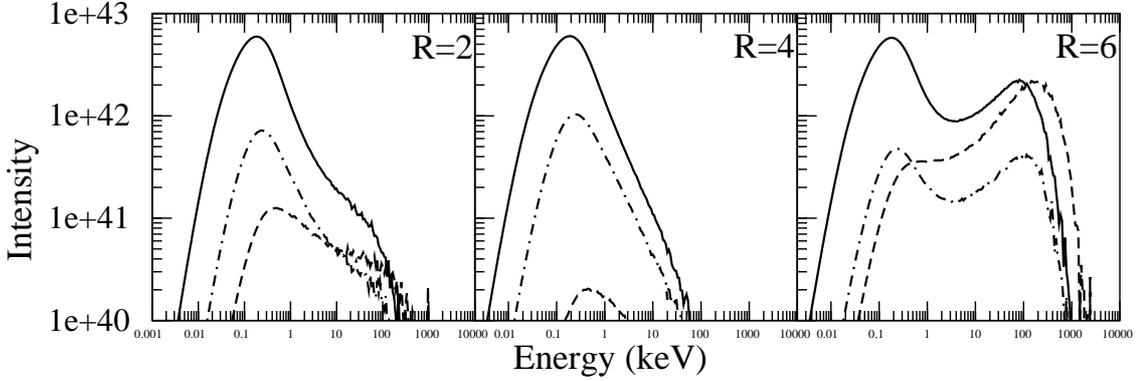}}
%\vskip -6.0cm
\caption{(a-c): Variation of the components of the
emerging spectrum with the shock strength (R).
The dashed curves correspond to the photons emerging
from the CENBOL region and the dash-dotted curves are for the photons
coming out of the jet region. The solid curve is the spectrum for all
the photons that have suffered scatterings (GGCL10). See, the text for details.
}
\label{fig2.7}
\end{figure}

In Fig. \ref{fig2.7}, we show the components of the emerging spectrum 
for all the three cases presented in Fig. \ref{fig2.6}. The solid 
curve is the intensity of all the photons which suffered at least one 
scattering. The dashed curve corresponds to the photons emerging
from the CENBOL region and the dash-dotted curve is for the photons
coming out of the jet region. We find that the spectrum from
the jet region is softer than the spectrum from the CENBOL. As $N_{jet}$
increases and  $N_{cenbol}$ decreases, the spectrum from the jet becomes
softer because of two reasons. First, the temperature of the jet is lesser
than the CENBOL, so the photons get lesser amount of energy from thermal
Comptonization. Second, the photons are
down-scattered by the outflowing jet which eventually make the spectrum
softer. We note that a larger number of photons are present in the spectrum
from the jet than the spectrum from the CENBOL, which shows the photons
have actually been down-scattered. The effect of down-scattering is larger 
when $R=4$. For $R=2$ also there is significant amount of
down scattered photons. But this number is very small for the case $R=6$
as $N_{cenbol}$ is much larger than $N_{jet}$. So most of the photons are
up-scattered. The difference between the total (solid) and the sum of the
other two regions gives an idea of the contribution from the sub-Keplerian
halo located in the pre-shock region. In our choice of geometry (half 
angles of the disk and the jet), the contribution of the pre-shock flow 
is significant. In general it could be much less. This is especially true 
when the CENBOL is farther out.

\begin{figure}
\centering{
\includegraphics[width=10truecm]{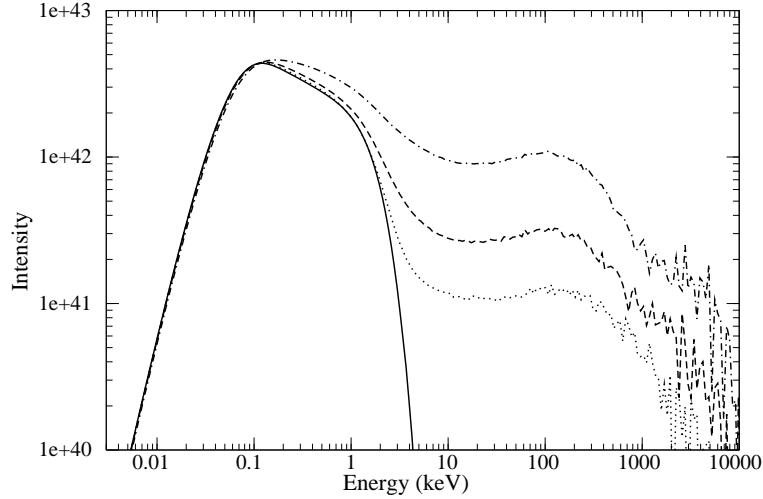}}
\caption{The spectrum which includes the effects of bulk motion Comptonization.
Solid (Injected), dotted ($\dot{M_h} = 0.5$), dashed ($\dot{M_h} = 1$),
dash-dotted ($\dot{M_h} = 1.5$). $\dot{M_d} = 1.5$ for all the cases.
Keplerian disk extends up to $3.1 r_g$. Table 2 summarizes the
parameter used and the simulation results for these cases (GGCL10).}
\label{fig2.8}
\end{figure}

\begin{figure}
\centering{
\vskip 2.0cm
\includegraphics[height=5truecm]{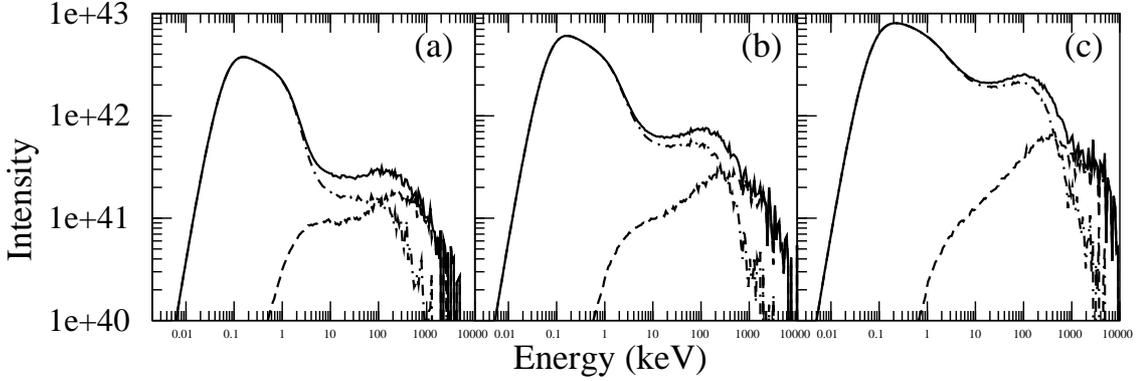}}
%\vskip -6.0cm
\caption{Components of the emerging spectrum for the Cases 2(a-c).
Solid curves are the spectra of all the photons that have suffered
some scattering. The dashed and dash-dotted curves are the spectra
of photons which are emitted from inside and outside of the
marginally stable orbit ($3 r_g$) respectively. The photons emitted
inside the marginally stable radius are Comptonized by the
bulk motion of the infalling matter. Here the jet is absent (GGCL10).
}
\label{fig2.9}
\end{figure}

In Fig. \ref{fig2.8}, the emerging spectra due to the bulk motion
Comptonization is shown when the halo rate is varied. The solid curve is the 
injected spectrum (modified blackbody). The dotted, dashed and 
dash-dotted curves are for $\dot{m}_h = 0.5, \ 1$ and $1.5$, respectively. 
$\dot{m_d} = 1.5$ for all the cases. The Keplerian disk extends up to 
$3 r_g$. Table 2 summarizes the parameters used and the results of the 
simulations. As the halo rate increases, the density of the CENBOL also
increases causing a larger number of scatterings. 
From Fig. \ref{fig2.5}(a), we notice that the bulk velocity variation 
of the electron cloud is the same for all the four cases. Hence, the 
case where the density is maximum, the photons got energized to a very 
high value due to repeated scatterings with that high velocity cold matter.
As a result, there is a hump in the spectrum around 100 keV energy for 
all the cases. We find the signature of two power-law regions in the 
higher energy part of the spectrum. The spectral indices are  given 
in Table 2. It is to be  noted that $\alpha_2$ is increased with 
$\dot{m}_h$ and becomes softer for higher $\dot{m}_h$.
Our geometry here at the inner edge is conical, unlike a sphere 
in Laurent and Titarchuk (2001). This may 
be the reason why our slope is not the same as in 
Laurent and Titarchuk (2001) where $\alpha_2=1.9$. In Fig. \ref{fig2.9},
we present the components of the emerging spectra. 
As in Fig. \ref{fig2.7}, solid curves are the spectra of all the 
photons that have suffered scattering. The dashed and dash-dotted 
curves are the spectra of photons emitted from inside and outside of the
marginally stable orbit ($3 r_g$), respectively. The photons emitted
inside the marginally stable radius are Comptonized by the 
bulk motion of the converging infalling matter and produce the 
power-law tail whose spectral index is given by $\alpha_2$ (Table 2).

%%%%%%%%%%%%%%%%% ADDED %%%%%%%%%%%%%%%%%%%%%%%%%%%%%
As mentioned earlier, to compute the spectral properties of the TCAF in this Chapter,
we have used the pseudo-Newtonian potential to describe the geometry around a
non-rotating black hole. On the other hand, if we choose a rotating (Kerr)
black hole, the space-time geometry will change which will affect the structure
of the accretion disk, which in turn may affect spectral properties. Computation
of topologies of global viscous transonic accretion flow has been done by
Chakrabarti (1996a, 1996b) by solving the general relativistic equations and by
% Chakrabarti S. K., 2006, MNRAS, 283, 325
% Chakrabarti S. K., 2006, ApJ, 471, 237 
Chakrabarti \& Mondal (2006) and Mondal \& Chakrabarti (2006) using pseudo-Kerr
% Chakrabarti S. K., Mondal S., 2006, MNRAS, 369,976
% Mondal S., Chakrabarti S. K., 2006, MNRAS, 371, 1418
potential. These works show that in general, shock in accretion flow
forms closer around rotating black holes. Thus, size of CENBOL
is smaller compared to that in case of a non-rotating black hole and we expect the
number of intercepted soft photons by the CENBOL to be lower. Also, the inner
edge of the Keplerian disk will extend to much closer to the black hole
when it is rotating for the same set of flow parameters. This may result in increase in number of soft photons, though
the overall luminosity may increase. However, as the
shock forms closer, post-shock region becomes hotter and may thus contribute
to very high energy X-rays and gamma-rays (Chakrabarti 1996a). Therefore, as a result
of all these effects, we expect overall softer spectra with powerlaw tail extending to
higher energies. However, detailed study of the spectral properties from the accretion
disk around a rotating black hole is necessary.
\clearpage
%\end{document}
%@@@@@@@@@@@@@@@@@@@@@@@@@@@@@@@@@@@@@@@@@@@@@@@@@@@@@

	\reseteqn
	\resetsec
	\resetfig
	\resettab

%~~~~~~~~~~~~~~~~~~~~~~~~~~~~~~~~~~~~~~~~~~~~~~~~~~~~~~~~~~~~~~~~~~~~~~~~~~~~~~
\alpheqn
\resec
\refig
\retab
%%%%%%%%%%%%%%%%%%%%%%%%%%%%%%%%%%%%%%%%%%%%%%%%%%%%%%%%%%%
% Chapter 3 : SPECTRAL PROPERTIES
%%%%%%%%%%%%%%%%%%%%%%%%%%%%%%%%%%%%%%%%%%%%%%%%%%%%%%%%%%%

\def\k{{\bf k}}
\def\aug{{\tilde{\cal H}}}
\def\ahat{{\mathaccent "7E {\cal A}}}
\def\bhat{{\mathaccent "7E {\cal B}}}
\def\chat{{\mathaccent "7E {\cal F}}}
\def\dhat{{\mathaccent "7E {\cal D}}}
\def\shat{{\mathaccent "7E {\cal S}}}
\def\proj{{\cal P}}
\def\ket{\vert \vert  \{ \emptyset \} \rangle}
\def\ket2{\vert \vert \otimes \{ R \} \rangle}
\def\sqr{$^{2}$}
\def\T{$T$}
\def\P{$P$}
\def\Q{$Q$}
\def\G{$G$}
\def\M{{\bf M}}
\def\H{{\bf H}}
\def\I{{\bf I}}
\def\Pr{{\bf P}}
\def\Tr{{\bf T}}
\def\eq{\enskip =\enskip}
\def\pls{\enskip+\enskip}
\def\mns{\enskip -\enskip}
\def\oper{O^{\bf k}_{R}}
\def\aug{\tilde{\cal H}}
\def\envfn{\chi^{\alpha}_{RL}}
\def\struc{S^{\alpha}_{RL,R^{\prime}L^{\prime}}}
\def\rR{r_{R}}

\newpage
\markboth{\it Time Dependent Radiation Hydrodynamic Simulation}
{\it Time Dependent Radiation Hydrodynamic Simulation}
\chapter{SPECTRAL PROPERTIES USING TIME DEPENDENT RADIATION HYDRODYNAMIC SIMULATION}
%~~~~~~~~~~~~~~~~~~~~~~~~~~~~~~~~~~~~~~~~~~~~~~~~~~~~~~~~~~~~~~~~~~~~~~~~~~~~~~~

In this Chapter, we discuss the development of a time dependent 
radiation hydrodynamic simulation code. In this code, we couple a time 
dependent hydrodynamic simulation code with the Monte Carlo code 
described in the previous Chapter (Ghosh, Garain, Giri \& Chakrabarti 2011,
hereafter GGGC11; Garain, Ghosh \& Chakrabarti 2012, 2013, hereaftre 
GGC12 and GGC13, respectively). Using this radiation hydrodynamic 
simulation code, we study the time variation of the hydrodynamics as 
well as the spectral properties of a TCAF.
% Ghosh H., Garain S. K., Giri K., Chakrabarti S. K., 2011, MNRAS, 416, 959
% Garain S. K., Ghosh H., Chakrabarti S. K., 2012, ApJ, 758, 114
% Garain S. K., Ghosh H., Chakrabarti S. K., submitted, MNRAS
 
\section {Hydrodynamic simulation code}

Here we present the equations which are solved in the 
hydrodynamic simulation code and outline the procedure to build this code. 
This code is an explicit, second order, 
finite difference code which uses the 
principle of total variation diminishing (TVD). Hence it is called
TVD code. The TVD scheme was originally developed by Harten (1983).
% Harten A., 1983, Jour. of Comp. Phys., 49, 357
This code is designed to solve the hyperbolic system of conservation 
equations, like the conserved hydrodynamic equations. It is a nonlinear 
scheme obtained by first modifying the flux function and then applying 
a non-oscillatory first-order accurate scheme to get a resulting second 
order accuracy (Ryu, Brown, Ostriker \& Loeb 1995 and references therein; MRC96). 
% Ryu, D., Brown, G. L., Ostriker, J. P., \& Loeb, A. 1995, ApJ, 452, 364
Thus the key merit of this scheme is to achieve the high resolution of a 
second-order accuracy while preserving the robustness of a non-oscillatory
first-order scheme.  In this code, the following time dependent 
hydrodynamic equations (Landau \& Lifshitz 1987)
% Landau L. D., Lifshitz , 1987, Fluid Mechanics, Pergamon Press, Oxford
are solved. In presence of gravity, these equations can be written as 
follows:
\begin{itemize}
\item[i.] Continuity equation:
$$
\frac{\partial \rho}{\partial t} + \boldsymbol{\nabla .}(\rho \boldsymbol{v}) = 0,
$$
\item[ii.] Euler equation:
$$
\frac{\partial \boldsymbol{v}}{\partial t} + \boldsymbol{(v.\nabla)v}+
{1\over \rho}\boldsymbol{\nabla}P = \boldsymbol{g},
$$
\item[iii.] Energy conservation equation:
$$
\frac{\partial E}{\partial t} +\boldsymbol{ \nabla.}(\boldsymbol{v} (E+P)) = \rho \boldsymbol{v.g}.
$$
\end{itemize}
Here, $\rho$ is the density, $\boldsymbol{v}$ is the velocity, $P$ is the pressure, 
$\boldsymbol{g}$ is the acceleration due to gravity and $E$ is the energy density 
(without potential energy) given as $E = P/(\gamma -1) + \rho v^2/2$,
where, $\gamma$ is the polytropic index.  

These equations, after some rearrangements, can be written in the vector 
form in cylindrical coordinate system [$R, \theta, z$] and assuming
axisymmetry, as (MRC96)
\begin{equation}\label{eqno3.1}
\frac{\partial \boldsymbol{q}}{\partial t} + {1\over R}\frac{\partial (R\boldsymbol{F_1})}{\partial R}
+\frac{\partial \boldsymbol{F_2}}{\partial R} + \frac{\partial \boldsymbol{G}}{\partial z} 
= \boldsymbol{S},
\end{equation}
where, the state vector is 
$$
\boldsymbol{q} = \left( \begin{array}{c}
                         \rho\\
                         \rho v_R\\
                         \rho v_\theta\\
                         \rho v_z\\
                         E
			\end{array}\right),
$$
the flux functions are 
$$
\boldsymbol{F_1} = \left( \begin{array}{c}
                         \rho v_R\\
                         \rho v^2_R\\
                         \rho v_R v_\theta\\
                         \rho v_R v_z\\
                         (E+P)v_R
			\end{array}\right),
\boldsymbol{F_2} = \left( \begin{array}{c}
                         0\\
                         P\\
                         0\\
                         0\\
                         0
			\end{array}\right),
\boldsymbol{G} = \left( \begin{array}{c}
                         \rho v_z\\
                         \rho v_R v_z\\
                         \rho v_\theta v_z\\
                         \rho v^2_z + P\\
                         (E+P)v_z
			\end{array}\right),
$$
and the source function is
$$
\boldsymbol{S} = \left( \begin{array}{c}
                         0\\
      \frac{\rho v^2_\theta}{R} - \frac{\rho R}{2(\sqrt{R^2+z^2}-1)^2\sqrt{R^2+z^2}}\\
             -\frac{\rho v_R v_\theta}{R}\\
      -\frac{\rho z}{2(\sqrt{R^2+z^2}-1)^2\sqrt{R^2+z^2}}\\
      -\frac{\rho (Rv_R+zv_z)}{2(\sqrt{R^2+z^2}-1)^2\sqrt{R^2+z^2}}\\
			\end{array}\right).
$$
In the above equations, we have used $\boldsymbol{g}=-\boldsymbol{\nabla}\phi$, 
where, $\phi$ is the pseudo-Newtonian potential (PW80) given by,
$$
\phi = -{1 \over 2(r-1)},
$$
with $r=\sqrt{R^2 + z^2}$.

With the state vector $\boldsymbol{q}$ and the flux functions 
$\boldsymbol{F(q)=F_1(q)+F_2(q)}$ 
and $\boldsymbol{G(q)}$, the Jacobian matrices, 
$\boldsymbol{A(q)=\partial F/\partial q}$
and $\boldsymbol{B(q)=\partial G/\partial q}$ are formed. The system 
of equations in (\ref{eqno3.1}) are called {\em hyperbolic} if all the eigenvalues
of the Jacobian matrix are real and the corresponding set of right
eigenvectors is complete. The eigenvalues and the left and right 
eigenvectors of $\boldsymbol{A(q)}$ and $\boldsymbol{B(q)}$ are used 
to build the TVD code.

In updating the vector $\boldsymbol{q^n}$ to $\boldsymbol{q^{n+1}}$ (here
$n$ represents the time step), there are two steps in this TVD code --
one is the {\em hydrodynamic} step and the other is {\em source} step.
In the hydrodynamic step, the eigenvectors and the eigenvalues, 
mentioned above, are used to compute the fluxes at the grid boundary 
along the $R$ and $z$ directions separately.
The source terms are taken care of separately in source step. 
The detail procedure to build up the hydrodynamic and source steps 
are given in Ryu, Ostriker, Kang \& Cen (1993) and Ryu et al. (1995).
% Ryu D., Ostriker J. P., Kang H., Cen R., 1993, ApJ, 414, 1

\section{Coupling of hydrodynamic and radiative transfer codes}

Using the TVD code, described above, time variation of the hydrodynamical 
configuration of an accretion disk is simulated (MRC96; RCM97;
Giri, Chakrabarti, Samanta \& Ryu 2010).
% Giri K., Chakrabarti S. K., Samanta M. M., Ryu D., 2010, MNRAS, 403, 516
In order to include the effects of radiative cooling on the hydrodynamics, 
we couple this hydrodynamic code with the Monte Carlo code, described 
in the previous Chapter. So far, however, only bremsstrahlung  or 
pseudo-Compton cooling has been added into the time-dependent flows 
(MSC96; CAM04; Proga 2007; Proga, Ostriker, Kurosawa 2008). Using the MC code, we 
% Proga D., 2007, ApJ, 661, 693
% \bibitem[]{} Proga D., Ostriker J. P., Kurosawa R., 2008, ApJ, 676, 101
compute the Comptonized spectrum from the Compton cloud. In the so 
called TCAF scenario (CT95), soft photons, emitted from the Keplerian 
disk, are intercepted and energized by the inverse-Compton scattering 
by the high energy electrons in the halo. In this process, the electrons 
lose their energy and become cooler. In the coupled code, we include the 
effects of this cooling and see how the hydrodynamics as well as the 
spectral properties of TCAF are affected because of this.
In the following, we describe the procedure of coupling the hydrodynamic
and Monte Carlo codes in detail (GGGC11; GGC12; GGC13).

\subsection{Photon packet}

In our simulations, soft photons are emitted from the surface of a standard Keplerian
disk (SS73). We divide the Keplerian disk in different annuli of width 
$D(R)=0.5 r_g$ ($r_g=2GM_{bh}/c^2$).
Each annulus having mean radius $R$ is characterized by its average
temperature $T(R)$. The number of photons emitted from the disk
surface of each annulus can be calculated using Eq. (\ref{eqno2.7}):
$$
dN(R) =  4\pi R D(R) n_{\gamma}(R) \mathrm{~s^{-1}}.
$$
The total number of photons $N$ from the whole disk surface can be 
computed by adding up the contributions from all the annuli.
This total number $N$ is huge (e.g. $\sim~10^{46}$ per second for a disk 
extending from 3$r_g$ to 300$r_g$ and disk accretion rate $\dot{m}_d = 0.1$).
In reality, one cannot inject these many number of photons in a Monte Carlo 
simulation because of the limitation of computation time. So we replace the 
large number of photons from each annulus [$dN(R)$] by a smaller number of 
bundles of photons, say, $dN_{comp}(R)$ and calculate a weightage factor,
$$
f_W = \frac{dN(R)}{dN_{comp}(R)}.
$$
$dN_{comp}(R)$ is computed following the distribution function $dN(R)$. 
Thus each photon in the Monte Carlo simulation is assumed to be a bundle
consisting of $f_W$ number of actual photons. All the photons in a bundle
is assumed to behave in the similar way. When an injected photon is 
inverse-Comptonized (or, Comptonized) by an electron in a volume element 
of size $dV$, we assume that $f_W$ number of photons have suffered 
similar scattering with the electrons inside the volume element $dV$. 
If the energy loss (gain) per electron in this scattering is $\Delta E$, 
we multiply this amount by $f_W$ and distribute this loss (gain) among 
all the electrons inside that particular volume element. This is 
continued for all the  $N_{comp}$ bundles of photons
and the revised energy distribution is obtained. Typically, we use
$N_{comp}\sim 10^7$ in all the simulations presented in this thesis.

\subsection{Computation of the temperature profile after cooling}

Since the hydrogen plasma considered here is ultra-relativistic 
($\gamma=\frac{4}{3}$ throughout the hydrodynamic simulation), thermal 
energy per particle is $3k_BT_e$, where, $k_B$ is Boltzmann constant and 
$T_e$ is the temperature of electron. The electrons are cooled (heated up) 
by the inverse-Comptonization (Comptonization) of the soft photons emitted 
from the Keplerian disk. The protons are cooled because of the Coulomb 
coupling with the electrons. Total number of electrons inside
any box with the center at grid location $(ir,iz)$ is given by,
$$
dN_e(ir,iz) = 4\pi rn_e(ir,iz)dRdz, 
$$
where, $n_e(ir,iz)$ is the electron number density at $(ir,iz)$ grid, and
$dR$ and $dz$ represent the grid size along $R$ and $z$ directions, respectively. 
So the total thermal energy in any box is given by, 
$$
3k_BT_e(ir,iz)dN_e(ir,iz) = 12\pi rk_BT_e(ir,iz)n_e(ir,iz)dRdz,
$$ 
where, $T_e(ir,iz)$ is the temperature at $(ir,iz)$ grid. We calculate 
the total energy loss (gain) $\Delta E$ of electrons inside the
box according to what is presented above and subtract that amount to 
get the new temperature of the electrons after each scattering inside 
that box as
$$
k_BT_{e,new}(ir,iz) = k_BT_{e,old}(ir,iz)-\frac{\Delta E}{3dN_e(ir,iz)}.
$$
%%%%%%%%%%%%%%%%%%%%%%%%%

\subsection{Coupling procedure}

Once a quasi steady state is achieved using the non-radiative
hydro-code, we compute the radiation spectrum using the
Monte Carlo code as described in the previous Chapter. This is the first 
approximation of the spectrum. To include the cooling in the coupled code, 
we follow these steps (GGGC11; GGC12; GGC13):\\
(i) we calculate the velocity, density and temperature profiles of the
electron cloud from the output of the hydro-code. \\
(ii) Using the Monte Carlo code, we calculate the spectrum. \\
(iii) Electrons are cooled (heated up) by the inverse-Compton (Compton) scattering.
We calculate the amount of heat loss (gain) by the electrons and the new 
temperature and energy distributions of the flow and \\
(iv) taking the new temperature and energy profiles as initial condition,
we run the hydro-code for a time interval much shorter than the local infall time.\\ 
Subsequently, we repeat the steps (i-iv). In this way, we see how the 
spectrum is modified as the 
time proceeds. Clearly, we proceed using a two-step process. Since the 
calculation of Compton process is via Monte Carlo code (which is highly non-linear)
while the hydrodynamic simulation is by finite difference method, they 
cannot be coupled directly.

\begin{figure}
\vskip 1cm
\centering{
\includegraphics[height=6cm]{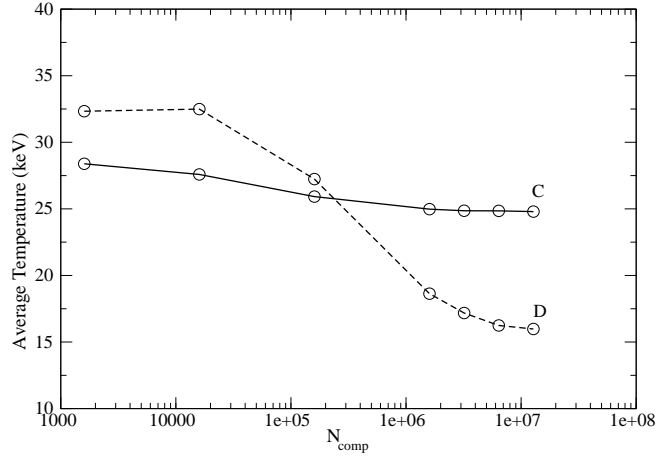}}
\caption{Variation of the average temperature (keV) of the electron 
cloud with bundle of photons $N_{comp}$ for two different Keplerian 
disk rates $\dot{m}_d$, keeping the halo rate fixed at $\dot{m}_h = 0.5$. 
Simulation cases (Table 3) are marked on each curve. Clearly, the 
temperature converges for $N_{comp} > 6\times 10^6$ (see also, GGC12).
}
\label{fig3.0}
\end{figure}

Before proceeding further, we first show that our choice of  
$N_{comp}~\sim~10^7$ produces converging results. 
Before the choice of $N_{comp} ~\sim~10^7$ was made, we varied $N_{comp}$ from 
$\sim 10^3$ to $\sim 1.5\times 10^7$ and conducted a series of runs 
to ensure that our final result does not suffer from any statistical 
effects. Fig. \ref{fig3.0} shows the average electron temperature 
of the cloud when $N_{comp}$ was varied. There is a clear convergence 
in our result for $N_{comp} > 6\times 10^6$ (see also, GGC12). 
%Thus, our choice of $N_{comp}~\sim~10^7$ is quite safe.

As mentioned earlier, we model soft photons emitted from Keplerian disks using $\rm {N_{comp}}$.
As the disk rate is increased, number of soft photons increases and to model
higher number of photons, we need higher value of $\rm {N_{comp}}$. In the present work,
we have shown that $\rm {N_{comp}} > 6\times 10^6$ models the number soft photons emitted for
the higher disk rates (Cases C and D) very well. Therefore, it is evident
that this result will remain valid for lower values of disk rates also (i.e., for other
cases presented in Table 3). For this
reason, we have shown the temperature convergence results for cases C and D only.
However, we conducted more tests in GGC12 and there we showed similar results for other disk rates.
Thus, we believe that using the value $\rm {N_{comp}} \sim 10^7$ will produce correct results for
moderate disk rates ($\sim$ 1 or lower).

%%%%%%%%%%%%%%%%%%%Fig.3.0%%%%%%%%%%%%%%%%%%%%%%%%%
%%%%%%%%%%%%%%%%%%%%%%%%%%%%%%%%%%%%%%%%%%%%%%%%%%%

\section{Spectral properties of TCAF using coupled code}

\begin{figure}
\centering{
\includegraphics[height=15truecm,angle=90]{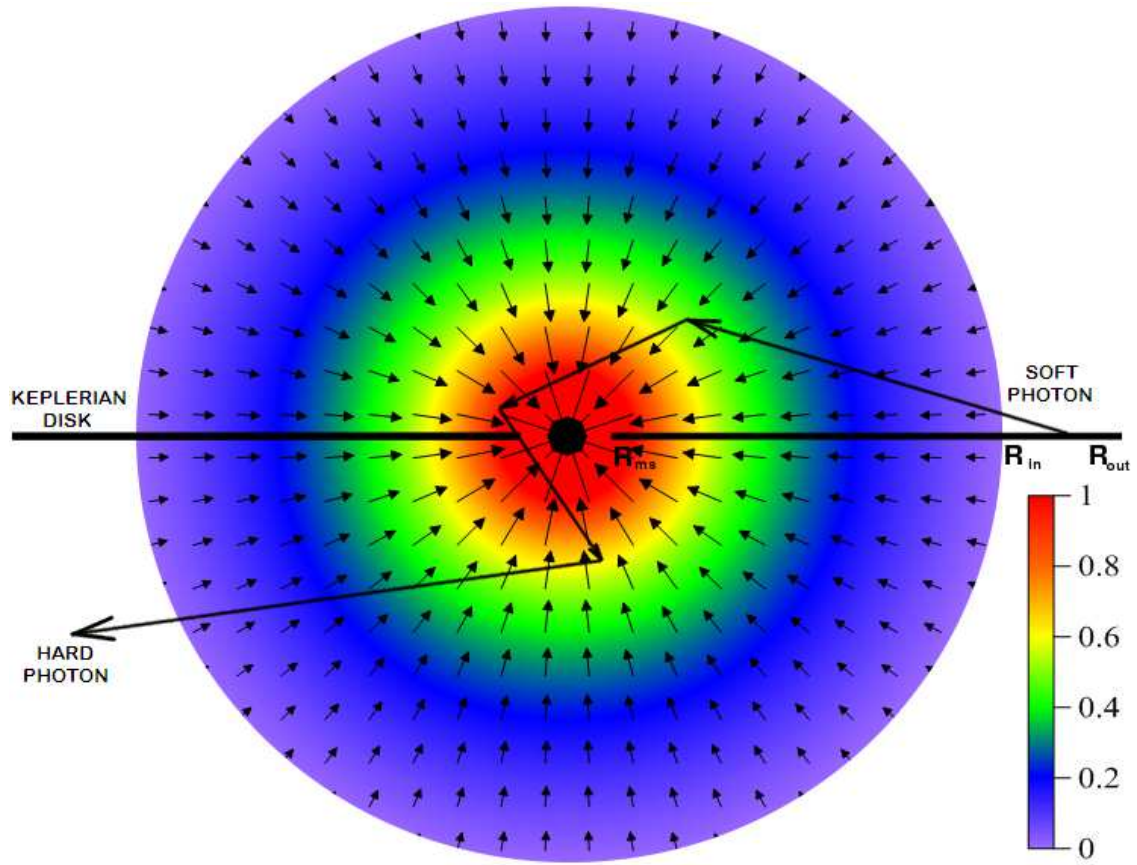}}
\vskip 2cm
\caption{Schematic diagram of the geometry of our Monte Carlo simulations 
for $\lambda=0$. The colors show the normalized density in a logarithmic
scale. Velocity vectors of the infalling electrons are also shown.
The zig-zag trajectory is the typical path followed by a photon inside 
the cloud (see also, GGGC11).}
\label{fig3.1}
\end{figure}

We study the spectral properties of the TCAF using the coupled radiation
hydrodynamic simulation code. In Fig. \ref{fig3.1}, we present the 
schematic diagram of the simulation set up (see also, GGGC11). 
%%%%%%%%%%%%%%%%%%%Fig.3.1%%%%%%%%%%%%%%%%%%%%%%%%%
%%%%%%%%%%%%%%%%%%%%%%%%%%%%%%%%%%%%%%%%%%%%%%%%%%%

We consider a spherically symmetric 
Compton cloud (halo) with zero angular momentum (specific angular momentum
$\lambda=0$) within a sphere of radius $R_{in} = 200r_g$. 
The Keplerian disk resides at the equatorial plane. The outer edge of 
this disk is  located at $R_{out} = 300r_g$ and it extends inside up 
to the marginally stable orbit $R_{ms} = 3r_g$. At the centre of the sphere, 
a non-rotating black hole of mass $10 M_{\odot}$ is kept. 
The matter is injected with spherical symmetry from the radius $R_{in}$. 
It intercepts the soft photons emerging out of the Keplerian disk and 
reprocesses them via Compton or inverse-Compton scattering.

\subsection{Density, velocity and temperature profiles inside the halo component}
A realistic accretion disk is three-dimensional. Assuming axisymmetry,
we reduce the problem to two dimensions and solve the hydrodynamic 
equations, written in two dimensions, numerically to simulate the 
accretion flow of the halo. The flow dynamics is calculated using the 
TVD code (discussed in the earlier Section).
At each time step, we carry out Monte Carlo simulation to obtain the 
cooling/heating due to Comptonization. We incorporate the cooling/heating 
of each grid while executing the next time step of hydrodynamic 
simulation, as discussed above. The numerical calculation for the 
two-dimensional flow has been carried out with $900 \times 900$ equi-spaced cells in a 
$200 r_g \times 200 r_g$ box (GGGC11). We choose the units in such a way that 
the outer boundary ($R_{in}$) is chosen to be unity and the 
matter density is normalized to become unity. 
We assume the black hole to be non-rotating and we use the 
pseudo-Newtonian potential $-\frac{1}{2(r-1)}$
(PW80) to calculate the flow geometry around the black hole
($r$ is in $r_g$). 

%%%%%%%%%%%%%%%%%%%Fig.3.2%%%%%%%%%%%%%%%%%%%%%%%%%
\begin{figure}
\centering{
\includegraphics[height=7.5truecm]{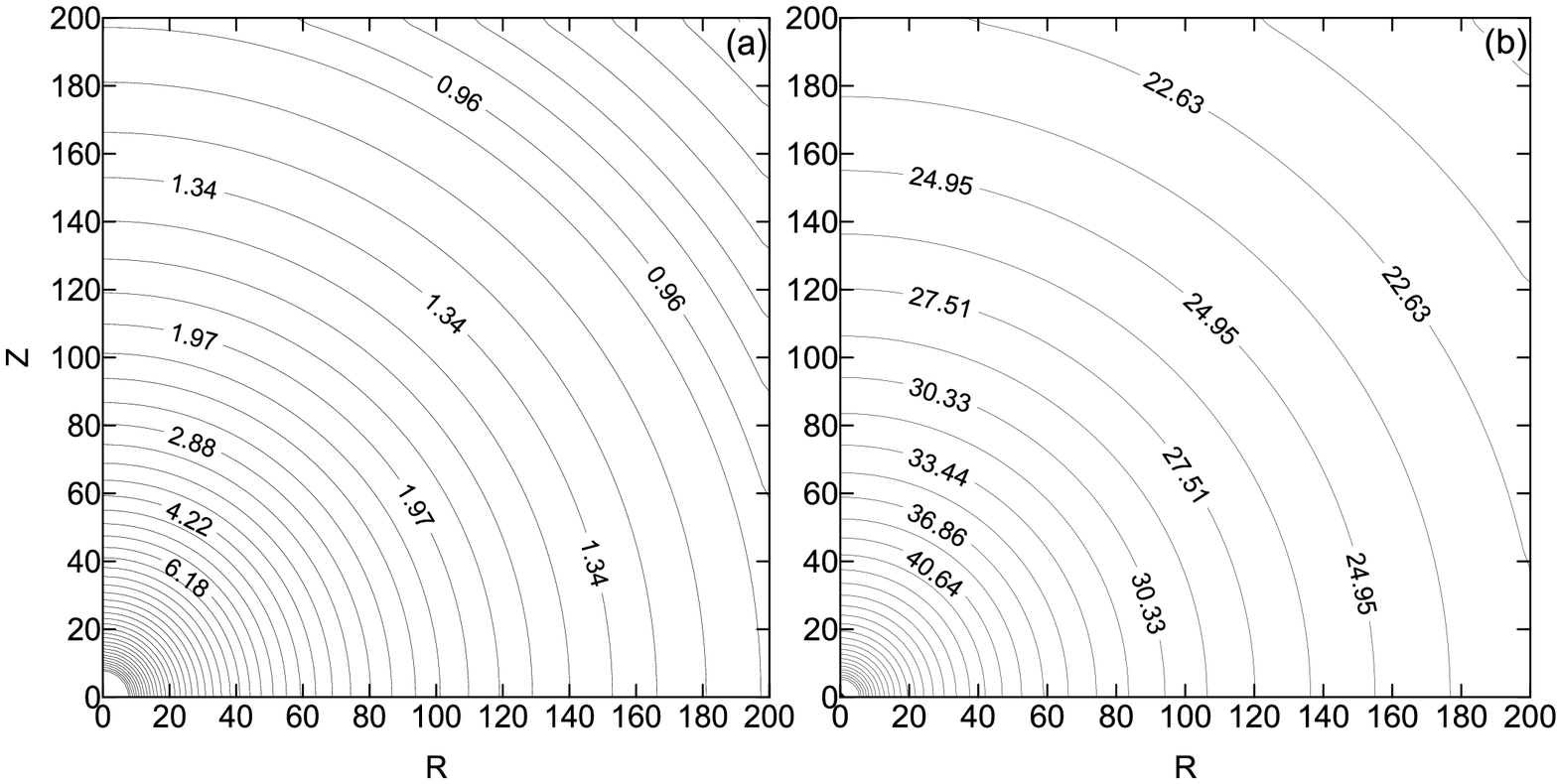}
\includegraphics[height=7.5truecm]{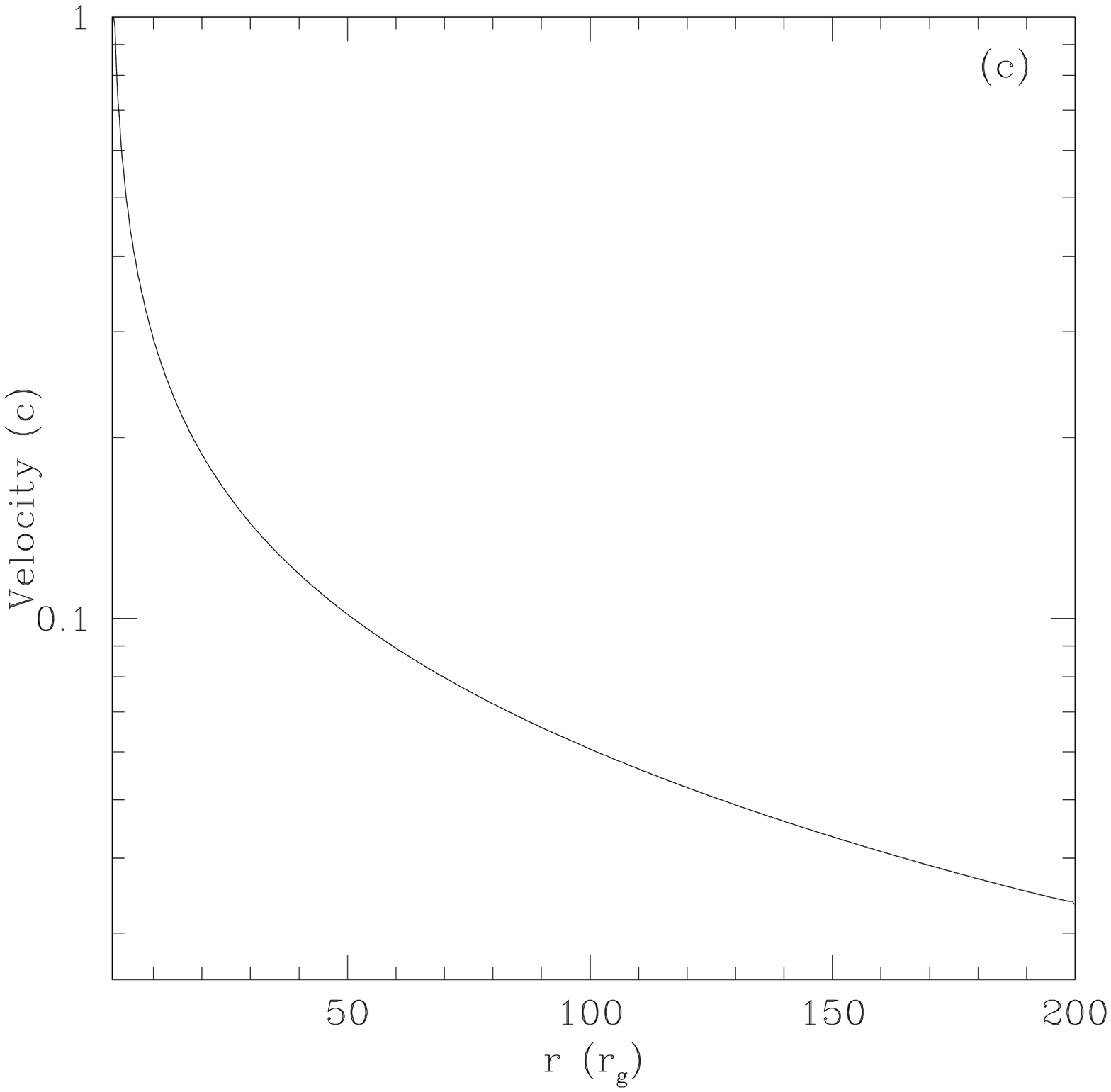}}
\caption{(a) Density and (b) temperature contours inside the spherically accreting
halo in the absence of Compton cooling. Here, densities are in normalized
units and temperatures are in keV (GGGC11).  The radial variation of the velocity
profile is also shown in (c). See text for details. }
\label{fig3.2}
\end{figure}
%%%%%%%%%%%%%%%%%%%%%%%%%%%%%%%%%%%%%%%%%%%%%%%%%%%

In Figs \ref{fig3.2}(a-b), we show the snapshots of the normalized density
and temperature (in keV) profiles obtained in a steady state purely from 
our hydrodynamic simulation. The density contour levels are drawn for 
$0.65 - 1.01$ (levels increasing by a factor of $1.05$) and $1.01 - 66.93$ 
(levels increasing by a factor of $1.1$). The temperature contour levels are 
drawn for $16.88 - 107.8$ keV (levels increasing by a factor of $1.05$).
The radial variation of the velocity profile is also shown in Fig.
\ref{fig3.2}(c). The velocity is measured in the unit of $c$. 
As it is seen from \ref{fig3.2}(c), the velocity $v \to c$ as $r \to 1$ (event horizon) and
that makes it important to consider gravity effects (light bending, redshift etc.) on emerging photon.
For the present simulations, we have taken into account the effect of redshift, although
the effects of light bending has not been taken into consideration. 
However, we are in the process to improve the code by cosidering the 
gravity effects as much as possible and in this regard, we have discussed 
the computational procedure to compute photon trajectory in Schwarzschild 
geometry in Appendix A.
%However, with time we are improving our
%simulation code by incorporating different effects.
%In this regard, we have developed a subroutine to compute
%photon trajectory in Schwarzschild geometry and included the computational
%procedure of the same as Appendix A in the thesis. Thus, the process to take
%into account the gravity effects as much as possible are on the way and we
%hope to produce these results in future.

\subsection{Keplerian disk}

We assume a steady Keplerian disk and use the standard Shakura-Sunyaev 
disk (SS73) as the source of soft photons for this case. The 
Keplerian disk is placed on the equatorial plane. For the 
present simulations, we neglect the reflection and/or absorption of 
photons by the Keplerian disk (GGGC11; GGC12; GGC13). 

The radial variations of the disk temperature and number of generated
photons are given in Eq. (\ref{eqno1.1}) and (\ref{eqno2.7}), respectively. 
In the Monte Carlo simulation, we incorporated the directional effects 
of photons coming out of the Keplerian disk with the maximum number 
of photons emitted in the $z$-direction and minimum number of photons 
are generated along the plane of the disk (GGGC11). Thus, in the absence of 
photon bending effects, the disk is invisible as seen edge on. 
The position of each emerging photon is randomized using the distribution 
function given by Eq. (\ref{eqno2.7}).

\subsection{Simulation procedure}

The spectral properties are studied using the coupled radiation
hydrodynamic simulation code, described in the previous Section. 
For a particular simulation, we assume a Keplerian disk rate 
($\dot{m}_d$) and a halo rate ($\dot{m}_h$). The specific energy 
($\epsilon$) of the halo provides the hydrodynamic (e.g., number 
density of the electrons and the velocity distribution) and the 
thermal properties of matter.

Before beginning the coupled code simulation, we simulate a steady 
state flow profile using the non-radiative hydro code (RCM97; 
Giri et al. 2010). This steady 
state flow is used as the initial condition for the coupled simulation
run. On this flow profile, we start the Monte Carlo simulation. 
The simulation procedure of the Monte Carlo part is exactly same 
as that is described in Section 2.1. When a photon interacts with an electron 
via Compton or inverse-Compton scattering, it loses or gains some energy 
($\Delta E$). At each grid point, we compute $\Delta E$ (PSS83). We update 
the energy of the flow at this grid by this amount and continue the 
hydrodynamic code with this modified energy. This, in turn, modify the 
hydrodynamic profile. In case the final state is 
steady, the temperature of the cloud would be reduced progressively
to a steady value from the initial state where no cooling was assumed.
If the final state is oscillatory, the solution would settle into a 
state with Comptonization. 

\subsection{Results and discussions}

In Table 3, we summarize all the cases for which the simulations have 
been presented in this Chapter. Similar cases have been presented for 
different sets of parameters in GGGC11. In Column 1, various case IDs are 
marked. Column 2 shows the specific energy ($\epsilon$) of the halo at 
the outer boundary. The Keplerian 
disk rate ($\dot{m}_d$) and the sub-Keplerian halo rate ($\dot{m}_h$) 
are listed in Columns 3 and 4. The number of soft photons, injected 
from the Keplerian disk ($N_{inj}$) for various disk rates, can be found 
in Column 5. Column 6 lists the number of photons ($N_{sc}$) that have 
suffered at least one scattering inside the electron cloud. The number 
of photons ($N_{unsc}$), escaped from the cloud without any scattering 
are listed in Column 7. Columns 8 and 9 give the percentages of injected 
photons that have entered into the black hole ($N_{bh}$) and suffered 
scattering ($p = \frac{N_{sc}}{N_{inj}}$), respectively. The cooling time 
($t_0$) of the system is defined as the expected time for the
system to lose all its thermal energy with the particular flow parameters 
(namely, $\dot{m}_d$ and $\dot{m}_h$). We calculate $t_0 = E/{\dot{E}}$ 
in each time step, where, $E$ is the total energy content of the system 
and ${\dot{E}}$ is the energy gain or loss by the system in that particular 
time step and present it in Column 10. We present the energy spectral index 
$\alpha, ~[I(E) \sim E^{-\alpha}]$ obtained from our simulations 
in the last column.
\begin{center}
{\footnotesize{
\begin {tabular}[h]{|c|c|c|c|c|c|c|c|c|c|c|}
\hline
\multicolumn{11}{|c|}{Table 3: Parameters used for the simulations and a summary of results.}\\
\hline Case & $\epsilon$ & $\dot{m}_d$ & $\dot{m}_h$ & $N_{inj}$ &
$N_{sc}$ & $N_{unsc}$ & $N_{bh}$ (\%) & $p$ (\%) & $t_0$ (s) & $\alpha$\\
\hline
A & 22e-4 & 1e-4 & 0.5 & 7.1e+41 & 9.6e+40 & 6.1e+41 & 0.184 & 13.511 & 196.4 & 1.41, 0.91 \\
B & 22e-4 & 1e-3 & 0.5 & 4.0e+42 & 5.4e+41 & 3.5e+42 & 0.186 & 13.5   & 24.4 & 1.44, 0.75 \\
C & 22e-4 & 1e-2 & 0.5 & 2.3e+43 & 3.1e+42 & 2.0e+43 & 0.179 & 13.351 & 4.0 & 1.61, 0.66 \\
D & 22e-4 & 1e-1 & 0.5 & 1.2e+44 & 1.5e+43 & 1.1e+44 & 0.149 & 12.683 & 0.2 & 2.11, 0.53 \\
E & 22e-4 & 1e-3 & 1   & 4.0e+42 & 9.8e+41 & 3.0e+42 & 0.327 & 24.377 & 6.0 & 1.15       \\
F & 22e-4 & 1e-3 & 2   & 4.0e+42 & 1.6e+42 & 2.4e+42 & 0.608 & 40.636 & 1.2 & 0.85       \\
G & 22e-4 & 1e-3 & 5   & 4.0e+42 & 2.7e+42 & 1.3e+42 & 1.224 & 66.507 & 0.2 & 0.52       \\
\hline
\end{tabular}
}}
\end{center}
%%%%%%%%%%%%%%%%%%%Fig.3.3%%%%%%%%%%%%%%%%%%%%%%%%%
\begin{figure}
\centering{
\includegraphics[height=15truecm,width=15truecm,angle=0]{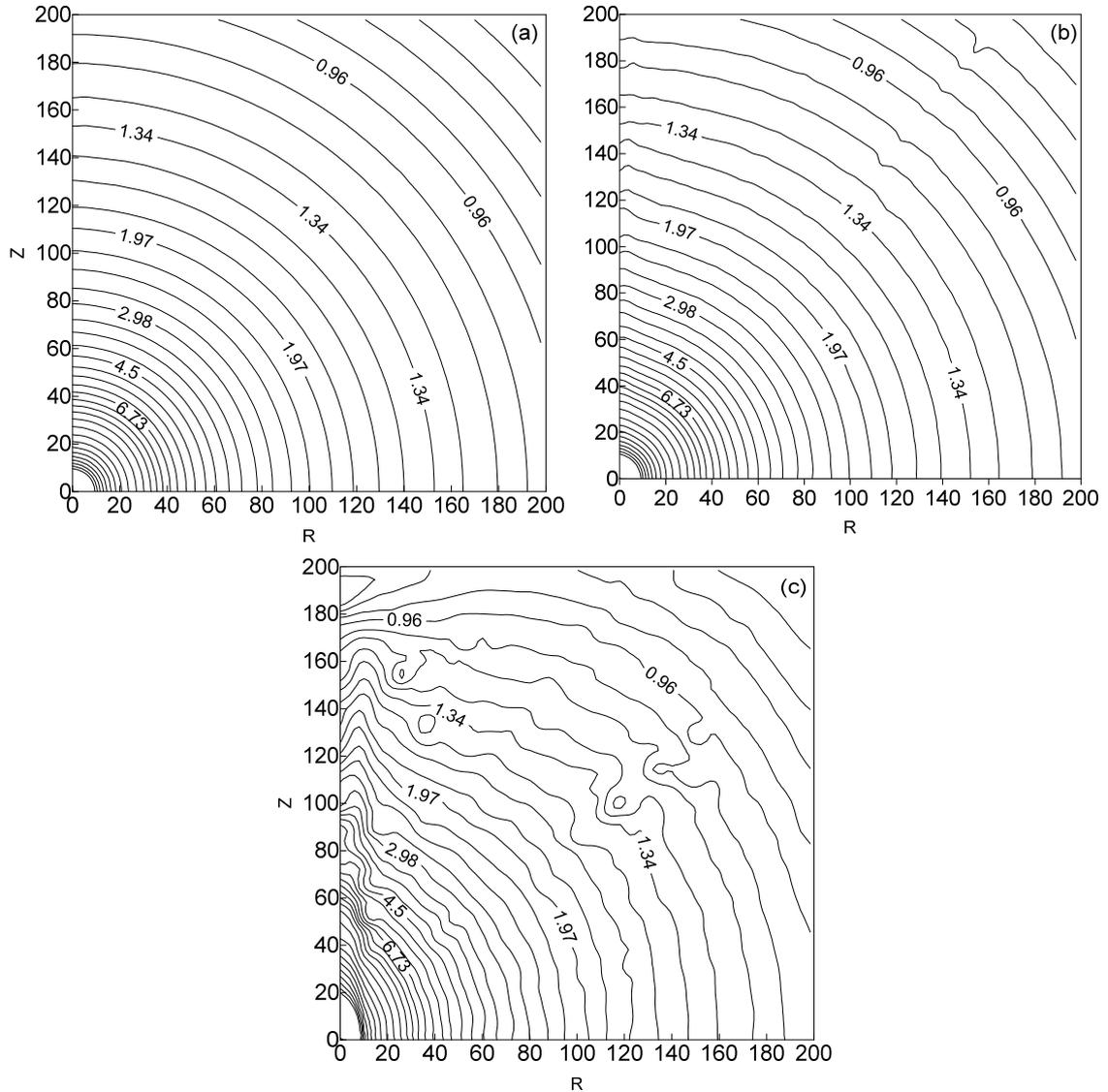}}
\caption{Changes in the density distribution in presence of cooling. 
$\dot{m}_h = 0.5$ for all the cases. Disk accretion rates $\dot{m}_d$ used
are (a) $10^{-4}$, (b) $10^{-3}$ and (c) $10^{-2}$, respectively (Cases A-C of Table 3).
The density contours are drawn using the same contour levels as in 
Fig. \ref{fig3.2}(a) (see also, GGGC11).}
\label{fig3.3}
\end{figure}
%%%%%%%%%%%%%%%%%%%%%%%%%%%%%%%%%%%%%%%%%%%%%%%%%

%%%%%%%%%%%%%%%%%%%Fig.3.4%%%%%%%%%%%%%%%%%%%%%%%%%
\begin{figure}
\centering{
\includegraphics[height=15truecm,width=15truecm,angle=0]{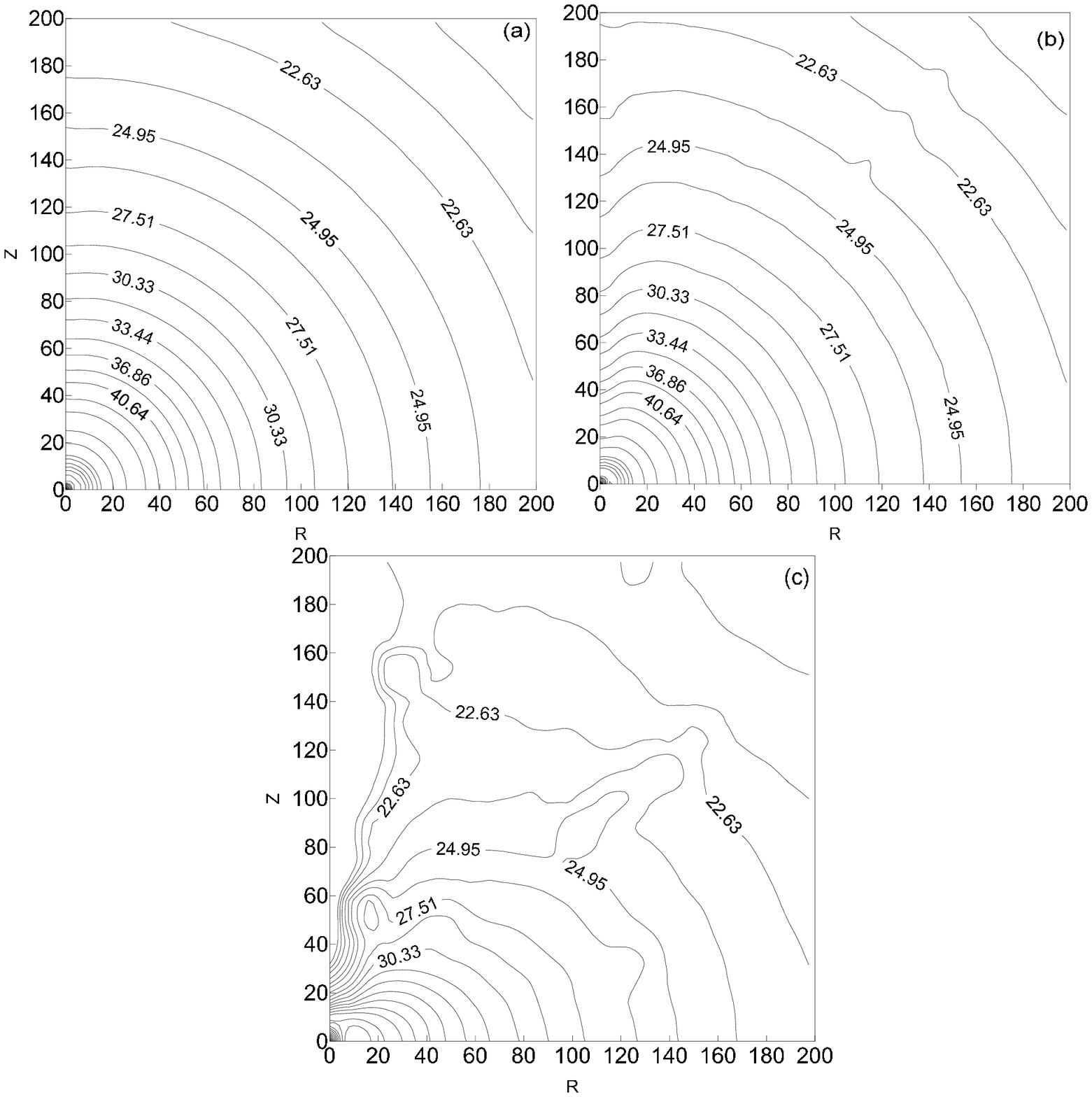}}
\caption{Changes in the temperature distribution in presence of cooling.
$\dot{m}_h = 0.5$ for all the cases. Disk accretion rates $\dot{m}_d$ are 
(a)$10^{-4}$, (b)$10^{-3}$ and (c)$10^{-2}$, respectively (Cases A-C of Table 3). 
Contours are drawn using the same levels as in Fig. \ref{fig3.2}(b) (see also, GGGC11).}
\label{fig3.4}
\end{figure}
%%%%%%%%%%%%%%%%%%%%%%%%%%%%%%%%%%%%%%%%%%%%%%%%%

In Figs \ref{fig3.3}(a-c), we present the changes in density distribution
as the disk accretion rates are changed: $\dot{m}_d =$ (a) $10^{-4}$, (b) $10^{-3}$
and (c) $10^{-2}$, respectively. We notice that as the accretion rate
of the disk is enhanced, the density distribution losses its spherical 
symmetry (see also, GGGC11). In particular, the density at a given radius is enhanced in a 
conical region along the axis. This is due to the cooling of the 
matter by Compton scattering. To show this, in Figs \ref{fig3.4}(a-c),
we show the contours of constant temperatures (marked on curves) for the 
same three cases. We notice that the temperature is reduced along the 
axis (where the optical depth as seen by the soft photons from the 
Keplerian disk is higher) drastically after repeated Compton scatterings.

%%%%%%%%%%%%%%%%%%%Fig.3.5%%%%%%%%%%%%%%%%%%%%%%%%%
\begin{figure}
\vspace {0.4cm}
\centering{
\includegraphics[height=6.5truecm,width=6.5truecm,angle=0]{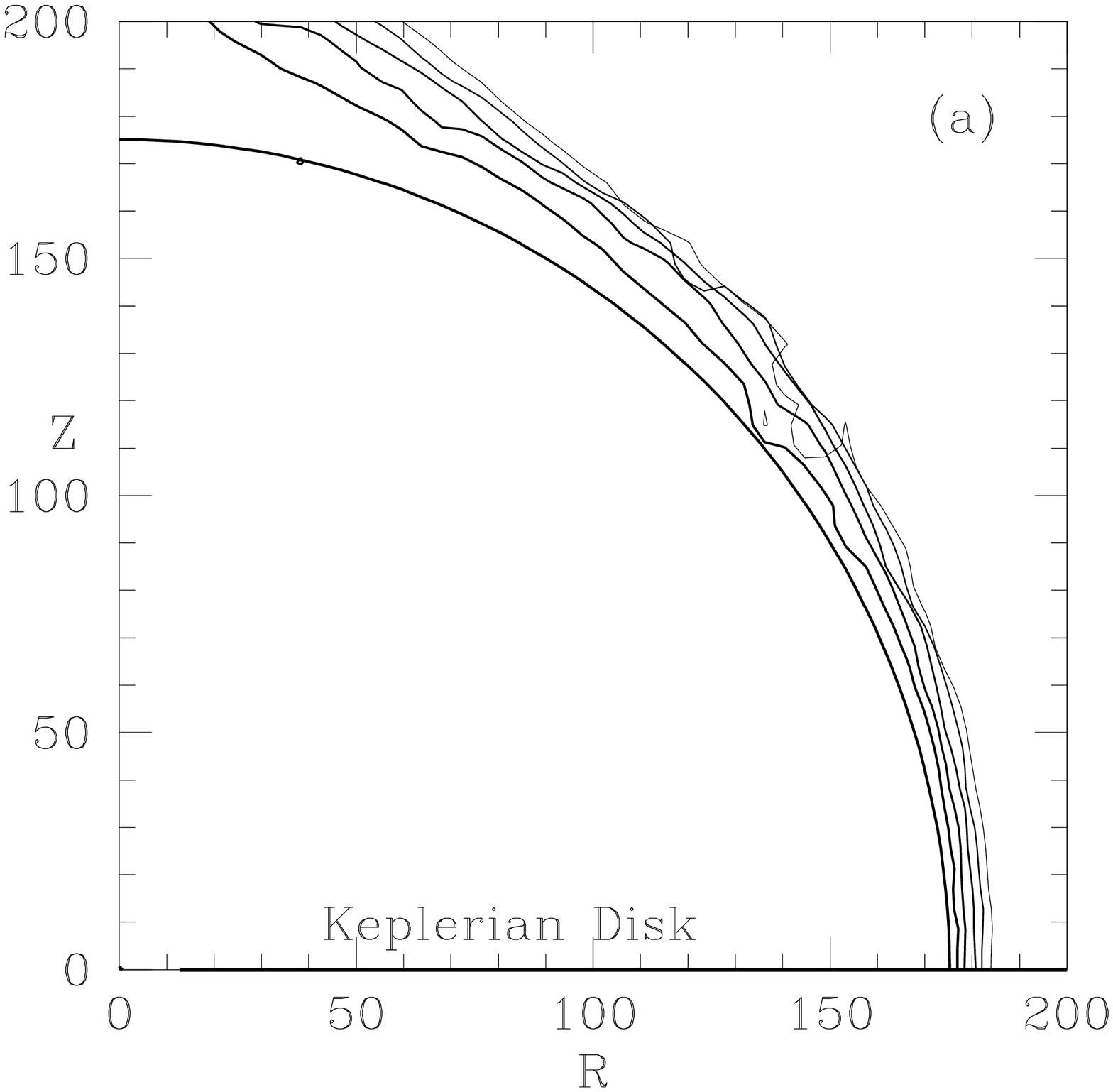}
\hskip 0.5cm
\includegraphics[height=6.0truecm,width=7.0truecm,angle=0]{Figure/Fig3.5b.eps}}\\
\vspace {0.7cm}
\centering{
\includegraphics[height=6.5truecm,width=6.5truecm,angle=0]{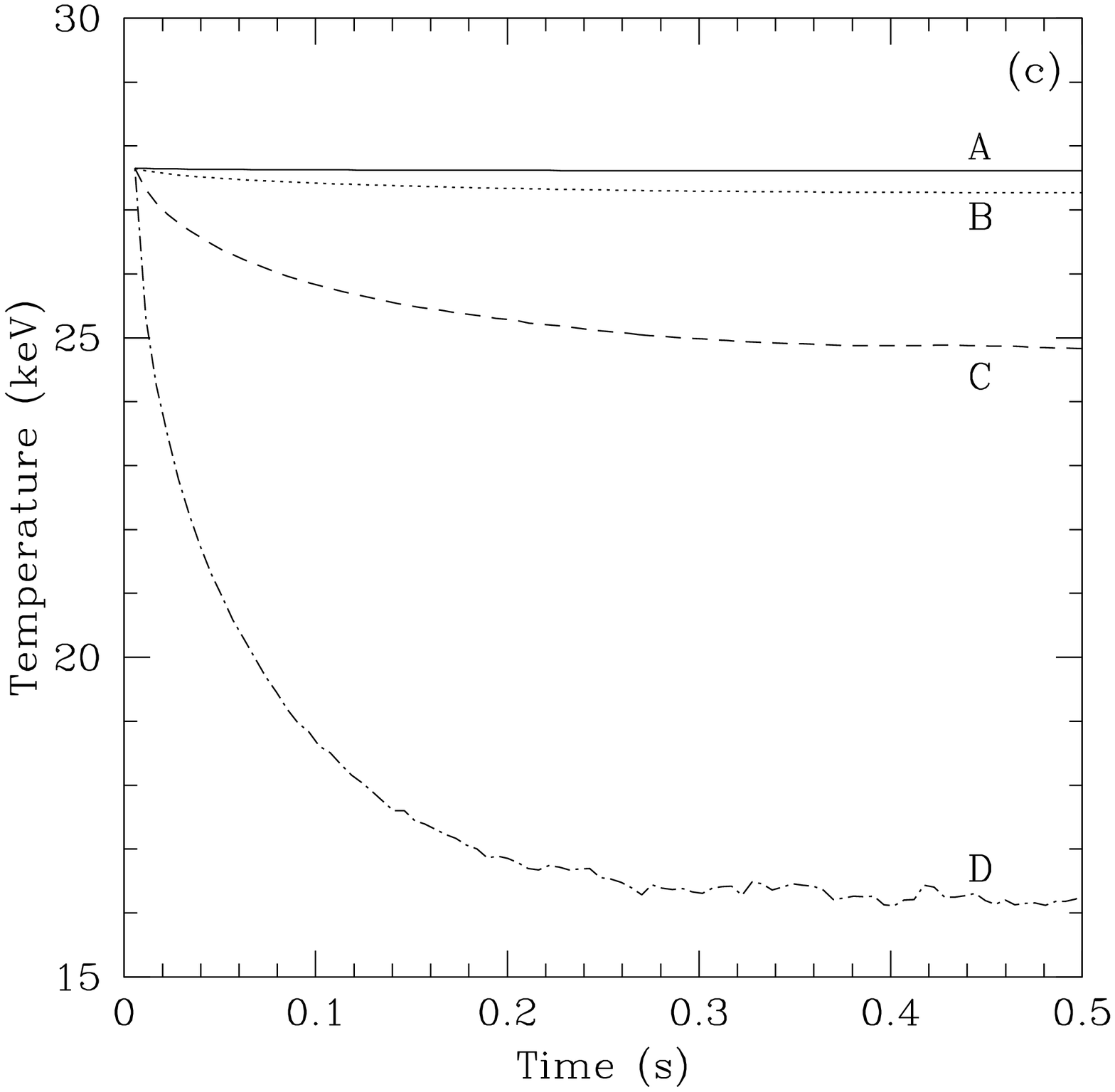}
\hskip 0.5cm
\includegraphics[height=6.0truecm,width=7.0truecm,angle=0]{Figure/Fig3.5d.eps}}
\caption{(a) Sonic surfaces at different stages of iterations. The 
outermost curve represents the final converged solution (see also, GGGC11). 
The initial spherical 
sonic surface become prolate spheroid due to cooling by the Keplerian 
disk at the equatorial plane. Parameters are for Case C (Table 3).
(b) Mach number variation as a function of distance after a complete 
solution of the radiative flow is obtained. Plot no. 1 corresponds to 
the solution from adiabatic Bondi flow. Plots 2-4 are the solutions 
along the equatorial plane, the diagonal and the axis of the disk, 
respectively (see also, GGGC11). Parameters are for Case C (Table 3).
(c) Variation of the average temperature of the Compton cloud as the 
iteration proceeds when the disk accretion rate is varied keeping the
halo rate constant at  $\dot{m}_h = 0.5$. The solid, dotted, dashed 
and dash-dotted plots are for $\dot{m}_d = 0.0001$, $0.001$, $0.01$ 
and $0.1$ respectively. Case numbers (Table 3) are marked.
With the increase of disk rate, the temperature of the Compton cloud 
converges to a lower temperature (see also, GGGC11).
(d) Variation of the spectrum with the increase of disk accretion rate. 
Parameters are the same as in (c). With the increase in $\dot{m}_d$, 
the intensity of the spectrum increases due to the increase in $N_{inj}$ 
(see, Table 3). The spectrum is softer for higher values of $\dot{m}_d$
(see also, GGGC11). The spectral slope for each of these spectra is listed in Table 3.}
\label{fig3.5}
\end{figure}
%%%%%%%%%%%%%%%%%%%%%%%%%%%%%%%%%%%%%%%%%%%%%%%%%

In Figs \ref{fig3.5}(a-d), we show the hydrodynamic and radiative 
properties. In Fig. \ref{fig3.5}(a), we show the sonic surfaces. The 
lowermost curve corresponds to theoretical solution for an adiabatic 
flow (e.g., C90). Other curves from the bottom to the 
top are the iterative solutions for the Case C mentioned above.
As the disk rate is increased, the cooling increases, and consequently, 
the Mach number increases along the axis (see also, GGGC11). Of course, there are other 
effects: the cooling causes the density to go up to remain in pressure 
equilibrium. In Fig. \ref{fig3.5}(b), the Mach number variation is shown. 
The lowermost curve (marked 1) is from the theoretical consideration. 
Plots marked 2-4 are the variation of Mach number with radial distance
along the equatorial plane, along the diagonal and along the vertical 
axis respectively (see also, GGGC11). In Fig. \ref{fig3.5}(c), the average temperature 
of the spherical halo is plotted as a function of the iteration time 
until almost steady state is reached. The cases are marked on the curves. 
We note that as the injection of soft photons increases, the average 
temperature of the halo decreases drastically (see also, GGGC11). In Fig. \ref{fig3.5}(d), 
we plot the energy dependence of the photon intensity.
We find that, as we increase the disk rate, keeping the halo rate fixed, 
number of photons coming out of the cloud in a particular energy bin 
increases and the spectrum becomes softer (see also, GGGC11). This is also clear from Table 3; 
$N_{inj}$ increases with $\dot{m}_d$, increasing $\alpha$. We find the
signature of double slopes in these cases. As the disk rate increases, 
the second slope becomes steeper. This second slope is the signature of 
bulk motion Comptonization (CT95). As $\dot{m}_d$ increases, the cloud becomes
cooler [Fig. \ref{fig3.5}(c)] and the power-law tail due to the bulk motion 
Comptonization (CT95; Chakrabarti 1997; GGCL10; GGGC11) becomes prominent.
% Chakrabarti S. K., 1997, ApJ, 484, 313

%%%%%%%%%%%%%%%%%%%Fig.3.6%%%%%%%%%%%%%%%%%%%%%%%%%
\begin{figure}
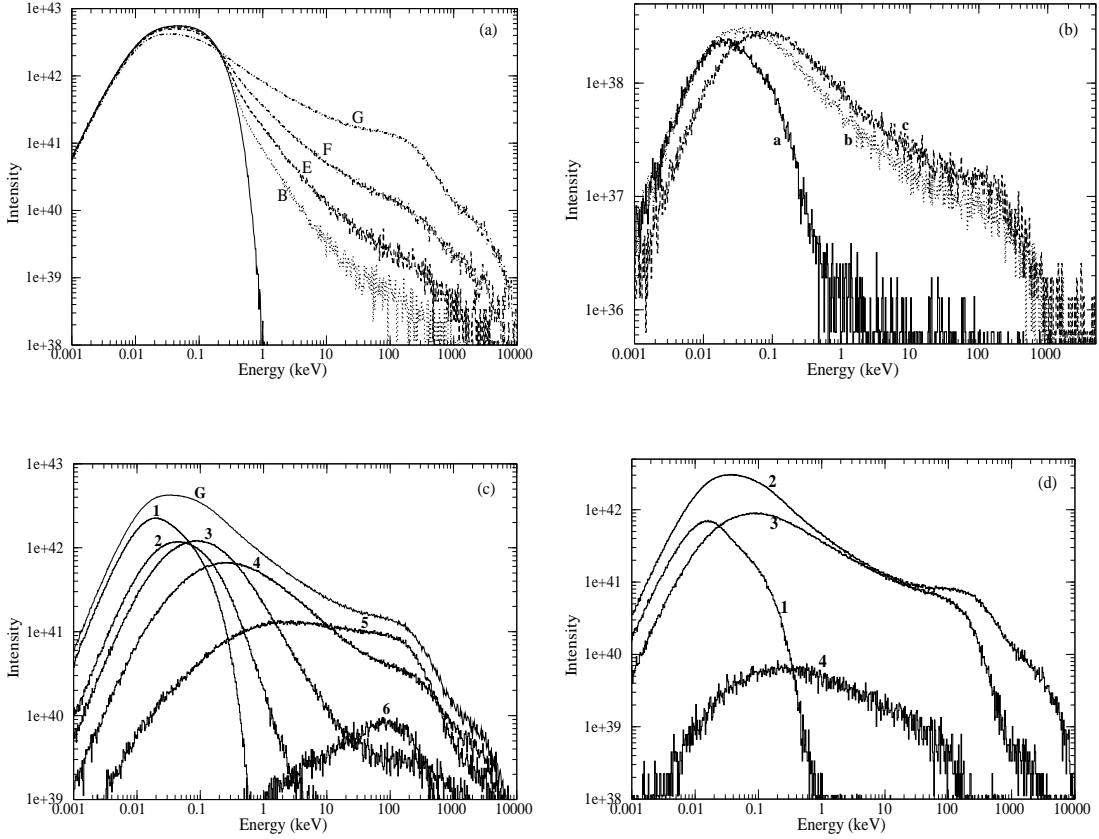

\vspace {0.4cm}
\centering{
\includegraphics[height=5truecm,width=7truecm,angle=0]{Figure/Fig3.6a.eps}
\hskip 0.2cm
\includegraphics[height=5truecm,width=7truecm,angle=0]{Figure/Fig3.6b.eps}}\\
\vspace {1.0cm}
\centering{
\includegraphics[height=5truecm,width=7truecm,angle=0]{Figure/Fig3.6c.eps}
\hskip 0.2cm
\includegraphics[height=5truecm,width=7truecm,angle=0]{Figure/Fig3.6d.eps}}
\caption{(a) Variation of the spectrum with the increase of the halo 
accretion rate, keeping the disk rate ($\dot{m}_d = 0.001$) fixed. 
The dotted, dashed, dash-dotted and double dot-dashed 
curves show the spectra for $\dot{m}_h = 0.5$, $1$, $2$ and $5$ respectively.
The injected multicolor blackbody spectrum supplied by the Keplerian 
disk is shown by solid line (see also, GGGC11).
(b) Directional dependence of the spectrum:  $\dot{m}_h = 5$, 
$\dot{m}_d = 0.001$  are the flow parameters. The solid, 
dotted and dashed curves are for observing angles $2^{\circ}$, 
$45^{\circ}$ and $90^{\circ}$ respectively (see also, GGGC11). All the angles are measured 
with respect to the rotation axis ($z$-axis). Intensity of spectra 
emerging from the cloud after suffering various number of scatterings 
(c) and at four different times (d) immediately after the injection 
of soft photons. Case G is assumed. The spectra of the photons 
suffering 0, 1, 2-3, 4-7, 8-15 and more than 16 scatterings are
shown by the plots marked as 1, 2, 3, 4, 5 and 6 (Fig. \ref{fig3.6}c) 
respectively, within the cloud. Curve G is the net spectrum for which 
these components are drawn. As the number of scattering increases, the
photons gain more and more energy from the hot electron cloud through 
inverse-Comptonization process (see also, GGGC11). The spectra of the photons spending 
0.01-20, 20-40, 40-100 and more than 100 ms time inside the electron 
cloud are marked by 1, 2, 3 and 4 (Fig. \ref{fig3.6}d) respectively (see also, GGGC11).}
\label{fig3.6}
\end{figure}
%%%%%%%%%%%%%%%%%%%%%%%%%%%%%%%%%%%%%%%%%%%%%%%%%

In Fig. \ref{fig3.6}(a), we show the variation of the energy spectrum with 
the increase of the halo accretion rate, keeping the disk rate fixed
($\dot{m}_d = 0.001$). The injected multi-color blackbody spectrum 
supplied by the Keplerian disk is shown (solid line). The dotted, 
dashed, dash-dotted and double dot-dashed curves show the spectra for 
$\dot{m}_h = 0.5$, $1$, $2$ and $5$, respectively. The spectrum 
becomes harder for higher values of $\dot{m}_h$ as it is difficult to 
cool a higher density matter with the same number of injected soft 
photons (CT95; GGGC11). In Fig. \ref{fig3.6}(b), we show the directional 
dependence of the spectrum for $\dot{m}_h = 5$, $\dot{m}_d = 0.001$ (Case G). 
The solid, dotted and dashed curves are for observing angles 
(a) $2^{\circ}$,  (b) $45^{\circ}$ and (c) $90^{\circ}$ respectively.
All the angles are measured with respect to the rotation axis ($z$-axis).
As expected, the photons arriving along the z-axis would be dominated 
by the soft photons from the Keplerian disk while the power-law would 
dominate the spectrum coming edge-on (see also, GGGC11). We now study the dependence of the
spectrum on the residence time of an injected photon inside the cloud. 
Depending on the number of scattering suffered and the length 
of path traveled, different photons spend different times inside the 
Compton cloud (GGGC11). The energy gain or loss by any photon depends on this 
time. Fig. \ref{fig3.6}(c) shows the spectrum of the photons suffering
different number of scatterings inside the cloud. Here, counts marked as 1, 2, 3, 4, 5 
and 6 show the spectrum for 6 different ranges of number of scatterings. 
Plot 1 shows the spectrum of the photons that have escaped from the 
cloud without suffering any scattering. This spectrum is as 
same as the injected spectrum, only difference is that it is Doppler 
shifted. As the number of scattering increases (spectrum 2, 3 and 4), 
the photons are more and more energized via inverse-Compton scattering 
with the hot electron cloud. For scatterings more than 8, the high 
energy photons start loosing energy through Compton scattering with 
the relatively lower energy electrons (see also, GGGC11). Components 5 and 6 show the 
spectra of the photons suffering 8-15 scatterings and the photons
suffering more than 16 scatterings respectively. Here the flow parameters are:
$\dot{m}_d = 0.001$ and $\dot{m}_h = 5$ (Case G, Table 3).

In Fig. \ref{fig3.6}(d), we plot the spectrum emerging out of the electron 
cloud at four different time ranges. In the simulation, the photons 
take $0.01$ to $150$ ms to come out of the system. We divide this time 
range into 4 suitable bins and plot their spectrum. Case G of Table 3 
is considered. We observe that the spectral slopes and intensities of 
the four spectra are different. As the photons spend more and more time 
inside the cloud, the spectrum gets harder (plots 1, 2 and 3). 
However, very high energy photons which spend maximum time inside the 
cloud lose some energy to the relatively cooler electrons before escaping 
from the cloud. Thus the spectrum 4 is actually the spectrum dominated by the 
Comptonized photons (see also, GGGC11).

%%%%%%%%%%%%%%%%%%%Fig.3.7%%%%%%%%%%%%%%%%%%%%%%%%%
\begin{figure}
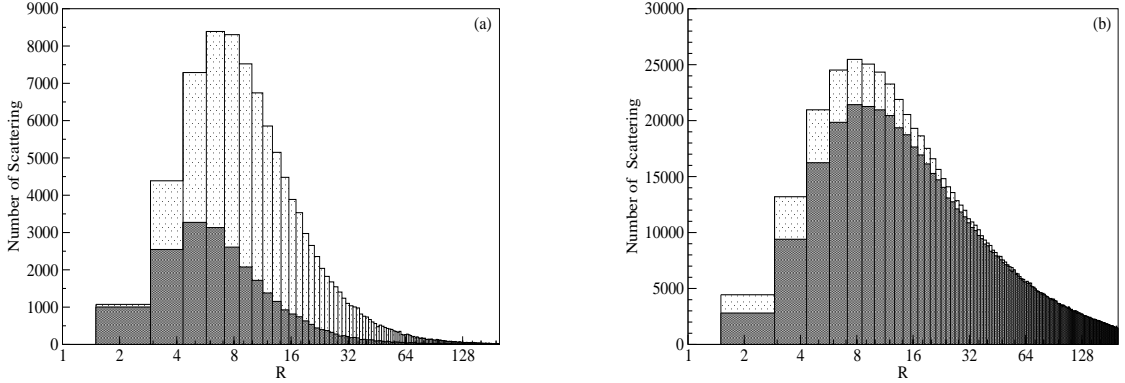

\vspace {1.0cm}
\centering{
\includegraphics[height=5truecm,width=6.5truecm,angle=0]{Figure/Fig3.7a.eps}
\hskip 1.5cm
\includegraphics[height=5truecm,width=6.5truecm,angle=0]{Figure/Fig3.7b.eps}}
\caption{Number of scatterings inside the spherical shell between $R$ 
to $R + \delta R$ ($ \delta R \sim 1.4$). The light and dark shaded 
histograms are for the cloud with and without bulk velocity, respectively (see also, GGGC11).
(a) Only the photons emerging from the cloud with energies $E$, 
where $50$ keV $< E <150$ keV, are considered here. (b) All the photons 
emerging from the cloud are considered here. Parameters used:
$\dot{m}_d=0.001$ and  $\dot{m}_h=5$ (Case G Table 3).}
\label{fig3.7}
\end{figure}
%%%%%%%%%%%%%%%%%%%%%%%%%%%%%%%%%%%%%%%%%%%%%%%%%
We observe that the emerging spectrum has a bump, especially at higher 
accretion rates of the halo, at around $100$ keV (e.g., the spectra 
marked F, G in Fig. \ref{fig3.6}(a)). A detailed analysis of the emerging 
photons having energies between 50 to 150 keV was made to see where in 
the Compton cloud these photons were produced (GGGC11). In Fig. \ref{fig3.7}(a), 
we present the number of scatterings inside different spherical shells 
within the electron cloud suffered by these photons ($50 < E < 150$ keV) 
before leaving the cloud. Parameters used: $\dot{m}_d=0.001$ and $\dot{m}_h=5$. 
The light and dark shaded histograms are for the cloud with and without 
bulk velocity components, respectively. We find that the presence of 
bulk motion of the infalling electrons pushes the photons towards the 
hotter and denser [Figs \ref{fig3.2}(a-b)] inner region of the cloud 
to suffer more and more scatterings (GGGC11). We find that the photons responsible 
for the bump suffered maximum number of scatterings at around $8 r_g$. From 
the temperature contours, we find that the cloud temperature around 
$8 r_g$ is $\sim 100$ keV. This explains the existence of the bump. In 
Fig. \ref{fig3.7}(b), we consider all the outgoing photons independent 
of their energies. The difference between the two cases is not so visible. 
This shows that the bulk velocity contributes significantly to produce 
the highest energy photons.
%%%%%%%%%%%%%%%%%%%Fig.3.8%%%%%%%%%%%%%%%%%%%%%%%%%
\begin{figure}
\vspace {0.5cm}
\centering{
\includegraphics[height=6truecm]{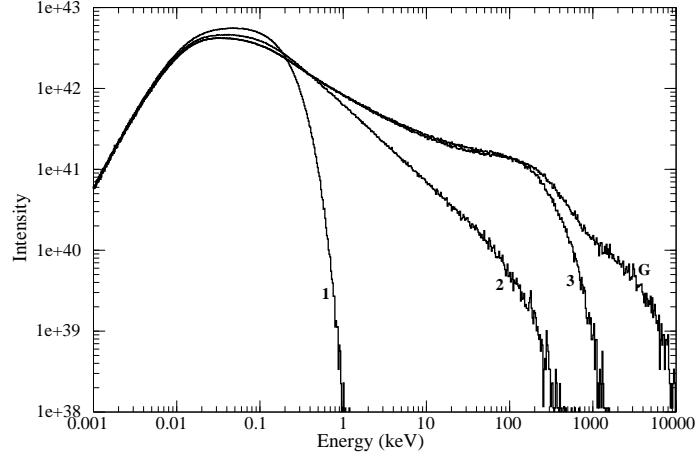}}
\caption{The spectrum for the Case G. The curves marked 2 and 3 give 
the spectra when the bulk velocity of the electron is absent for the 
whole cloud and for the cloud inside $3 r_g$, respectively.
The curve marked 1 gives the injected spectrum. The bulk motion
Comptonization of the photons inside the $3 r_g$ radius creates the 
hard tail. The bump near 100 keV is due to a combined effect of the temperature 
and bulk velocity of the rest of the cloud (see also, GGGC11).}
\label{fig3.8}
\end{figure}

In Fig. \ref{fig3.8}, we explicitly show the effects of the bulk 
velocity on the spectrum. We note that the bump disappears when the 
bulk velocity of the electron cloud is chosen to be zero (Curve marked 2). 
This fact shows that the region around $8 r_g$ in presence of the bulk 
motion behaves more like a blackbody emitter, which creates the bump 
in the spectrum (see also, GGGC11). Since the photons are suffering a large number of 
scatterings near this region ($8 r_g$), most of them emerge from the 
cloud with the characteristic temperature of the region. The effect 
of bulk velocity in this region is to force the photons to suffer larger
number of scatterings (GGGC11). This bump vanishes for lower density cloud 
(low $\dot{m}_h$) as the photons suffer lesser number of scatterings.
The photons which are scattered close to the black hole horizon and 
escape without any further scattering, produce the high energy tail in 
the output spectrum. Curve 3 of Fig. \ref{fig3.8} shows the intensity 
spectrum of Case G (Table 3), when there are zero bulk velocity inside 
$3 r_g$. We find that in the absence of bulk velocity inside $3 r_g$, 
the high energy tail in the Curve G vanishes. This is thus a clear 
signature of the presence of bulk motion Comptonization near the black 
hole horizon (CT95; Chakrabarti 1997; GGGC11).

\clearpage
%~~~~~~~~~~~~~~~~~~~~~~~~~~~~~~~~~~~~~~~~~~~~~~~~~~~~~~~~~~~~~~~~~~~~~~~~~~~~~~~

	\reseteqn
	\resetsec
	\resetfig
	\resettab

%~~~~~~~~~~~~~~~~~~~~~~~~~~~~~~~~~~~~~~~~~~~~~~~~~~~~~~~~~~~~~~~~~~~~~~~~~~~~~~
\alpheqn
\resec
\refig
\retab
%%%%%%%%%%%%%%%%%%%%%%%%%%%%%%%%%%%%%%%%%%%%%%%%%%%%%%%%%%%
% Chapter 4 : EFFECTS OF COMPTON COOLING ON OUTFLOW
%%%%%%%%%%%%%%%%%%%%%%%%%%%%%%%%%%%%%%%%%%%%%%%%%%%%%%%%%%%

\def\k{{\bf k}}
\def\aug{{\tilde{\cal H}}}
\def\ahat{{\mathaccent "7E {\cal A}}}
\def\bhat{{\mathaccent "7E {\cal B}}}
\def\chat{{\mathaccent "7E {\cal F}}}
\def\dhat{{\mathaccent "7E {\cal D}}}
\def\shat{{\mathaccent "7E {\cal S}}}
\def\proj{{\cal P}}
\def\ket{\vert \vert  \{ \emptyset \} \rangle}
\def\ket2{\vert \vert \otimes \{ R \} \rangle}
\def\sqr{$^{2}$}
\def\T{$T$}
\def\P{$P$}
\def\Q{$Q$}
\def\G{$G$}
\def\M{{\bf M}}
\def\H{{\bf H}}
\def\I{{\bf I}}
\def\Pr{{\bf P}}
\def\Tr{{\bf T}}
\def\eq{\enskip =\enskip}
\def\pls{\enskip+\enskip}
\def\mns{\enskip -\enskip}
\def\oper{O^{\bf k}_{R}}
\def\aug{\tilde{\cal H}}
\def\envfn{\chi^{\alpha}_{RL}}
\def\struc{S^{\alpha}_{RL,R^{\prime}L^{\prime}}}
\def\rR{r_{R}}

\newpage
\markboth{\it Effects of Compton Cooling on Outflows}
{\it Effects of Compton Cooling on Outflows}
\chapter{EFFECTS OF COMPTON COOLING ON OUTFLOWS}

%\section{Introduction}
The winds and jets in a compact binary system containing black holes 
are generally believed to be originated from the disk itself (C96; 
C99; Das et al. 2001; Das \& Chattopadhyay 2008). There are
% Das S., Chattopadhyay I., 2008, New Astron., 13, 549 
several hydrodynamical models of the formation of outflows from the 
disks ranging from the twin-exhaust model of Blandford \& Rees (1974), 
% Blandford, R. D., \& Rees, M. J. 1974, MNRAS, 169, 395
to the self-similar models of Blandford \& Payne (1982) and
% Blandford, R. D., \& Payne, D. G. 1982, MNRAS, 199, 883 
% Blandford, R. D., \& Begelman, M. C. 1999, MNRAS, 303, L1
Blandford \& Begelman (1999). Assuming that the outflows are transonic 
in nature, Fukue (1983) and Chakrabarti (1986) computed the velocity 
% Fukue, J. 1983, PASJ, 35, 539
% Chakrabarti, S. K. 1986, ApJ, 303, 582
distribution without and with rotational motion in the flow and showed 
that the flow could become supersonic close to the black hole.
Camenzind and his group extensively worked on the magnetized jets and 
showed that the acceleration and collimation of the jets could be 
achieved (e.g., Appl \& Camenzind 1993). In a subsequent two component 
% Appl, S., \& Camenzind, M. 1993, A\&A, 270, 71
transonic flow model, CT95 pointed out that the jets could be formed 
only from the inner part of the disk i.e., the CENBOL.

While the general picture of the outflow formation is thus understood
and even corroborated by the radio observation of the base of the powerful 
jet, such as in M87 (Junor, Biretta \& Livio 1999), that the base of the jet is 
% Junor, W., Biretta, J. A. \& Livio, M. 1999, Nature, 401, 891
only a few tens of $r_g$, a major question still remained: what fraction 
of the matter is driven out from the disk and what are the flow 
parameters on which this fraction depends?  In a numerical simulation
using SPH code, MLC94 showed that the outflow rates from an inviscid 
accretion flow strongly depends on the outward centrifugal force and 
15-20 percent matter can be driven out of the disk. Chakrabarti (1998c), C99, 
Das et al. (2001), Chattopadhyay, Das \& Chakrabarti (2004) estimated the ratio 
% Chakrabarti, S. K. 1998, arXiv:astro-ph/9801079
% Chakrabarti, S. K. 1999, A\&A, 351, 185
% Chattopadhyay, I., Das, S., \& Chakrabarti, S. K. 2004, MNRAS, 348, 846
of the outflow rate to the inflow rate analytically and found that the shock 
strength determines the ratio. For very strong and very weak shocks, 
the outflow rates are very small, while for the shocks of intermediate 
strength, the outflow rate is significant. This is in line with the 
observations (Gallo, Fender \& Pooley 2004; Fender, Belloni, Gallo 2004; 
Fender, Gallo \& Russell 2010) that the spectrally soft states 
have less outflows. 
% Gallo, E., Fender, R. P., \& Pooley, G. 2004, NuPhS, 132, 363
% Fender, R. P., Belloni, T. M., \& Gallo, E. 2004, MNRAS, 355, 1105
% Fender, R. P., Gallo, E., \& Russell, D. 2010, MNRAS, 406, 1425

In this Chapter, we concentrate on the formation of outflows from the 
accretion disk around black holes by numerical simulation and study 
the effects of Compton cooling on it using
the time dependent radiation hydrodynamic simulation code (GGC12). 
While computing the time variation of the velocity components, density 
and temperature, we also compute the temporal dependence of the spectral 
properties. As a result, not only we compute the outflow properties, 
we correlate them with the spectral properties (GGC12). 
Not surprisingly, we find that whenever the Compton cloud or the CENBOL 
is cooled down and the spectrum becomes softer, the outflow, originating 
from CENBOL, losses its drive and its rate is greatly reduced.
In the following, we discuss the simulation procedure and the results 
of our study. Same conclusions are found in GGC12 where simulations are 
done using different sets of input parameters on similar simulation set up. 

\section{Simulation set up}

\begin{figure}
\centering{
\includegraphics[height=12truecm]{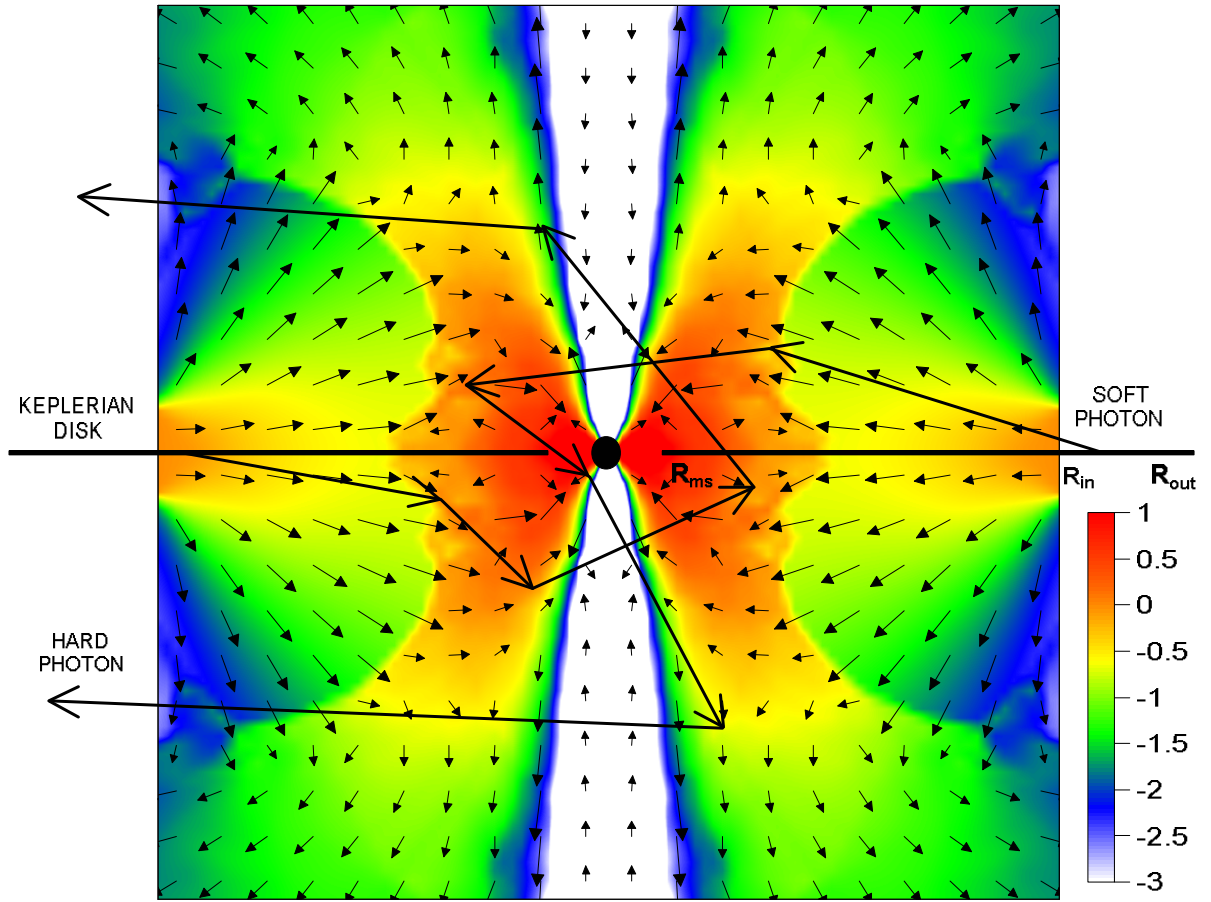}}
\vskip 2cm
\caption{Schematic diagram of the geometry of our Monte Carlo simulations.  
The colors represent the normalized density in logarithmic scale. The zig-zag 
trajectories are the typical paths followed by the photons. The velocity
vectors of the infalling matter inside the cloud are shown. The velocity 
vectors are plotted for $\lambda=1.75$ (see also, GGC12).}
\label{fig4.1}
\end{figure}

In Fig. \ref{fig4.1}, we present the schematic diagram of our simulation 
set up for the Compton cloud with a specific angular momentum 
$\lambda=1.75$ (see also, GGC12). The sub-Keplerian matter 
enters the simulation box through the 
outer boundary at $R_{in} = 100 r_g$ ($r_g=2GM_{bh}/c^2$). The Keplerian 
disk resides at the equatorial plane 
of the cloud. The outer edge of this disk is assumed to be at $R_{out} = 200 r_g$ 
and it extends inside up to the marginally stable orbit $R_{ms} = 3 r_g$. 
At the centre, a non-rotating black hole of mass $M_{bh} = 10M_\odot$ is 
located. The soft photons emerging out of the Keplerian disk are 
intercepted and reprocessed via Compton or inverse-Compton scattering 
by the sub-Keplerian matter. An injected photon may undergo no scattering 
at all or a single or multiple scatterings with the hot electrons in between its 
emergence from the Keplerian disk and its escape from the sub-Keplerian halo. 
The photons which enter the black hole are absorbed.
%%%%%%%%%%%%%%%%%%%Fig.1%%%%%%%%%%%%%%%%%%%%%%%%%
%%%%%%%%%%%%%%%%%%%%%%%%%%%%%%%%%%%%%%%%%%%%%%%%%
\subsection{Density, velocity and temperature profiles inside the halo component}

The inflowing halo matter has some angular momentum with respect
to the black hole. Therefore, as it approaches the black hole, the outward 
centrifugal force becomes comparable to the gravitational force at  acertain
radius. As a result, the matter slows down and we find the formation 
of shocks in the incoming supersonic flow (Chakrabarti 1989a,b; C90). 
Subsequently, the matter accelerates and becomes supersonic again as it 
approaches the black hole. In the post-shock region, the density and 
the temperature of the halo increases to much higher value. This region 
is the CENBOL of the TCAF model (CT95). 

Assuming axisymmetry, we calculate the flow dynamics using the 
TVD code in a similar way as described in the previous Chapter.
At each time step, we carry out Monte Carlo simulation to obtain the 
cooling/heating due to Comptonization. We incorporate the cooling/heating 
of each grid while executing the next time step of hydrodynamic simulation.
For the present case, the numerical simulation for the two-dimensional 
flow has been carried out with $512 \times 512$ equi-spaced cells in a 
$100 r_g \times 100 r_g$ box (GGC12; GGC13). We choose the units in a way that
the outer boundary ($R_{in}$) is unity and the matter density at the 
outer boundary is also normalized to unity (MRC96; Giri et al. 2010; GGGC11). 
We assume the black hole 
to be non-rotating and we use the pseudo-Newtonian potential 
$-\frac{1}{2(r-1)}$ (PW80) to calculate the flow geometry around a black 
hole. 
%%%%%%%%%%%%%%%%%%%%%%%%%%%%%%%%%%%%%%%%%%%%%%%%%

\section{Simulation procedure}

All the simulations have been carried out using the time dependent radiation
hydrodynamic simulation code described in the previous Chapter. Before
starting the simulation, we generated a steady state flow profile
using the non-radiative hydro code. This steady state is used as the 
initial condition for the radiation hydrodynamic simulation (GGGC11; GGC12).
The Keplerian disk at the equatorial plane supplies soft photons.
These photons interact with the high energy electrons of the halo and
they exchange their energy through Compton or inverse-Compton scattering. 
%**************************
%The amount of exchanged energy is suitably adjusted against the total
%energy of the electrons as well as the photons.  (WHAT IS THIS LINE?????)

We use a standard Keplerian disk (SS73) as the source of  soft
photons. The emission is of blackbody type characterized by the local
surface temperature. The radial variations of the surface temperature 
and the number of generated photons from the disk surface are given
in Eq. (\ref{eqno1.1}) and (\ref{eqno2.7}), respectively. We also 
incorporate the directional effects as described in Chapter 3. In 
these simulations, we neglect the reflection and/or absorption of the 
soft photons by the Keplerian disk. 

For a particular simulation, we use the Keplerian disk rate ($\dot{m}_d$) 
and the sub-Keplerian halo rate ($\dot{m}_h$) as parameters. The specific 
energy ($\epsilon$) and the specific angular momentum ($\lambda$) determine 
the hydrodynamics (shock location, number density and velocity variations 
etc.) and the thermal properties of the sub-Keplerian matter (GGC12; GGC13).
We assume the absorbing boundary condition at $r = 1.5$ since any inward 
pointing photon at that radius would be sucked into the black hole.

%%%%%%%%%%%%%%%%%%%%%%%%%%%%%%%%%%%%%%%%%%%%%%%%
\section{Results and discussions}
\begin{center}
\begin {tabular}[h]{cccc}
\multicolumn{4}{c}{Table 4: Parameters used for the simulations.}\\
\hline Case & $\epsilon, \lambda$ &  $\dot{m}_h$ & $\dot{m}_d$ \\
\hline
Ia & 0.0021, 1.80 & 0.1 & No Disk \\
Ib & 0.0021, 1.80 & 0.1 & 1e-4     \\
Ic & 0.0021, 1.80 & 0.1 & 2e-4    \\
Id & 0.0021, 1.80 & 0.1 & 5e-4    \\
\hline
IIa & 0.0021, 1.75 & 0.1 & No Disk  \\
IIb & 0.0021, 1.75 & 0.1 & 1e-4      \\
IIc & 0.0021, 1.75 & 0.1 & 2e-4     \\
IId & 0.0021, 1.75 & 0.1 & 5e-4     \\
\hline
\end{tabular}
\end{center} 

In Table 4, we list various Cases with all the simulation parameters 
used in this Chapter. The specific energy ($\epsilon$) and specific 
angular momentum ($\lambda$) of the sub-Keplerian halo at the outer 
boundary are given in Column 2. These parameters are chosen from the 
region of the parameter space for which shock formation is possible
in an accretion flow in vertical equilibrium (Chakrabarti \& Das 2001). 
% Chakrabarti S. K., Das S., 2001, MNRAS, 327, 808 
Columns 3 and 4 give the halo ($\dot{m}_h$)
and the disk ($\dot{m}_d$) accretion rates. The corresponding cases 
are marked in Column 1. In Cases Ia and IIa, no Keplerian disk was 
placed in the equatorial plane of the halo. These are non-radiative 
hydrodynamical simulations and no Compton cooing is included. To show 
the effects of Compton cooling on the hydrodynamics of the flow, the 
Cases I(b-d) and II(b-d) are run till the same time as the Cases Ia and IIa (GGC12).
%Similar results  for different sets of parameter are already presented in
%(GGC12).  CHECK THIS ****************
%as w ****************** WHAT?   Put "See also, ****" as the refs. if you are 
%not incorporating the figures of the papers.

All of the simulations, presented in this Chapter, have been run for $\sim$ 2 sec.
For these cases, the time averaged values of the infall time scale
for the matter is found to be $\sim$ 0.12 s for $\lambda = 1.80$ cases and
$\sim$ 0.1 s for $\lambda = 1.75$ cases. These numbers have been computed from the
numerical simulation results. Therefore, it is seen that the simulations have been
run for $\sim$ 20 dynamical times. It can be seen that accretion flow configuration
reaches the equilibrium in about 0.1 s and the flow does not change if the
accretion rate parameters are not changed with time. 
Thus, we believe that all the results are the stable.

\subsection{Properties of the shocks in presence of cooling}

\begin{figure}
\centering{
\includegraphics[width=7.0cm]{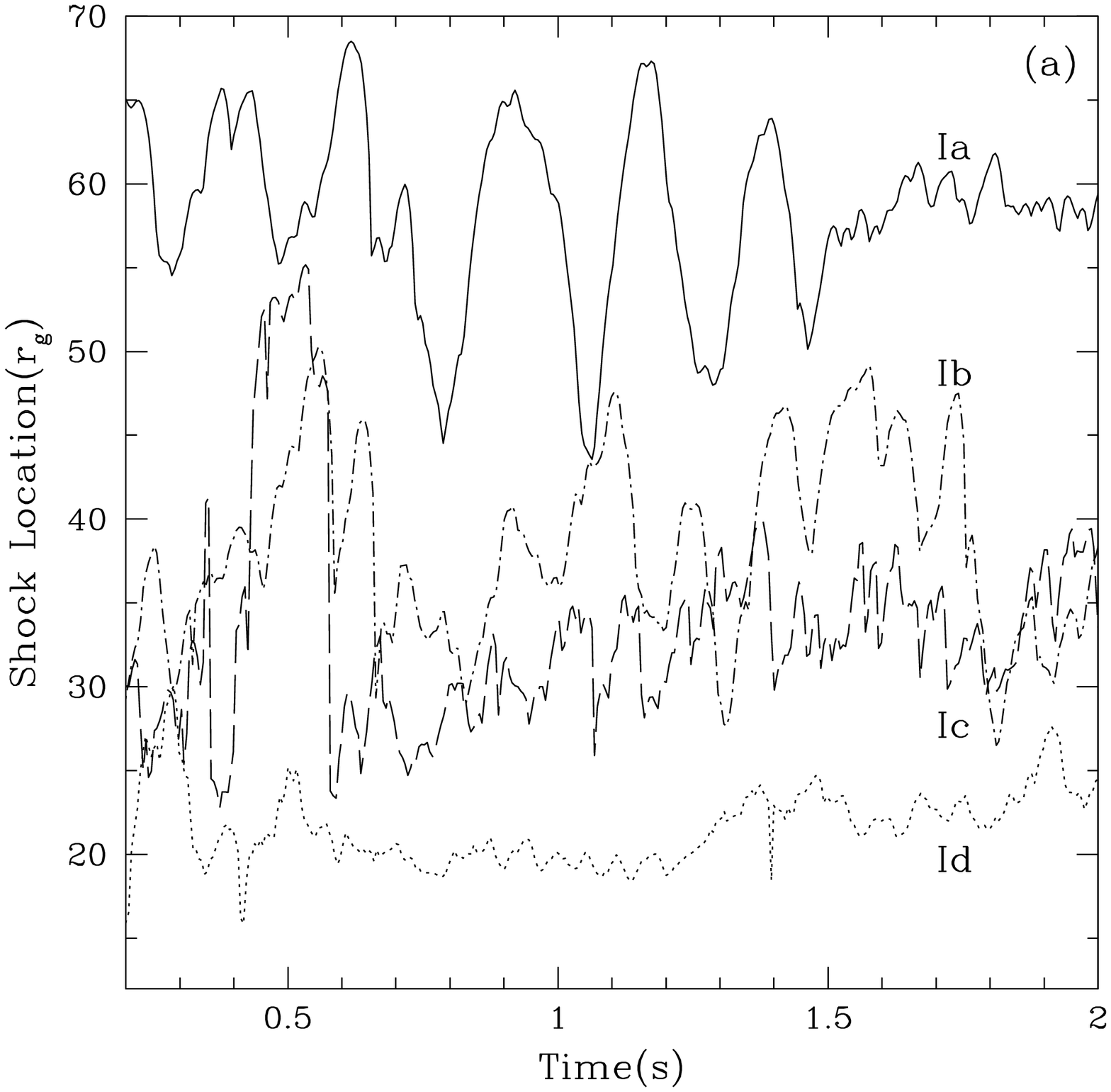}
\includegraphics[width=7.0cm]{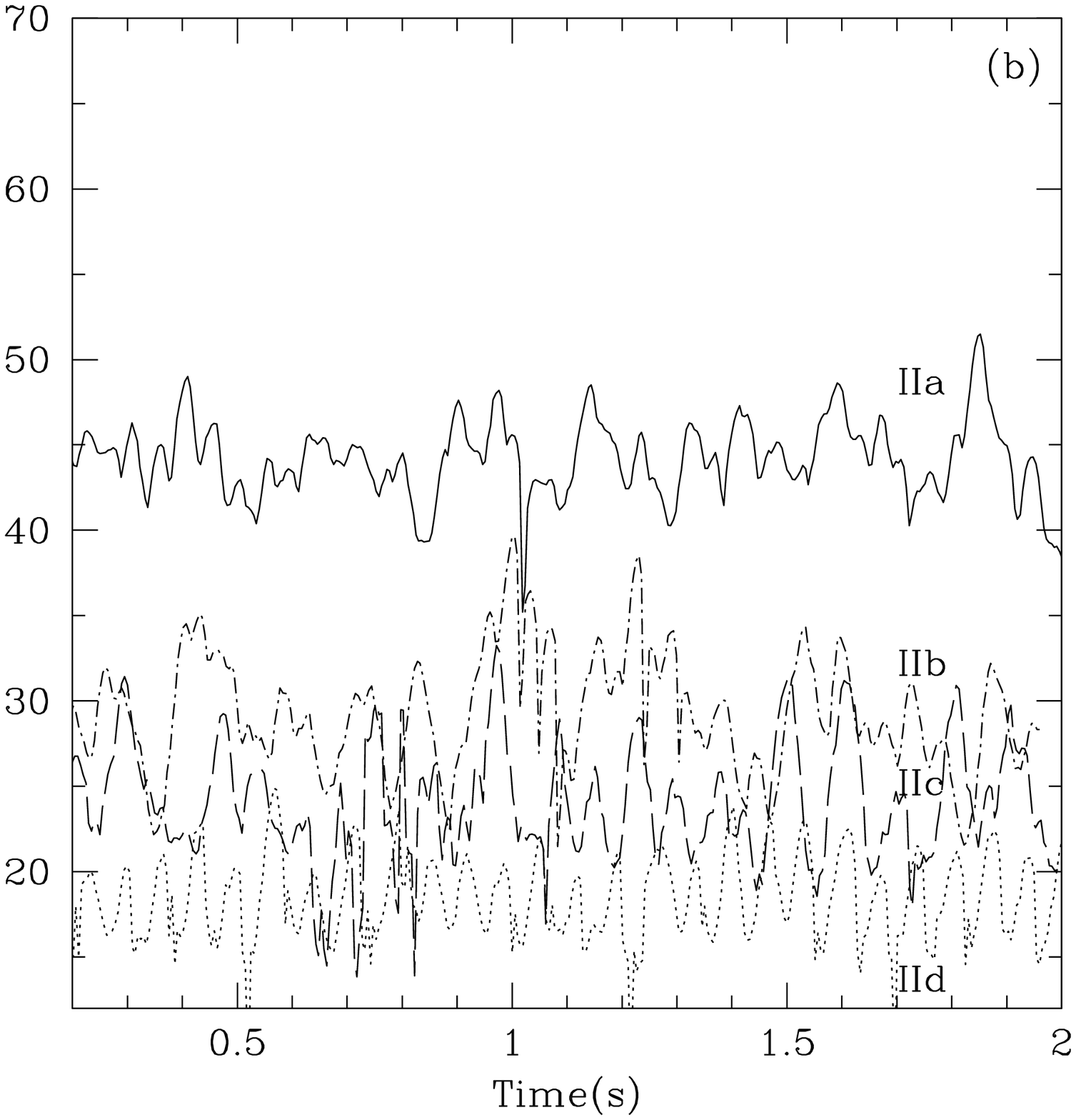}}
\caption{The variation of shock location (in $r_g$) at the equatorial 
plane with time (in sec) for different Keplerian disk rates $\dot{m}_d$, 
keeping the halo rate fixed at $\dot{m}_h = 0.1$. Simulation Cases are 
marked on each curve. (a) $\lambda = 1.80$ and (b) $\lambda = 1.75$. 
Cooling decreases the average shock location (see also, GGC12).}
\label{fig4.2}
\end{figure}
In Figs \ref{fig4.2}(a) and (b), we present the time variation of 
the shock location (in units of $r_g$) for various Cases (marked on 
each curve) given in Table 4. All the solutions  exhibit oscillatory 
shocks.  For no cooling, the higher angular momentum produces shocks 
at a higher radius, which is understandable, since the shock is primarily 
centrifugal force supported. However, as the cooling is increased the 
average shock location decreases since the cooling reduces the post-shock 
thermal pressure and the shock could not be sustained till higher thermal 
pressure is achieved at a smaller radius. The corresponding oscillations 
are also suppressed. The average shock location is found to be almost 
independent of the specific angular momentum at this stage (see also, GGC12). 

\begin{figure}
\centering{
\includegraphics[height=16cm]{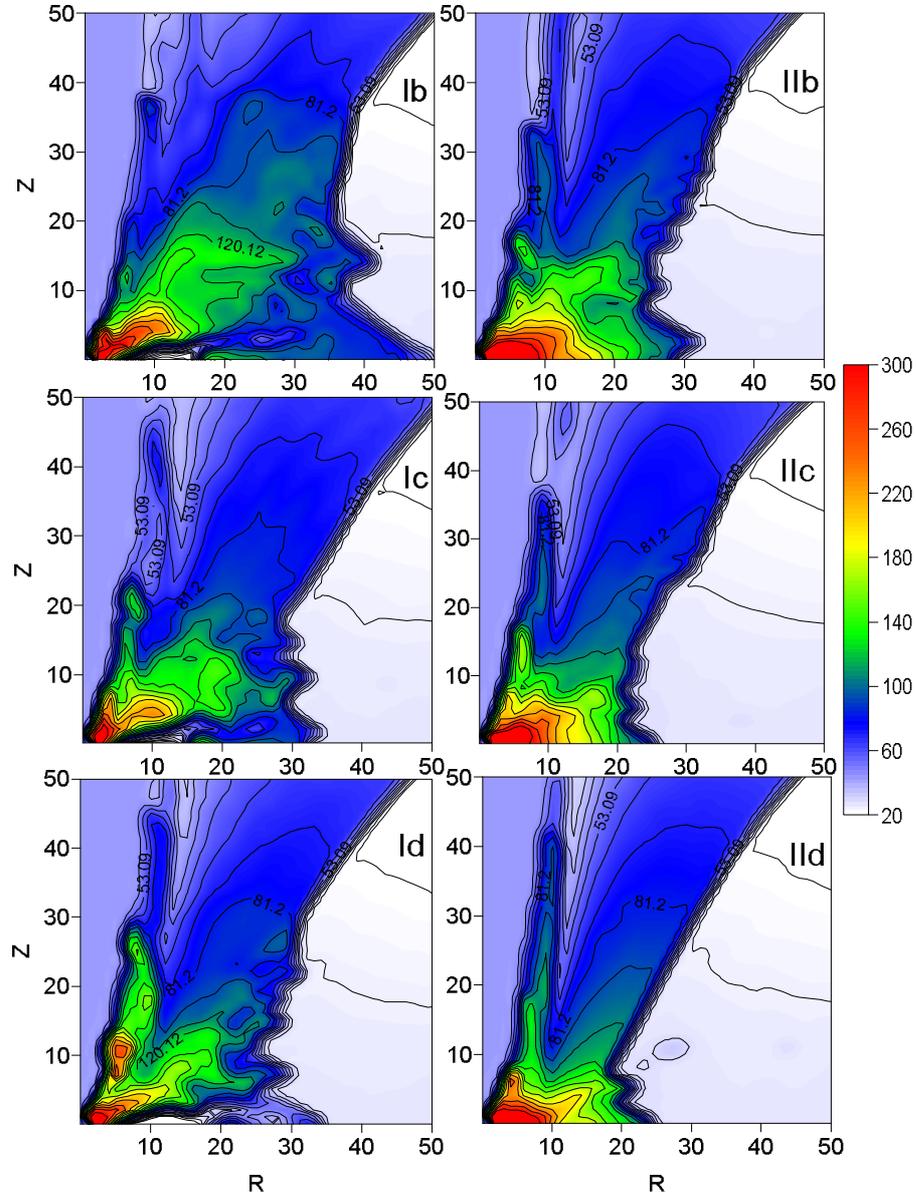}}
\caption{Colour map of final temperature distributions in the region 
($50 r_g \times 50 r_g$) of the accretion disk for different disk rates 
are shown. The left panel is for $\lambda = 1.80$ and the right is for 
$\lambda = 1.75$. As $\dot{m}_d$ is increased, we find that, the high 
temperature region (dark red) shrinks (see also, GGC12).}
\label{fig4.3}
\end{figure}
In Fig. \ref{fig4.3}, we show the colour map of the temperature distribution 
at the end of our simulation. We zoomed the region $50 r_g \times 50 r_g$. 
The specific angular momentum is $1.80$ in the left panel and $1.75$ in 
the right panel. Cases are marked. We note the collapse of the post-shock 
region as $\dot{m}_d$ is increased gradually (see also, GGC12). 

\begin{figure}
\centering{
\includegraphics[width=7.0cm]{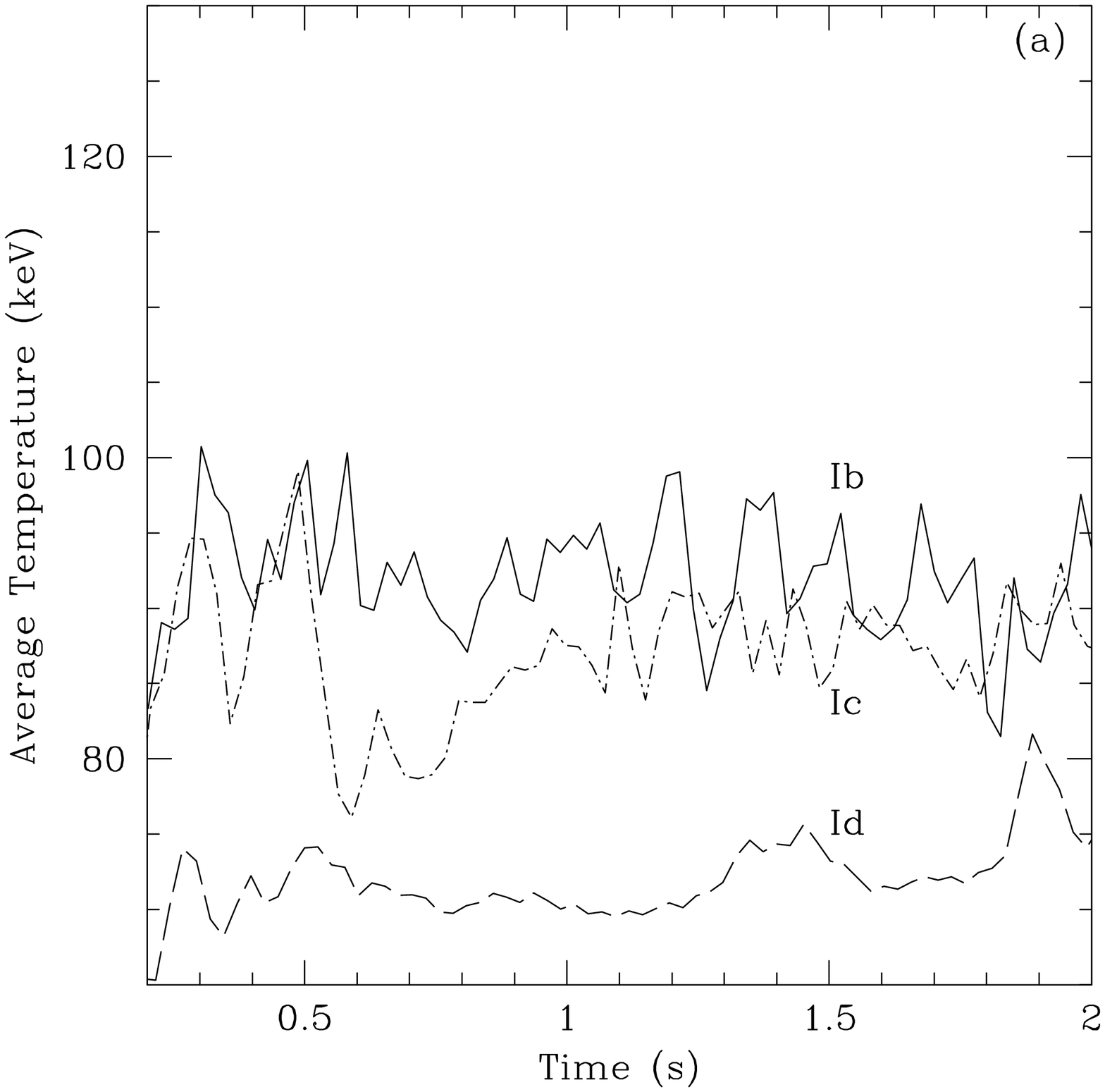}
\includegraphics[width=7.0cm]{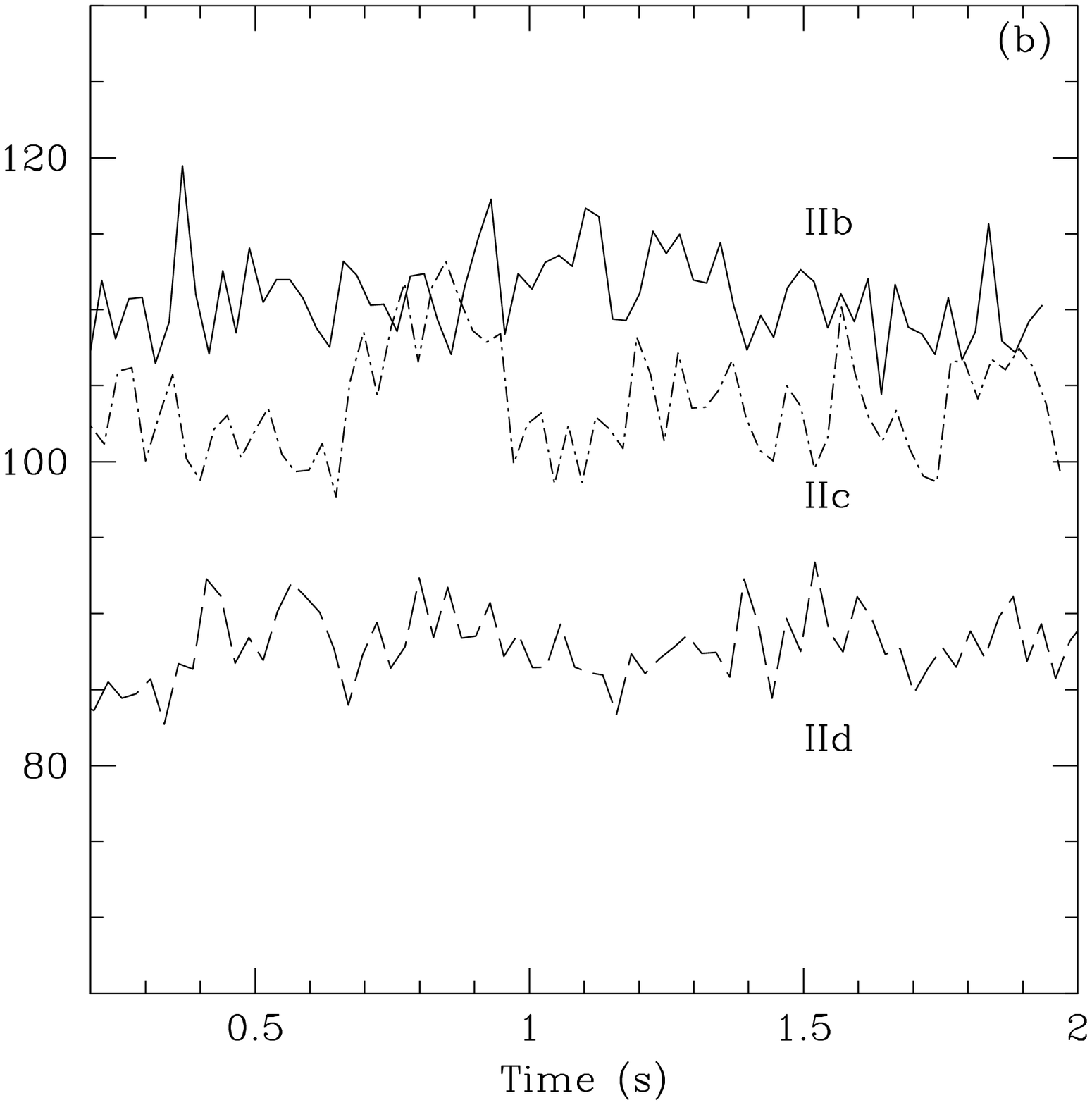}}
\caption{Variation of the average temperature (keV) of the post shock 
region with time (sec) for different Keplerian disk rates $\dot{m}_d$, 
keeping the halo rate fixed at $\dot{m}_h = 0.1$ (see also, GGC12). Parameters are the 
same as in Figs \ref{fig4.2}(a) and (b).}
\label{fig4.4}
\end{figure}

We take the post-shock region in each of these cases, and plot in Figs 
\ref{fig4.4}(a) and (b) the average temperatures of the post-shock region 
only for those cases where the cooling due to Comptonization is included. 
The average temperature was obtained by the optical depth weighted 
averaging procedure prescribed in CT95. The average temperature in the 
post-shock region is reduced rapidly as the supply of the soft photons 
is increased (see also, GGC12). 

\subsection{Effects of Comptonization on the outflow}

We now concentrate on how the outflow rate is affected by the Comptonization. 
Outflows move to very large distances and thus must not be bound to the 
system, i.e., the specific energy should be positive. Matter should also 
be of higher entropy as it is likely to be relativistic. Because of this,  
we wish to concentrate on the behaviour of matter which have highest 
energy and entropy (GGC12). Though we injected matter at the outer edge with a 
constant specific energy, the energy of matter in the post-shock region
is redistributed due to turbulence, Compton cooling and shock heating. 
Some entropy is generated as well. The high energy and high entropy matter 
escape in the form of a hollow cone around the axis. It is thus expected 
that if the post-shock region itself is collapsed due to Comptonization, 
the outflows will also be quenched (GGC12). We show this effect in our result below. 

\begin{figure}
\centering{
\includegraphics[height=18cm]{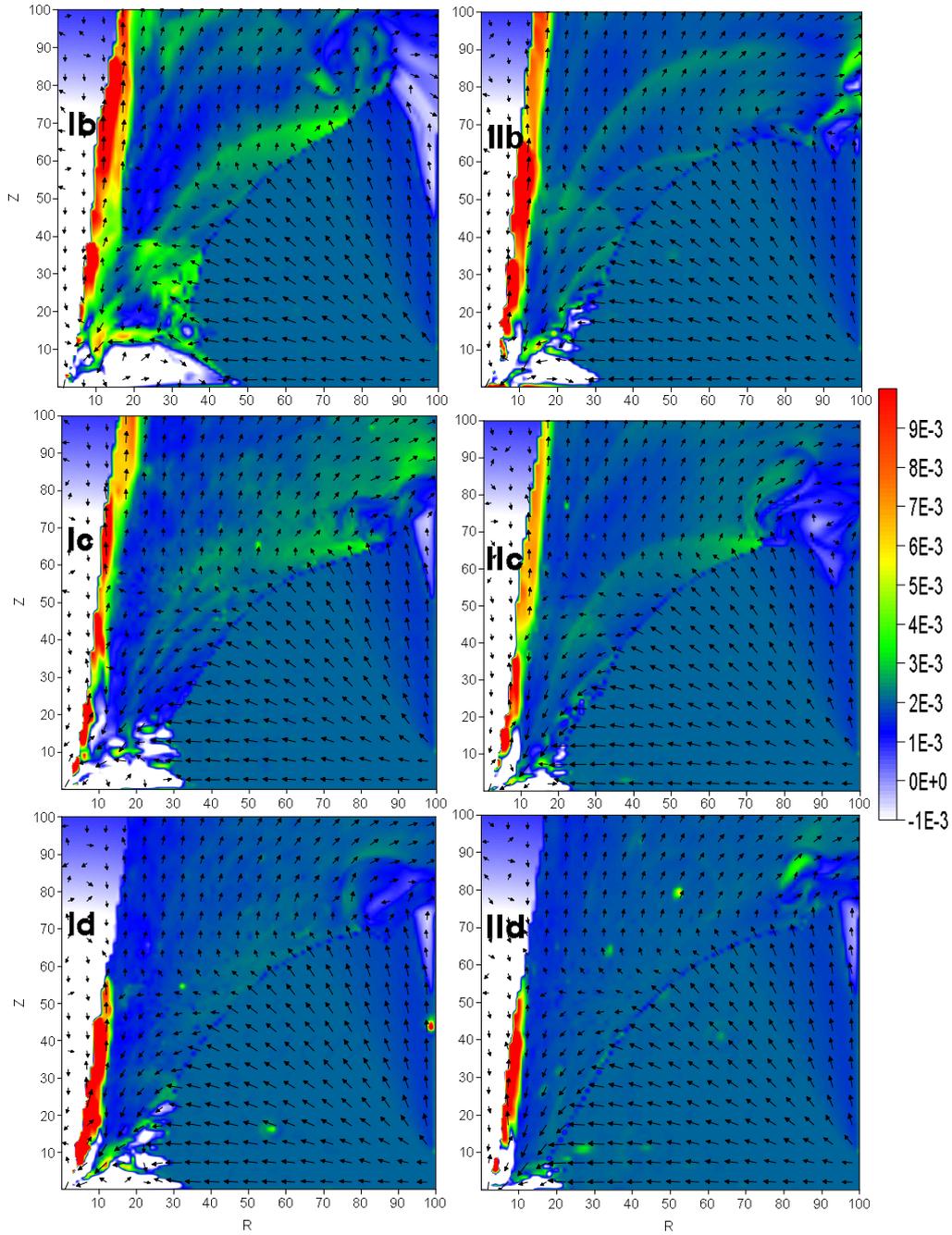}}
\caption{Colour map of final specific energy distribution inside the 
accretion disk for different disk rates. The high energy matter (dark red) 
are ejected outward as a hollow jet. The matter with a high energy 
flow decreases with the increase in disk rate. Velocity vectors at the 
injection boundary on the  right is of length $0.05$ (see also, GGC12).}
\label{fig4.5}
\end{figure}

In Fig. \ref{fig4.5}, we present the specific energy distribution for 
both the specific angular momenta (left panel for $\lambda=1.80$ and 
right panel for $\lambda=1.75$) for all the cooling Cases (marked in each box)
at the end of our simulation. The velocity vectors are also plotted. 
The scale on the right gives the specific energy. First we note that 
the jets are stronger for higher angular momentum. This is because the 
post-shock region (between the shock and the inner sonic point close to 
the horizon) is hotter. Second, lesser and lesser amount of matter has 
higher energy as the cooling is increased (see also, GGC12). 

\begin{figure}
\centering{
\includegraphics[height=18cm]{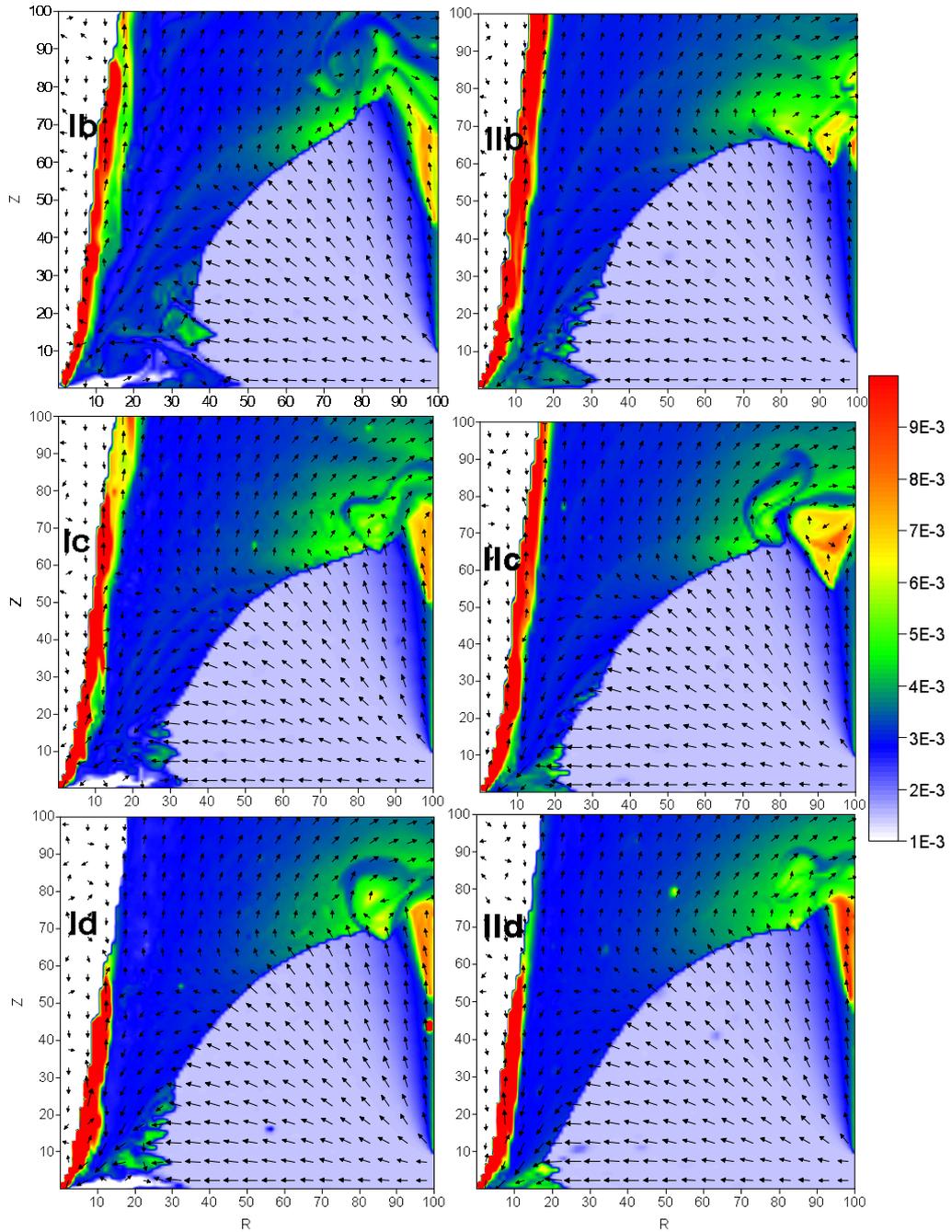}}
\caption{Color map of the final entropy ($K = \frac{P}{\rho^{\gamma}}$) 
distribution. Other parameters are as in Fig. \ref{fig4.5}. The 
high entropy flow decreases as the disk rate increases (see also, GGC12).}
\label{fig4.6}
\end{figure}

A similar observation could be made from Fig. \ref{fig4.6}, where 
the entropy distribution is plotted. The jet matter having upward 
pointing vectors have higher entropy. However, this region shrinks 
with the increase in Keplerian rate, as the cooling becomes significant 
the outward thermal drive is lost (GGC12). Here, the velocity vectors are of 
length $0.05$ at the outer boundary on the right and others are scaled accordingly. 

In order to quantify the decrease in the outflow rates with cooling, 
we define two types of outflow rates (GGC12). One is $\dot{M}_{out}$ which is 
defined to be the rate at which the outward pointing flow leaves the 
computational grid. This will include both the high and low energy 
components of the flow. 

\begin{figure}
\centering{
\includegraphics[width=7.0cm]{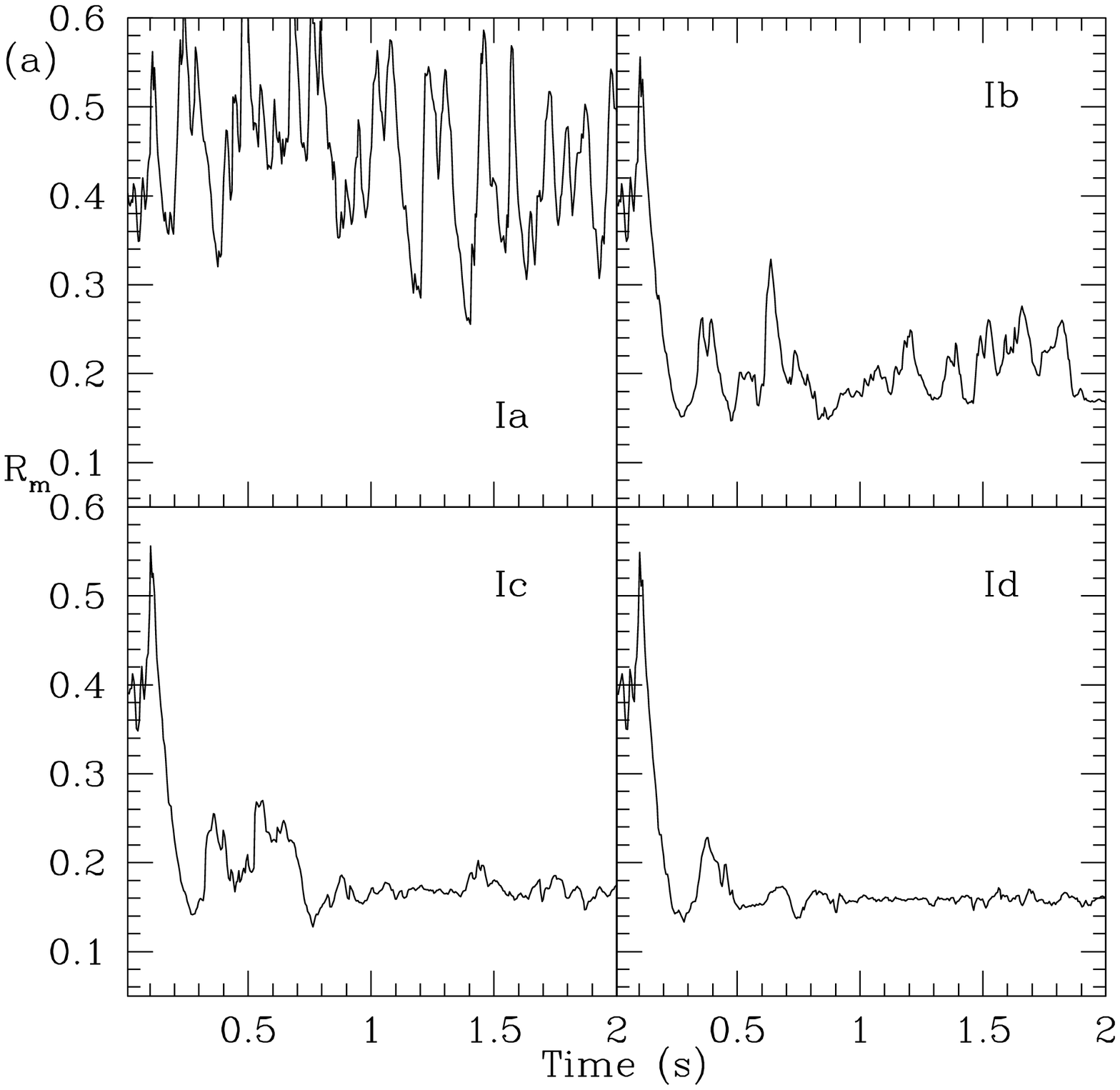}
\includegraphics[width=7.0cm]{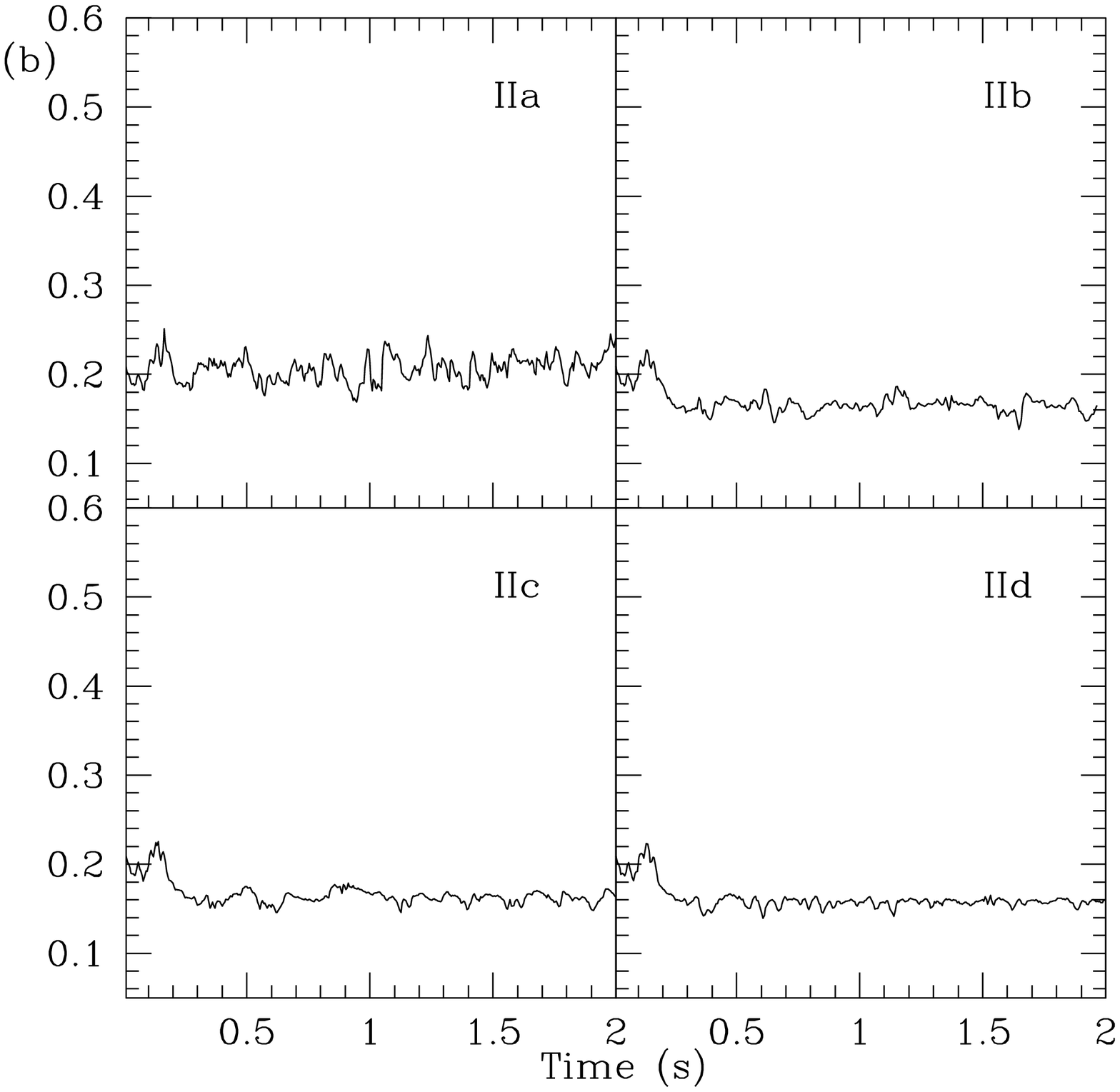}}
\caption{Variations of $R_{\dot{m}} (=\frac{\dot{M}_{out}}{\dot{M}_{in}})$ 
with time for different $\dot{m_d}$ are shown here. (a) $\lambda = 1.80$ 
and (b) $\lambda = 1.75$. The Cases are marked in each panel. The outflow 
rate is the lowest for the highest Keplerian disk accretion rate 
(Cases are Id and IId) (see also, GGC12).} 
\label{fig4.7}
\end{figure}

In Figs \ref{fig4.7}(a) and (b), we show the results of time variation 
of the ratio $R_{\dot{m}}$ (${\dot{M}}_{out}$ / ${\dot{M}}_{in}$) for the 
four cases (marked in each box), ${\dot{M}}_{in}$ being the constant 
injection rate on the right boundary. While the ratio is clearly a time 
varying quantity, we observe that with the increase in cooling, the 
ratio is dramatically reduced and indeed become almost saturated.% as soon as some cooling is introduced. 
Our rigorous findings once again verified 
what was long claimed to be the case, namely, the spectrally soft states
(those having a relatively high Keplerian rate) have weaker jets because 
of the presence of weaker shocks (Chakrabarti 1998b; C99; Das et al. 2001). %SUDIP ALSO CITE THE 
%SPECTRAL SOFTENING DUE TO OUTFLOWS ... my IJP paper ~ 1998-1999??? 
% Chakrabarti, S. K. 1998, Indian Journal of Physics, V. 72B, p. 565-569 (astro-ph/9810412)
% \bibitem[]{} Chakrabarti, S. K. 1999, A\&A, 351, 185
% Das, S., Chattopadhyay, I., Nandi, A., \& Chakrabarti, S. K. 2001, A\&A, 379, 683

Another measure of the outflow rate would be to concentrate only on the 
matter which has high positive energy and high entropy (GGC12). For 
concreteness, we concentrate only on the matter outflowing within $R=20r_g$ 
at the upper boundary of our computational grid. We define this to be
$J_{\dot{m}}$ ($= \frac{\dot{M}_{jet}}{\dot{M}_{in}}$). 

\begin{figure}
\centering{
\includegraphics[width=7.0cm]{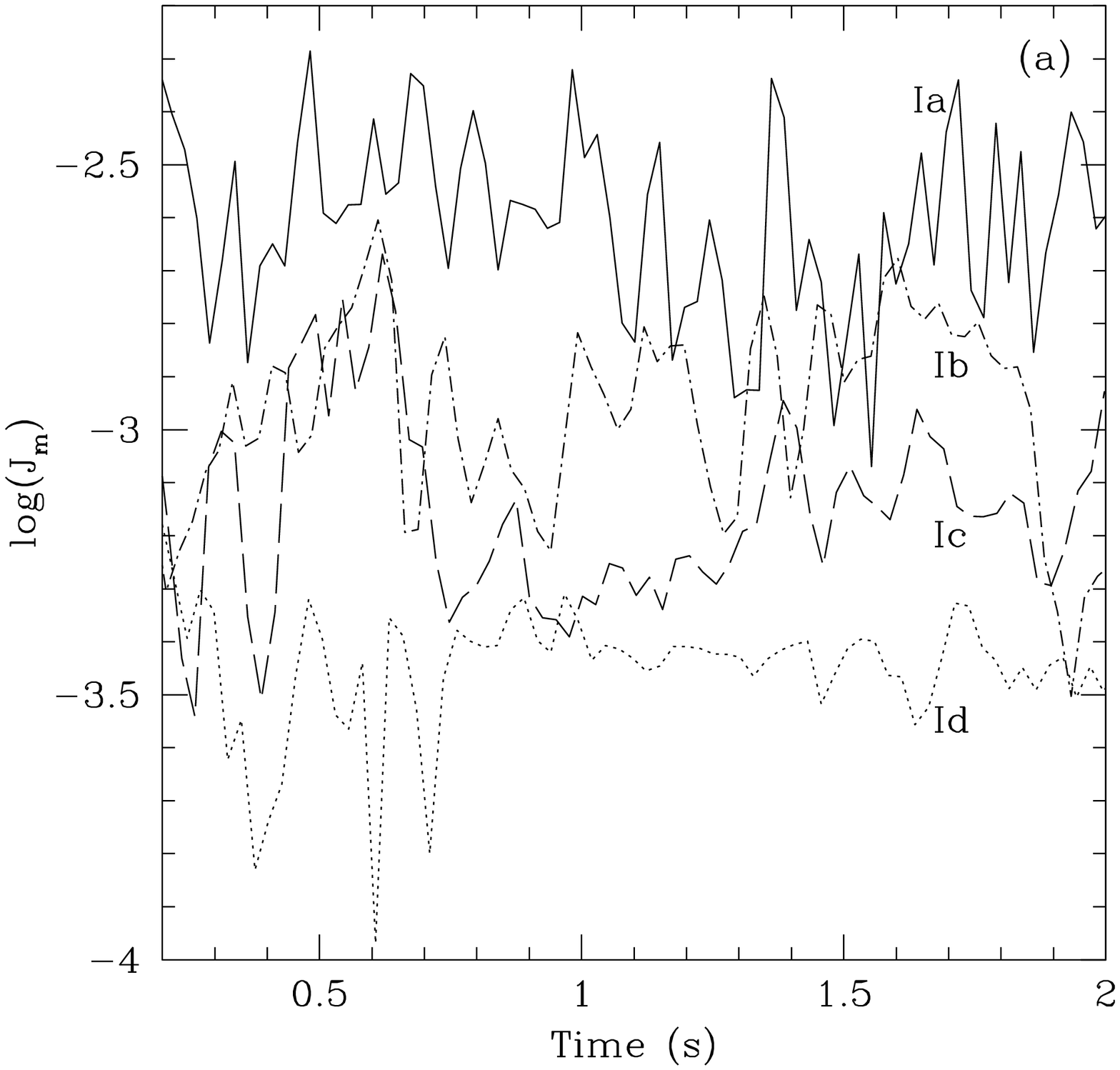}
\includegraphics[width=7.0cm]{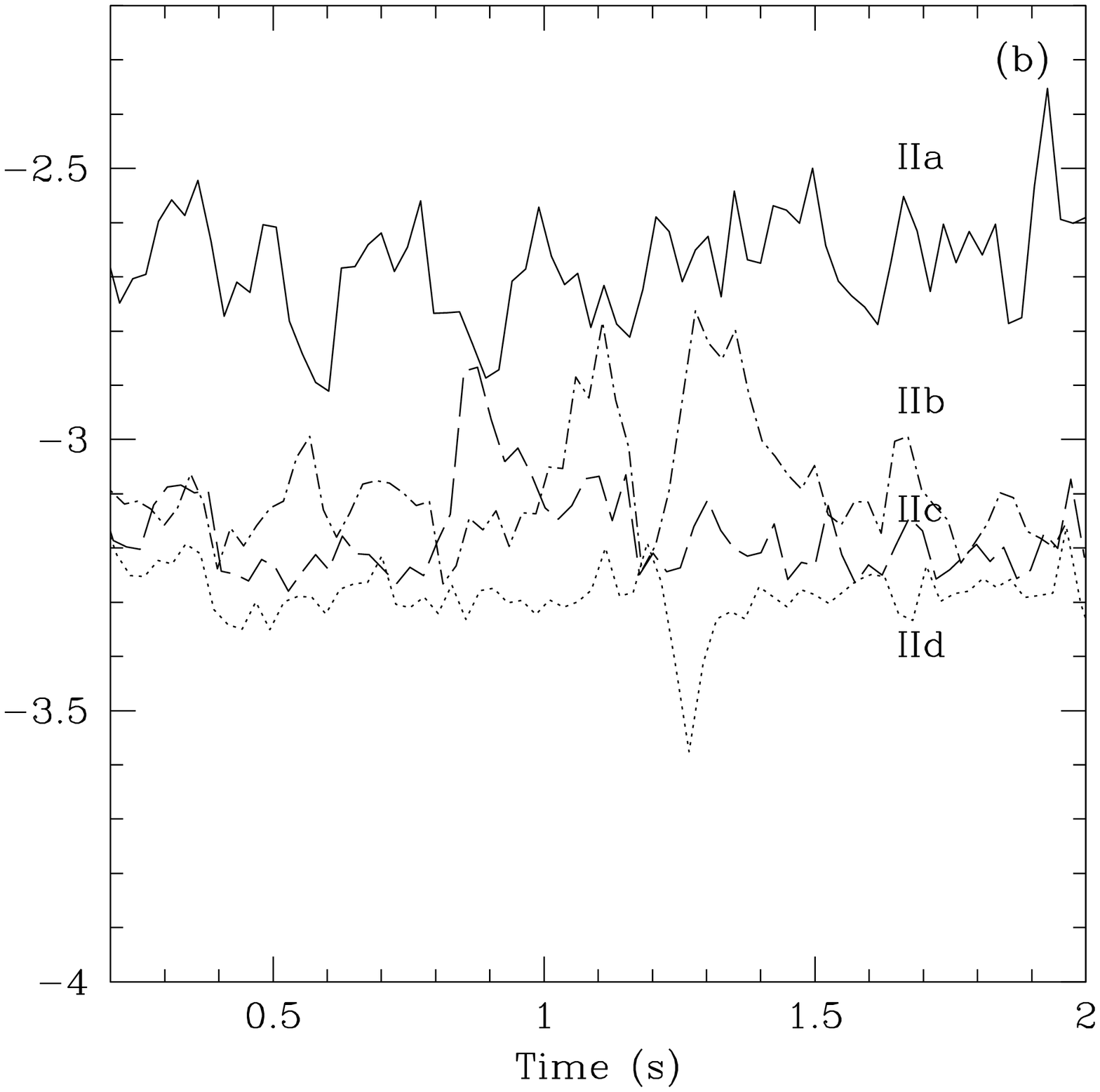}}
\caption{Variations of $J_{\dot{m}}$ ($= \frac{\dot{M}_{jet}}{\dot{M}_{in}}$) 
with time for different $\dot{m_d}$ are shown here (see also, GGC12). Here, $\dot{M}_{jet}$ 
and $\dot{M}_{in}$ are the high entropy (also, high energy) outflow and 
inflow rates, respectively. The left panel is for $\lambda = 1.80$ and 
the right panel is for $\lambda = 1.75$. The Cases are marked in each curve.}
\label{fig4.8}
\end{figure}

Figures \ref{fig4.8}(a) and (b) show time variation of $J_{\dot{m}}$. 
The different cases are marked on the curves. This outflow rate 
fluctuates with time. We easily find that the cooling process reduces 
this high energy component of matter drastically. Thus both the slow 
moving outflows and fast moving jets are affected by the Comptonization 
process at the base (see also, GGC12). 

\subsection{Spectral properties of the disk-jet system}

In each simulation, we also store the photons emerging out of the 
Computational box after exchanging energy and momentum with the free 
electrons in the disk matter (Ghosh et al. 2009; GGCL10). When the 
Keplerian disk rate is increased, 
the number of injected soft photons go up, cooling every electron in the 
sub-Keplerian halo component. Thus, the relative availability of the soft 
photons and the hot electrons in the disk and the jet dictates whether 
the emergent photons would be spectrally soft or hard.

\begin{figure}
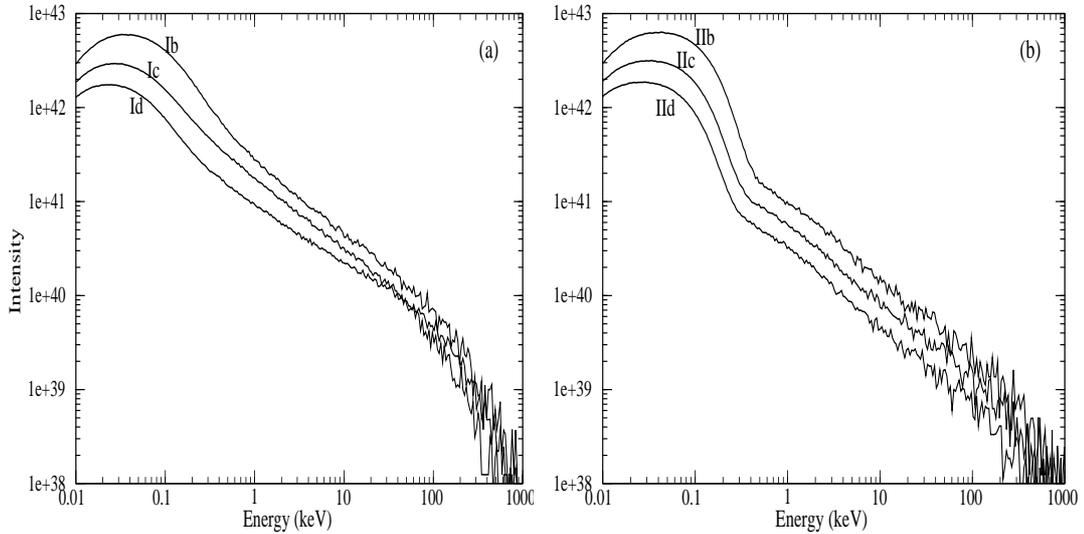
 
\vskip 1cm
\centering{
\includegraphics[width=7.0cm,height=7cm]{Figure/Fig4.9a.eps}
\includegraphics[width=7.0cm,height=7cm]{Figure/Fig4.9b.eps}}
\caption{The final emitted spectra for different disk rates are shown 
for (a) $\lambda = 1.80$ and (b) $1.75$. Corresponding Cases are marked 
on each curve. The spectrum appears to become softer with the increase 
in $\dot{m_d}$ (see also, GGC12).}
\label{fig4.9}
\end{figure}

In Figs \ref{fig4.9}(a) and (b), we show three spectra for each of 
the specific angular momentum: (a) $\lambda=1.80$ and (b) $\lambda=1.75$. 
The Cases are marked. We see that a spectrum is essentially made up of 
the soft bump (injected multi-colour blackbody spectrum from the 
Keplerian disk which are unscattered), and a Comptonized 
spectrum with an exponential cutoff -- 
the cutoff energy being dictated by the electron cloud temperature. 
%In these spectra, the soft bump is peaking at extremely low energies
As the disk rates are very small ($\sim$ 0.0001 Eddington rate), 
the temperature of the disk is small and the resulting spectra 
show peaks at extremely low energies in the above Figure.
If we define the energy spectral index  $\alpha$ to be 
$I(E) \propto E^{-\alpha}$ in the region $2-20$ keV, we note that 
$\alpha$ increases, i.e., the spectrum softens with the increase in 
$\dot{m}_d$. This is consistent with the predictions made by the 
static models of CT95 and Chakrabarti (1997). 

\begin{figure}
\vskip 2cm
\centering{
\includegraphics[width=14cm,height=7cm]{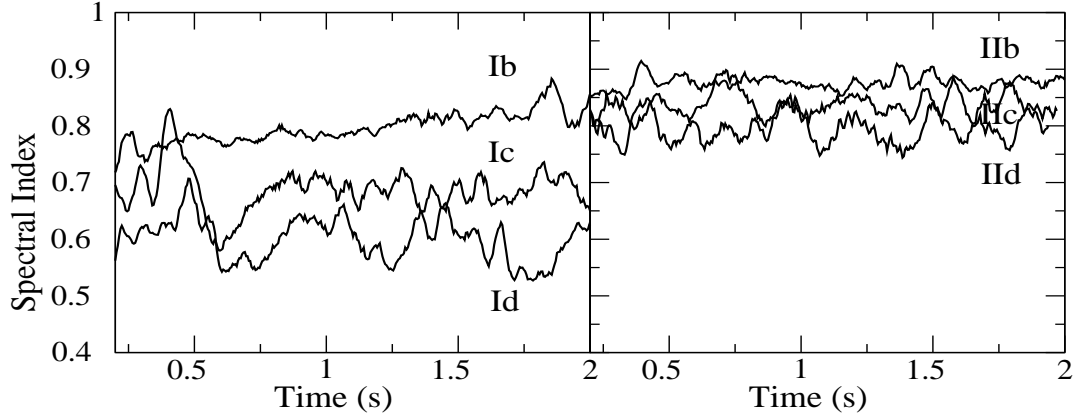}}
\caption{Time variation of the spectral index [$\alpha$, $I(E) \propto E^{-\alpha}$] 
for different disk rates. Different Cases are marked. We note that as the 
accretion rate goes up, the average $\alpha$ increases, i.e., the spectrum 
softens (see also, GGC12). 
}
\label{fig4.10}
\end{figure}

In reality, since the disk is not stationary, the spectrum also varies 
with time, and so is $\alpha$. In Fig. \ref{fig4.10}, we present the 
time variation of the spectral index for the different Cases. We clearly 
see that the spectral index goes up with the increase in the disk 
accretion rate (see also, GGC12). Thus, on an average, the spectrum softens. 

In the above works, we have not considered the effects of magnetic field.
In presence of magnetic field, synchrotron photons may be emitted from the
post-shock region which affect the spectral shape (Mandal \& Chakrabarti 2005a, 2005b;
Chakrabarti \& Mandal 2006; Mandal \& Chakrabarti 2008). Synchrotron emission may occur
% Mandal S., Chakrabarti S. K., 2005a, A\&A, 434, 839
% Mandal S., Chakrabarti S. K., 2005b, Ap\&SS, 297, 169
from the Maxwell-Boltzmann electrons of the pre-shock as well as the CENBOL and jet regions
and from the shock accelerated power-law electrons. These photons may get inverse-Comptonized
in the CENBOL and produce high energy X and gamma rays extending upto a few MeV,
as shown in the above references. These effects are not included in our radiative-hydro calculation.
\clearpage

	\reseteqn
	\resetsec
	\resetfig
	\resettab

%~~~~~~~~~~~~~~~~~~~~~~~~~~~~~~~~~~~~~~~~~~~~~~~~~~~~~~~~~~~~~~~~~~~~~~~~~~~~~~
\alpheqn
\resec
\refig
\retab
%%%%%%%%%%%%%%%%%%%%%%%%%%%%%%%%%%%%%%%%%%%%%%%%%%%%%%%%%%%
% Chapter 5 : Quasi Periodic Oscillations in Radiative Transonic Flow
%%%%%%%%%%%%%%%%%%%%%%%%%%%%%%%%%%%%%%%%%%%%%%%%%%%%%%%%%%%

\def\k{{\bf k}}
\def\aug{{\tilde{\cal H}}}
\def\ahat{{\mathaccent "7E {\cal A}}}
\def\bhat{{\mathaccent "7E {\cal B}}}
\def\chat{{\mathaccent "7E {\cal F}}}
\def\dhat{{\mathaccent "7E {\cal D}}}
\def\shat{{\mathaccent "7E {\cal S}}}
\def\proj{{\cal P}}
\def\ket{\vert \vert  \{ \emptyset \} \rangle}
\def\ket2{\vert \vert \otimes \{ R \} \rangle}
\def\sqr{$^{2}$}
\def\T{$T$}
\def\P{$P$}
\def\Q{$Q$}
\def\G{$G$}
\def\M{{\bf M}}
\def\H{{\bf H}}
\def\I{{\bf I}}
\def\Pr{{\bf P}}
\def\Tr{{\bf T}}
\def\eq{\enskip =\enskip}
\def\pls{\enskip+\enskip}
\def\mns{\enskip -\enskip}
\def\oper{O^{\bf k}_{R}}
\def\aug{\tilde{\cal H}}
\def\envfn{\chi^{\alpha}_{RL}}
\def\struc{S^{\alpha}_{RL,R^{\prime}L^{\prime}}}
\def\rR{r_{R}}

\newpage
\markboth{\it Quasi Periodic Oscillations in a Radiative Transonic Flow}
{\it Quasi Periodic Oscillations in a Radiative Transonic Flow}
\chapter{QUASI PERIODIC OSCILLATIONS IN A RADIATIVE TRANSONIC FLOW}

%\section{Introduction}
Quasi-periodic oscillations (QPOs) of observed X-rays are very important features for the study 
of accreting stellar mass black holes. Observations of QPOs in black hole candidates have been 
reported quite extensively in the literature (e.g., Remillard \& McClintock 2006; Chakrabarti et al. 2008 and the references 
therein). They are believed to be the manifestations of some regular movements 
of the underlying accretion flows and happen to be closely connected to the spectral 
states (Remillard \& McClintock 2006; Titarchuk, Shaposhnikov \& Arefiev 2007).  
%S. K. Chakrabarti, D. Debnath, P.S. Pal, A. Nandi, R. Sarkar, M. M. Samanta,
%P.J. Wiita, H. Ghosh and D. Som, Quasi periodic oscillations due to axisym-
%metric and non-axisymmetric shock oscillations in black hole accretion,
%Proc. 11th Marcel Grossman Meeting on General Relativity, Eds. H. Kleinert, R.
%T. Jantzen & R. Ruffini, World Scientific (pp 569-588) (2008)
%Remillard, Ronald A., McClintock, Jeffrey E., 2006,ARA&A, 44, 49
% Titarchuk L., Shaposhnikov N., Arefiev V., 2007, ApJ, 660, 556
X-ray transient sources in our galaxy exhibit various types of QPOs with frequencies ranging from 
$~ 0.001 - 0.01$ Hz to a few hundreds of Hz. (Morgan, Remillard \& Greiner 1997; Paul et al. 1998; 
Yadav et al. 1999; Remillard \& McClintock 2006). 
%Morgan, E. H., Remillard, R. A. & Greiner, J. 1997, ApJ 482, 993
%B. Paul, P. C. Agrawal, A. R. Rao, M. N. Vahia, and J. S. Yadav, 1998, ApJ 492, L63
%Yadav, J. S.; Rao, A. R.; Agrawal, P. C.; Paul, B.; Seetha, S.; Kasturirangan, K. 1999, ApJ, 517, 935
However, the  quasi-periodic behavior is not always observed. More common feature is the erratic 
variation of photon count rates even in a timescale of seconds (GGC13).

It has also been reported that there should exist a correlation between 
the QPO frequency and the spectral index (Remillard \& McClintock 2006; 
Chakrabarti, Debnath, Nandi \& Pal 2008; Chakrabarti, Dutta \& Pal 2009; Debnath, 
Chakrabarti \& Nandi 2010; Stiele, Belloni, Kalemci \& Motta 2013). These authors reported 
a rise in photon index with increasing centroid frequency of QPOs 
for various black hole candidates during the rising phase of the outbursts. 
It is also reported that during the decay phase of the outburst, the trend 
is time reversed, i.e., a fall in photon index with the decrease of the QPO frequency. 
The correlation does not depend much on any individual outburst, 
rather they follow a general trend. 
%S. K. Chakrabarti, D. Debnath, A. Nandi, and P. S. Pal, A&A 489, L41–L44 (2008)
%Sandip K. Chakrabarti, Broja G. Dutta and P. S. Pal, MNRAS, 394, 1463–1468 (2009)
%Debnath, D.; Chakrabarti, S. K.; Nandi, A., 2010A&A...520A..98
%Stiele, H.; Belloni, T. M.; Kalemci, E.; Motta, S. 2013MNRAS.429.2655S

The TCAF model (CT95) which explains the spectral properties quite satisfactorily,
has been applied to explain the QPOs. In this model, QPOs are the result of the 
oscillation of the centrifugal pressure supported shocks. Judging from the rapidity in which 
the spectral properties change, a realistic model would be where the 
`corona' itself moves at a shorter timescale. According to the TCAF model, 
the CENBOL (i.e., the post-shock region) which forms primarily because of the centrifugal barrier
in the sub-Keplerian flow, intercepts some of the soft photons emerging 
from the Keplerian disk and redistributes them in the higher energy bands.
In this model, the oscillation of X-ray intensity is caused by the
the oscillation of the CENBOL region (MSC96; RCM97; CM00; CAM04) which reprocesses 
different amounts of soft photons at different phases of oscillation. The numerical 
simulation of low-angular momentum accretion flows including the thermal cooling 
(MSC96; CAM04) or dynamical cooling (through outflows, e.g., RCM97) shows clearly 
that the shocks oscillate with frequencies similar to the observed QPO frequencies.

So far, while the shock-oscillation was generally proposed to be the 
cause of low frequency QPOs, and the numerical simulations generally showed that this 
could very well be the case (MSC96; CAM04), there was no attempt to 
include the Keplerian disk into the simulation and only enhanced 
power-law cooling was used as a proxy to inverse-Comptonization. 
In the work presented in this Chapter, we use the time dependent 
radiation hydrodynamic simulation code which includes both the source 
of seed photons and actual computation of the production of hard photons 
through the Comptonization process (GGC13). We analyze the resulting light curves 
to find the QPOs and other timing properties using the NASA HEAsoft package, 
as though our output corresponds to observed photon flux. We clearly find a 
correlation between the flow parameters and the QPO frequencies (GGC13). 
We vary the accretion rates and study the dependence.  

\section{Simulation set up and procedure}

%%%%%%%%%%%%%%%%%%%Fig.5.1%%%%%%%%%%%%%%%%%%%%%%%%%
\begin{figure}
\centering{
\includegraphics[height=12cm,angle=0]{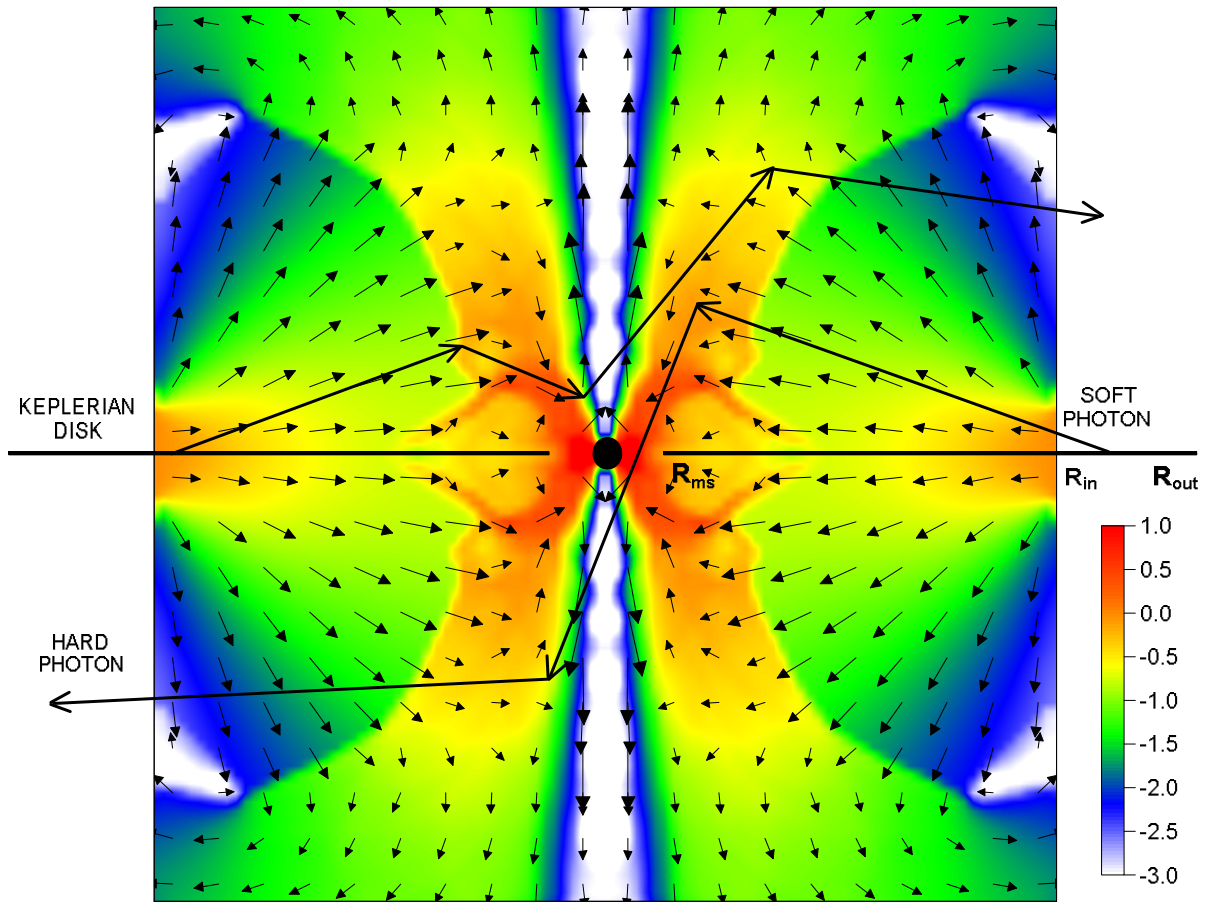}}
\vskip 2cm
\caption{The schematic diagram of our simulation set up. The velocity 
vectors of the infalling matter are shown. The colors show the normalized 
density in a logarithmic scale. The zig-zag trajectories are the typical 
paths followed by the photons. Velocity vectors are drawn for 
$\lambda=1.73$ (GGC13).
}
\label{fig5.1}
\end{figure}
%%%%%%%%%%%%%%%%%%%%%%%%%%%%%%%%%%%%%%%%%%%%%%%%%%%

In Fig. \ref{fig5.1}, we present a schematic diagram of our simulation 
set up (GGC13). The simulation set up is similar to the one used in 
Chapter 4. We consider a TCAF where the standard 
Keplerian disk is flanked by the sub-Keplerian flow (CT95). The soft photons, 
emerging out of the Keplerian disk, are intercepted and reprocessed via 
Compton or inverse-Compton scattering by the sub-Keplerian matter, 
both in the post-shock and the pre-shock regions.  The sub-Keplerian 
matter enters the simulation box through the outer boundary at $R_{in}$. The outer edge 
of the Keplerian disk is assumed to be at $R_{out}$ and it extends inside 
up to the marginally stable orbit $R_{ms}$. A black hole of mass 
$M_{bh}$ is located at the center.

\subsection{Sub-Keplerian and Keplerian flows}

The sub-Keplerian flow is simulated in two dimensions using the TVD code 
as is done in Chapters 3 and 4. To model the initial injection  
of matter, we consider an axisymmetric flow of gas in the pseudo-Newtonian 
gravitational field (PW80) of a non-rotating black hole of mass $M_{bh}=10M_\odot$, located
at the center in the cylindrical coordinate system $[R,\theta,z]$. 
Our computational box occupies one quadrant of the $R-z$ plane with
$0 \leq R \leq 100$ and $0 \leq z \leq 100$. 
All the calculations were performed with $512 \times 512$ equi-spaced 
cells (GGC12; GGC13). Thus, each grid has a size of $0.19$ in units of the 
Schwarzschild radius ($r_g=2GM_{bh}/c^2$). 
The incoming gas enters the box through the outer boundary, located 
at $R_{in} = 100 r_g$. We have chosen the normalized density of the incoming gas 
${\rho}_{in} = 1$ for convenience.  We provide sound speed $a$ (i.e., temperature) and velocity $v$ of the incoming
gas at outer boundary points. Values of $a$ and $v$ are computed in such a way that
total specific energy at outer boundary points remains the same as that of
the injected energy. In order to mimic the horizon of the black hole at $r_g$, we place an 
absorbing inner boundary at $r = 1.5 r_g$, inside which all material 
is completely absorbed into the black hole. For the sake of concreteness, 
we consider the sub-Keplerian flow with a specific angular momentum 
$\lambda=1.73$ and specific energy $\epsilon=0.0021$. These parameters
are chosen so that we find a stable shock solution (Chakrabarti \& Das 2001).

The Keplerian disk is the standard Shakura-Sunyaev (SS73) disk. The soft 
photons are produced from both the surfaces of the Keplerian disk (GGCL10; GGGC11; GGC12). 
The inner edge of the disk has been kept fixed at the marginally stable 
orbit $R_{ms} = 3 r_g$, while the outer edge is assumed to be at $R_{out}= 200 r_g$. 
The source of the soft photon has a multicolor blackbody 
spectrum and the emission is blackbody type with the local surface 
temperature $T(R)$ and the disk between radius $R$ to $R+\delta R$ 
produces $dN(R)$ number of soft photons. The form of $T(R)$ and $dN(R)$ 
are given in Eq. (\ref{eqno1.1}) and (\ref{eqno2.7}), respectively. 

\subsection{Simulation Procedure}
All the simulations have been performed using the time dependent
radiation hydrodynamic simulation code described in Chapter 3 and 4. The 
simulation procedure for these cases are similar to the 
procedure used for the simulations described in Chapter 4 and in 
GGGC11 and GGC12. 

For a particular simulation, we use the Keplerian disk rate ($\dot{m}_d$) 
and the sub-Keplerian halo rate ($\dot{m}_h$) as parameters. The specific 
energy ($\epsilon$) and the specific angular momentum ($\lambda$) 
determine the hydrodynamic (shock location, number density and velocity 
variations etc.) and the thermal properties of the sub-Keplerian matter. 
In the next Section, we present the results and discuss the possible 
implications (GGC13).

%%%%%%%%%%%%%%%%%%%%%%%%%%%%%%%%%%%%%%%%%%%%%%%
\section{Results and discussions}

\begin{center}
\begin {tabular}[h]{|c|c|c|c|c|c|c|c|}
\hline
\multicolumn{8}{|c|}{Table 5: Parameters used for the simulations and a summary of results (GGC13).}\\
\hline Case ID & $\epsilon, \lambda$ & $\dot{m}_d$ & $\dot{m}_h$ & ${\rm <R_{sh}>}$ & $\nu_{QPO}$ & $<\alpha>$ & $t_{in}\over t_{cool}$\\
\hline
C1 & 0.0021, 1.73 & 1e-4 & 0.1 & 25.75 & No QPO & 0.826 & 0.694\\
C2 & 0.0021, 1.73 & 2e-4 & 0.1 & 22.82 & 10.63 & 0.853 & 0.844\\
C3 & 0.0021, 1.73 & 3e-4 & 0.1 & 20.39 & 12.34 & 0.868 & 0.954\\
C4 & 0.0021, 1.73 & 4e-4 & 0.1 & 18.62 & 14.63 & 0.873 & 0.944\\
C5 & 0.0021, 1.73 & 5e-4 & 0.1 & 18.33 & 22.74 & 0.901 & 1.071\\
C6 & 0.0021, 1.73 & 1e-3 & 0.1 & 15.02 &  18.2 & 1.074 & 0.993\\
C7 & 0.0021, 1.73 & 1e-2 & 0.1 &  3.4  & No QPO & 1.139 & 3.558\\
C8 & 0.0021, 1.73 & 1e-1 & 0.1 &  3.6  & No QPO  & 1.102 & 33.236\\
C9 & 0.0021, 1.73 & 3e-4 & 0.05 & 23.53 & 10.74 & 0.998 & 0.859\\
C10 & 0.0021, 1.73 & 3e-4 & 0.15 & 18.37 & 22.80 & 0.797 & 1.031\\
C11 & 0.0021, 1.73 & 3e-4 & 0.2 & 16.78 & No QPO & 0.757 & 1.049\\
\hline
\end{tabular}\\
\end{center}

In Table 5, we show the parameters used for the simulations and a summary 
of the results (GGC13). In Column 2, we list the specific energy ($\epsilon$) 
and the specific angular momentum ($\lambda$) of the sub-Keplerian flow 
at the outer boundary.  Columns 3 and 4 show the Keplerian disk rate 
($\dot{m}_d$) and sub-Keplerian halo rate ($\dot{m}_h$), respectively. 
Both are in units of Eddington rate. In Column 5, we present the time 
averaged shock location in $r_g$ near the equatorial plane. For some 
combinations of $\dot{m}_d$ and $\dot{m}_h$, we find QPOs, about which 
we shall comment later. In Column 6, we list the QPO frequencies in Hz. 
The time averaged spectral slope [$\alpha,~I(E)\propto E^{-\alpha}$] 
is given in Column 7. In the last Column, we show the ratio of the infall 
time scale ($t_{in}$) and the cooling time scale ($t_{cool}$). The case 
IDs are given in the first Column.

\subsection{Spectral properties}
%%%%%%%%%%%%%%%%%%%Fig.5.2%%%%%%%%%%%%%%%%%%%%%%%%%
\begin{figure}
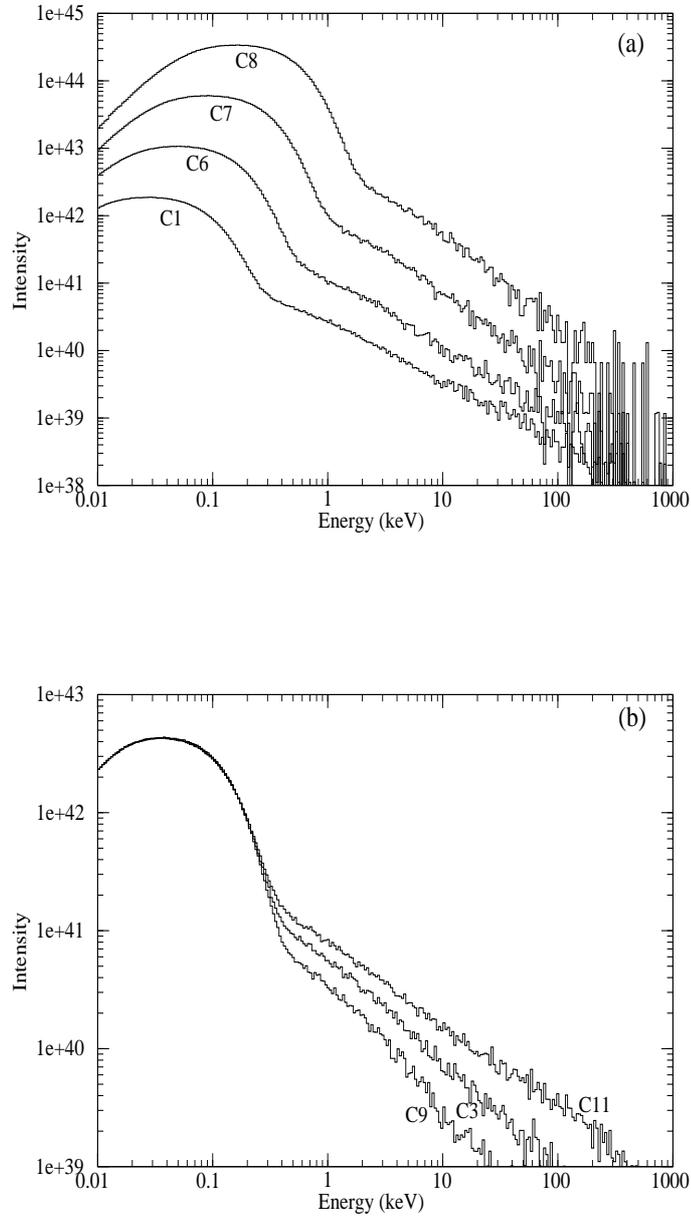

\centering{
\includegraphics[width=9cm,height=7cm]{Figure/Fig5.2a.eps}
\vskip 2cm
\includegraphics[width=9cm,height=7cm]{Figure/Fig5.2b.eps}}
\caption{
a) Variation of the shape of the spectrum when $\dot{m}_d$ is increased 
by a factor of 10 starting from $\dot{m}_d = 0.0001$ to $0.1$. 
Case IDs are marked for each plot. The spectrum becomes softer as 
$\dot{m}_d$ is increased.
b) Variation of the spectra when the halo rate $\dot{m}_h$ is increased 
keeping the disk rate constant at $\dot{m}_d = 0.0003$. The spectrum becomes 
harder as $\dot{m}_h$ is increased (GGC13). 
}
\label{fig5.2}
\end{figure}
%%%%%%%%%%%%%%%%%%%%%%%%%%%%%%%%%%%%%%%%%%%%%%%%%%%

In Figs \ref{fig5.2}(a) and (b), we show the variation of the shape of 
the final emergent spectra (GGC13). In Fig. \ref{fig5.2}(a), the Keplerian disk 
rate $\dot{m}_d$ is increased by a factor of 10 starting from 
$\dot{m}_d = 10^{-4}$, keeping the sub-Keplerian halo rate constant at 
$\dot{m}_h = 0.1$ Eddington rate. In Fig. \ref{fig5.2}(b), the 
sub-Keplerian rate $\dot{m}_h$ is increased keeping the Keplerian rate 
constant at $\dot{m}_d = 3$e-4. The case IDs are marked
on each plot. As $\dot{m}_d$ is increased, the relative intensity 
increases. This is understandable since increasing $\dot{m}_d$ increases 
the number of soft photons in a given energy band. At the same time, 
number of scatterings among the photons and the electrons increases 
and hence, the centrifugal pressure dominated inner region (including 
the post-shock region, when present) is cooled faster and the region 
collapses. Thus the volume of the post-shock region as well as the number 
of available hot electrons reduce with the increase of $\dot{m}_d$ (see, Chapter 4). 
Since the high energy power-law part of the spectrum is determined by 
the number of available hot electrons as well as the size of the post-shock 
region, the spectra become softer with the increase of $\dot{m}_d$. 
On the other hand, the spectra become harder as $\dot{m}_h$ is increased. 
As $\dot{m}_h$ is increased, the density of post-shock region increases 
and hence, the optical depth in this region is enhanced. This increases 
the number of scatterings and hence, more photons get inverse-Comptonized. 
Therefore, the number of high energy photons in a given energy band increases. 
Thus, the spectra become harder (GGC13). 

For both the cases (increase in either $\dot{m}_d$ or $\dot{m}_h$, 
keeping the other constant), the interaction between the photons and the 
electrons is enhanced and as a result, the electrons get cooler. Therefore, the 
shock location moves towards the black hole (see, Table 5), as was shown 
for thermal bremsstrahlung earlier (MSC96) and analytically by 
Das \& Chakrabarti (2004), Mondal \& Chakrabarti (2013)
who include both bremsstrahlung and Comptonization.
% Das, Santabrata; Chakrabarti, Sandip K., 2004, IJMPD, 13, 1955
% Mondal S., Chakrabarti S. K., 2013, MNRAS, 431, 2716

%%%%%%%%%%%%%%%%%%%Fig.5.3%%%%%%%%%%%%%%%%%%%%%%%%%
\begin{figure}
\centering{
\includegraphics[width=10cm,height=8cm]{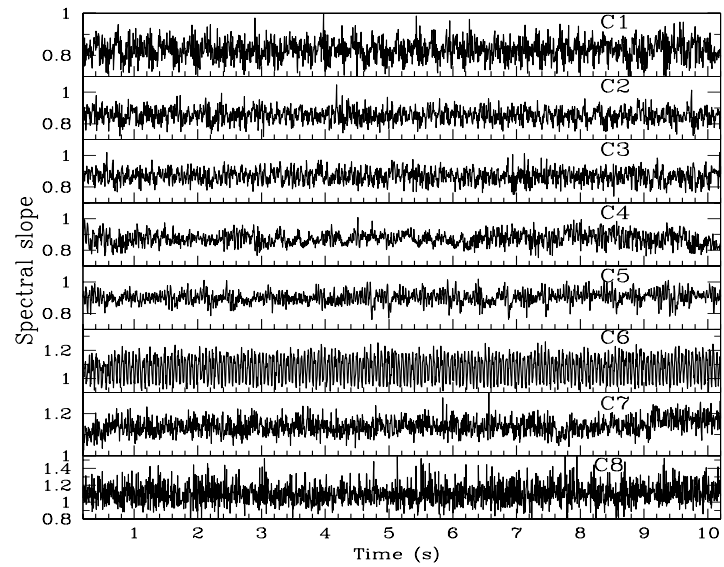}}
\caption{Time variations of the spectral slope of the power-law part of the spectra
are shown. Effects of variations of the disk rate $\dot{m}_d$ is shown. Case
IDs are marked on each panel (GGC13).
}
\label{fig5.3}
\end{figure}
%%%%%%%%%%%%%%%%%%%%%%%%%%%%%%%%%%%%%%%%%%%%%%%%%%%

%%%%%%%%%%%%%%%%%%%Fig.5.4%%%%%%%%%%%%%%%%%%%%%%%%%
\begin{figure}
\centering{
\includegraphics[width=10cm,height=8cm]{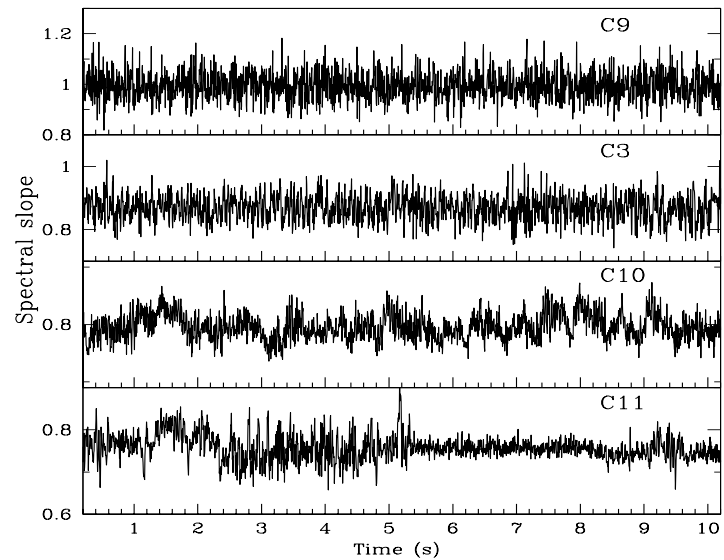}}
\caption{Same as Fig. \ref{fig5.3}, but the halo rate $\dot{m}_h$ is varied keeping 
$\dot{m}_d$ constant (GGC13).
}
\label{fig5.4}
\end{figure}
%%%%%%%%%%%%%%%%%%%%%%%%%%%%%%%%%%%%%%%%%%%%%%%%%%%

In Figs \ref{fig5.3} and \ref{fig5.4}, we show the time variation of 
the spectral slope $\alpha$ [$I(E)\propto E^{-\alpha}$] for all the cases 
presented in Table 5. The case IDs are marked on each panel. The time 
averaged values of the spectral slopes are given in Table 5. However, 
when we plot the time variations, we find an interesting behaviour, 
namely, the rocking of the spectrum between hard and soft states (GGC13). 
The spectral slope oscillates around or near the value 1. We find this 
effect for many cases e.g., C1, C6, C8 and C9. Among these cases, 
we find low frequency QPOs for C6 and C9 (see below).

\subsection{Timing properties}

We compute the time variations of the photon count rates for all the 
cases in order to generate simulated light curves (GGC13). 
In the coupled simulation run, a Monte Carlo simulation is run after 
each hydrodynamic time step. The photons are emitted continuously during 
this time of hydrodynamical evolution. However, in our two-step process, 
we process these photons through the Monte Carlo simulation step. In the 
Monte Carlo simulation, we track each photon and as soon as it leaves the 
disk, it is saved in an energy bin as well as in a time bin. We divide 
the energy range $0.01$ keV to $5000$ keV in $300$ logarithmic energy bins. 
The time resolution is of $5$ ms. After the run is completed, we rebin 
these photons with a time bin of $0.01$ s to compute the lightcurves. 
To count the photons in a given energy band, we add up the photon numbers 
in each energy bins which fall in the required energy band. 

%%%%%%%%%%%%%%%%%%%Fig.5.5%%%%%%%%%%%%%%%%%%%%%%%%%
\begin{figure}
\centering{
\includegraphics[width=10cm,height=8cm]{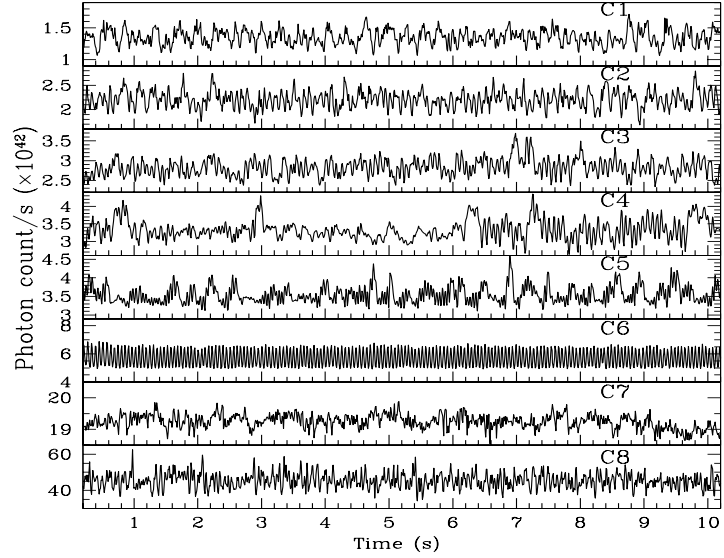}}
\caption{The light curves in $0.5-100$ keV range are shown.
Here, $\dot{m}_d$ is increased keeping $\dot{m}_h=0.1$ constant
(GGC13). See text for the detailed computational procedure of the light curves.
}
\label{fig5.5}
\end{figure}
%%%%%%%%%%%%%%%%%%%Fig.5.5%%%%%%%%%%%%%%%%%%%%%%%%%

%%%%%%%%%%%%%%%%%%%Fig.5.6%%%%%%%%%%%%%%%%%%%%%%%%%
\begin{figure}
\centering{
\includegraphics[width=10cm,height=8cm]{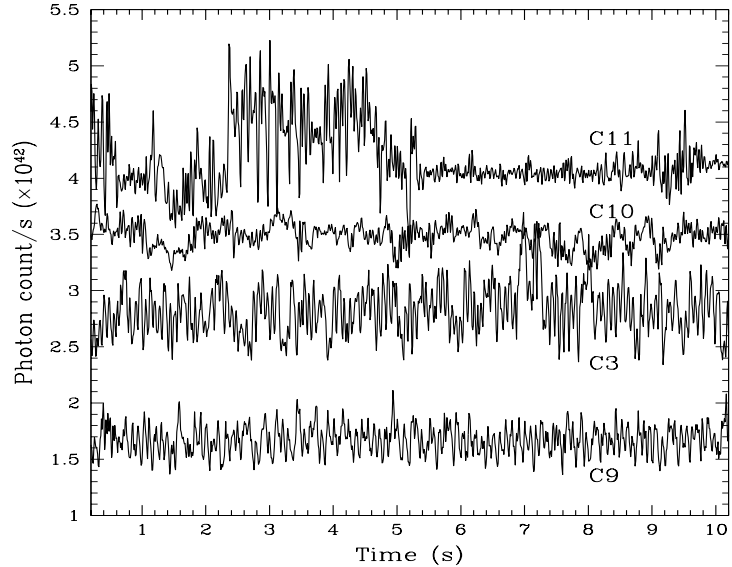}}
\caption{Same as Fig. \ref{fig5.5}, but $\dot{m}_h$ is increased keeping 
$\dot{m}_d$ constant at $0.0003$ Eddington rate (GGC13).}
\label{fig5.6}
\end{figure}
%%%%%%%%%%%%%%%%%%%%%%%%%%%%%%%%%%%%%%%%%%%%%%%%%%%

In Figs \ref{fig5.5} and \ref{fig5.6}, we plot light curves  
in the energy band 0.5 keV to 100 keV (for C7 and C8, 
2 keV$~<~E~<~100$ keV) (GGC13). The photons in this energy range are mostly 
the inverse-Comptonized photons. The case IDs are marked on each panel. 
In Fig. \ref{fig5.5}, the Keplerian rate $\dot{m}_d$ is 
increased keeping the sub-Keplerian rate $\dot{m}_h$ constant, whereas 
in Fig. \ref{fig5.6}, $\dot{m}_h$ is increased keeping $\dot{m}_d$ constant. 
In both the Figures, we see that the count rate increases with the increase 
of the variable parameter (e.g., $\dot{m}_d$ or $\dot{m}_h$). For Fig. 
\ref{fig5.5}, it is understandable since increasing $\dot{m}_d$ increases 
the number of available soft photons. On the other hand, when we increase 
$\dot{m}_h$, the optical depth of the post-shock region increases and 
hence, interception of the photons by the electrons increases. Thus the 
number of Comptonized photons increases and that explains the increase 
of count rates with $\dot{m}_h$ in Fig. \ref{fig5.6}. The variations 
in the lightcurves are arising because of the variations in the 
hydrodynamic and thermal properties of the post-shock region. 

%%%%%%%%%%%%%%%%%%%Fig.5.7%%%%%%%%%%%%%%%%%%%%%%%%%
\begin{figure}
\centering{
\includegraphics[width=9cm,height=9cm]{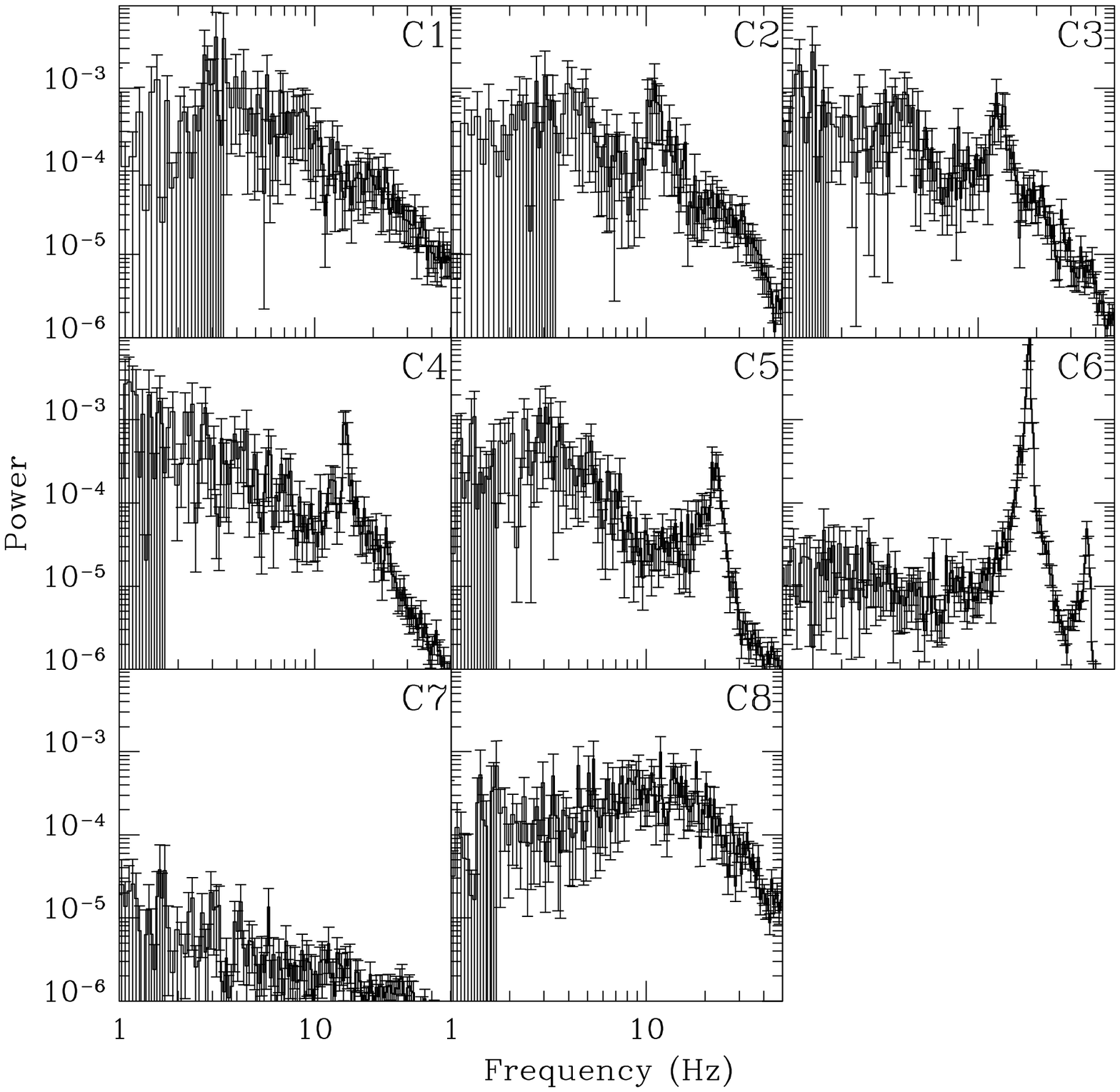}}
\caption{Power Density Spectra (PDS) of the all cases presented in 
Fig. \ref{fig5.5}. QPO frequency increases with the increase of 
$\dot{m}_d$ (GGC13).}
\label{fig5.7}
\end{figure}
%%%%%%%%%%%%%%%%%%%%%%%%%%%%%%%%%%%%%%%%%%%%%%%%%%%

%%%%%%%%%%%%%%%%%%%Fig.5.8%%%%%%%%%%%%%%%%%%%%%%%%%
\begin{figure}
\centering{
\includegraphics[width=8cm,height=8cm]{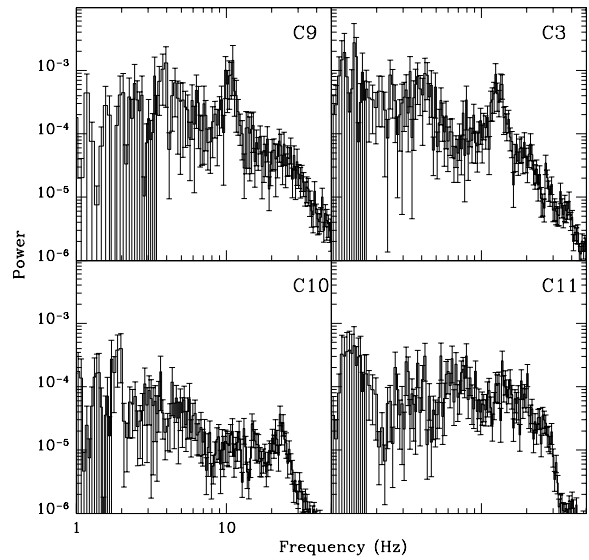}}
\caption{Power Density Spectra (PDS) of the all cases presented in 
Fig. \ref{fig5.6}. QPO frequency increases with the increase of 
$\dot{m}_h$ (GGC13).}
\label{fig5.8}
\end{figure}
%%%%%%%%%%%%%%%%%%%%%%%%%%%%%%%%%%%%%%%%%%%%%%%%%%%

In Figs \ref{fig5.7} and \ref{fig5.8}, we show the Power Density 
Spectra (PDS) for all the cases (GGC13). The case IDs are marked in each panel. 
We find low frequency quasi-periodic oscillations (LFQPO) for some cases. 
The frequencies are listed in Table 5. We find that LFQPO frequencies 
increase with the increase of both $\dot{m}_d$ and $\dot{m}_h$.  We have 
seen that the spectra become softer with the increase of $\dot{m}_d$ 
and with the decrease of $\dot{m}_h$. Therefore, from Figs \ref{fig5.3} 
and \ref{fig5.7}, we find that the LFQPO increases as the object transits 
from the harder state to the softer states. We find the opposite behavior 
in Figs \ref{fig5.4} and \ref{fig5.8}. 

It has been argued in the literature (MSC96; CM00; CAM04; Debnath et al. 2010) 
that the oscillation in the centrifugal barrier dominated post-shock region
(CENBOL) is responsible for the LFQPO observed in the black hole candidates. LFQPO 
arises when the infall time scale of post-shock matter roughly matches 
with the cooling time scale. For the present simulations, we find the 
shape of this hot post-shock region to be mostly paraboloid with the 
base to be oblate spheroid sometimes. Turbulence and backflow of matter 
are present sometimes in the post-shock region. Therefore, it is not 
always easy to calculate the exact infall time scale. In our grid based 
hydrodynamic simulation, we compute the infall time scale in the following 
way. We take the average of the radial velocity component over 20 vertical 
grids starting from the equatorial plane at each radius and compute a 
radial velocity profile near the equatorial plane: 
$$
v_{avg}(R)\sim\frac{\sum_{iz=1}^{20} v_R(R,iz)}{20},
$$
where, $v_R(R,1)$ represents the radial component of the velocity on the
first grid (i.e., on the equatorial plane), $v_R(R,2)$ represents the 
same on the second grid and so on. Then we calculate the infall time scale 
as (GGC13), 
$$
t_{in} =\sum_{R=R_{sh}}^{1.5}\frac{dR}{v_{avg}(R)}.
$$
The cooling timescale $t_{cool}$ is easy to compute as it includes only 
the scalar quantities: 
$$
t_{cool}=\frac{E_{th}}{\dot{E}}.
$$ 
Here, $E_{th}$ is the total thermal energy in the post-shock region and 
$\dot{E}$ is the cooling rate in the same region, which we calculate 
directly from the Monte Carlo simulation. In Table 5, we present the 
ratio $t_{in}\over t_{cool}$ in the last column. We see that this ratio 
is nearly 1 for all the cases when LFQPOs are seen. Thus the proposal 
of LFQPOs arising out of resonance oscillation (MSC96; CAM04) of the 
post-shock region appears to be justified. However, so far, only the 
power-law cooling was used as the proxy to the Compton cooling.
In the present work, actual Comptonization has been used. 

In this Chapter, we have presented results on LFQPO of $>$ 10 Hz. However,
LFQPO in this range 1-10 Hz can also be simulated using this simulation procedure.
As mentioned earlier, LFQPOs are observed when the resonance between the infall and the cooling time
scales occurs (MLC94; CAM04) and the frequency of the LFQPO can be found to be
$\nu \propto R_{sh}^{-3/2}$, where $R_{sh}$ is the shock location (CM00; Chakrabarti et al. 2008; Debnath et al. 2010;
Nandi, Debnath, Mandal, Chakrabarti 2012; Debnath, Chakrabarti, Mondal 2014).
% Nandi A., Debnath D., Mandal S., Chakrabarti S. K, A\&A, 542, 11
% Debnath D., Chakrabarti S. K., Mondal S., 2014, arXiv:astro-ph/1402.0989
Thus, when the shock forms closer, we find LFQPOs on higher side ($\sim$ 10 Hz) and
when shock forms far way, LFQPOs of lower side (1-10 Hz or even lower) is found.
In the present work, we have done simulations for accretion disk with angular
momentum $\lambda = 1.73$ for which the shocks from at a radial distance $\sim 20 R_g$. For
higher $\lambda$ with same disk-halo rate parameters, the shock is expected to form
at larger radius and we may get 1-10 Hz LFQPOs if the resonance condition is satisfied.

\clearpage
%~~~~~~~~~~~~~~~~~~~~~~~~~~~~~~~~~~~~~~~~~~~~~~~~~~~~~~~~~~~~~~~~~~~~~~~~~~~~~~~

	\reseteqn
	\resetsec
	\resetfig
	\resettab

%%%%~~~~~~~~~~~~~~~~~~~~~~~~~~~~~~~~~~~~~~~~~~~~~~~~~~~~~~~~~~~~~~~~~~~~~~~~~~~~~~
\alpheqn
\resec
\refig
\retab

%%%%%%%%%%%%%%%%%%%%%%%%%%%%%%%%%%%%%%%%%%%%%%%%%%%%%%%%%%%
% Chapter 6 : Conclusions and Discussions
%%%%%%%%%%%%%%%%%%%%%%%%%%%%%%%%%%%%%%%%%%%%%%%%%%%%%%%%%%%

\def\k{{\bf k}}
\def\aug{{\tilde{\cal H}}}
\def\ahat{{\mathaccent "7E {\cal A}}}
\def\bhat{{\mathaccent "7E {\cal B}}}
\def\chat{{\mathaccent "7E {\cal F}}}
\def\dhat{{\mathaccent "7E {\cal D}}}
\def\shat{{\mathaccent "7E {\cal S}}}
\def\proj{{\cal P}}
\def\ket{\vert \vert  \{ \emptyset \} \rangle}
\def\ket2{\vert \vert \otimes \{ R \} \rangle}
\def\sqr{$^{2}$}
\def\T{$T$}
\def\P{$P$}
\def\Q{$Q$}
\def\G{$G$}
\def\M{{\bf M}}
\def\H{{\bf H}}
\def\I{{\bf I}}
\def\Pr{{\bf P}}
\def\Tr{{\bf T}}
\def\eq{\enskip =\enskip}
\def\pls{\enskip+\enskip}
\def\mns{\enskip -\enskip}
\def\oper{O^{\bf k}_{R}}
\def\aug{\tilde{\cal H}}
\def\envfn{\chi^{\alpha}_{RL}}
\def\struc{S^{\alpha}_{RL,R^{\prime}L^{\prime}}}
\def\rR{r_{R}}

\newpage
\markboth{\it Conclusions and future plans}
{\it Conclusions and future plans}
\chapter{CONCLUSIONS AND FUTURE PLANS}

Here we summarize the main conclusions of the work. This will be 
followed by the current activities and the future plans.

\subsubsection{Summary and conclusions}

Here we present the summary of the works that are presented in different
Chapters and write down the main conclusions drawn. Most of the findings
regarding the spectral and timing properties of a black hole were
conjectured (or, even shown) earlier through analytical or numerical 
methods. These were: State transitions are possible by variations 
of the disk and halo accretion rates in a TCAF (CT95; Chakrabarti 1997); 
A direct correlation between the outflow rates and the spectral 
states must be present (Chakrabarti 1998b; C99; Das et al. 2001);
Low Frequency Quasi-Periodic Oscillations (LFQPOs) are the results of 
oscillation of centrifugal pressure dominated shocks in a transonic flow 
and the oscillations are due to resonance between the cooling and infall 
time scales (MSC96; CAM04). In this thesis, by incorporating the effects 
of radiative cooling (through Monte Carlo method) on the hydrodynamic 
solutions, we rigorously prove that indeed the spectral and timing 
properties of a TCAF are exactly as conjectured before. 

In Chapter 1, we gave a brief introduction about the theoretical 
models of the accretion disk around a black hole and described the
relevant radiative processes. After that, we discussed about the
developments of the numerical techniques to study the dynamics as 
well as the radiative processes inside an accretion flow.  

In Chapter 2, we described our simulation procedure for computing the 
Comptonized spectrum from a two component advective flow in presence 
of an outflow. The simulation was carried out using a Monte Carlo code and to
reduce the computational time, we parallelized this code.
We computed the effects of the thermal and the bulk motion 
Comptonization on the soft photons emitted from a Keplerian disk by 
the CENBOL, the pre-shock, sub-Keplerian disk and the outflowing jet (GGCL10).
We studied the emerging spectrum when the converging inflow and the 
diverging outflow (generated from the CENBOL) are simultaneously present. 
The converging inflow up-scatters the photons and the diverging outflow
down-scatters them. The interplay between the up-scattering 
and down-scattering effects determines the overall shape of the emerging 
spectrum. The outflow parameters strongly depend on the inflow 
parameters and hence, for a given inflow and outflow geometry, the 
strength of the shock can also determine whether the net scattering
by the jet would be significant or not. It is also found that sometimes
the halo can Comptonize and harden the spectrum by bulk motion 
Comptonization even without the CENBOL.

In Chapter 3, we described the development of the time dependent 
radiation hydrodynamic simulation code (GGGC11; GGC12; GGC13). 
In this code, we coupled the
Monte Carlo code with a time dependent hydrodynamic simulation code. 
The details of the hydrodynamic code and the coupling procedure are
given. Using this code, we studied the spectral and timing 
properties of the Two Component Advective Flows or TCAF. 
The accreting halo is assumed to be of zero 
angular momentum and spherically symmetric. It intercepts the soft 
photons coming out from the Keplerian disk residing on the equatorial
plane. We found that in presence of the axisymmetric disk, an originally
spherically symmetric accreting Compton cloud could become axisymmetric. 
This happens because, due to higher optical depth, there is a significant
cooling near the axis of the intervening accreting halo between the
disk and the axis. We also found the emitted spectrum to be direction
dependent. The spectrum along the axis shows a large soft bump, while
the spectrum along the equatorial plane is harder. The photons which
spend more time (up to 100ms in the case considered) inside the halo
are found to produce the harder part of the spectrum as they suffer
more scatterings. However, if they spend more than 100ms, they transfer
their energy to the relatively cooler electrons before escaping. These
results would be important while interpreting the timing properties
of the radiations from the black hole candidates. We also explored 
the effects of the bulk motion of the halo. We found that the 
inflowing matter push the photons towards inner radius and hence
force them to suffer more scatterings resulting in harder spectrum. 

In Chapter 4, we studied the effects of the Compton cooling on the
outflow in a TCAF using the time dependent radiation hydrodynamic
simulation code (GGC12). In this case, the accreting halo is assumed to have
some angular momentum with respect to the central black hole 
and a shock is found to form in the accreting halo. 
The outflow is found to produce from the CENBOL region. 
By simulating several cases for different inflow
parameters, we showed that the outflow rate is reduced for higher 
Keplerian disk rate. This happens because the number of injected soft 
photons increases with increasing disk rate, which cool the electrons 
of the post-shock region faster. This enhanced cooling reduces the 
thermal pressure of this region and the post-shock/CENBOL region shrinks. Therefore, the
spectrum also becomes softer as the high energy power law part of the 
spectrum is determined by the number of hot electrons as well as the
size of the post-shock region. We, thus, found a direct correlation
between the outflow rates and the spectral states of the accreting
black holes.    

In Chapter 5, we studied the quasi-periodic oscillations in the radiative
transonic accretion flows (GGC13). We considered TCAF as the flow configuration
and assumed the
accreting halo to have some angular momentum. We simulated 
several cases by varying the disk and the halo rates. The transition 
from a hard state to a soft state is found to be determined by 
the mass accretion rates of the disk and the halo. Low frequency QPOs
are found for several combinations of disk and halo rates. We found 
that the QPO frequency increases and the spectrum becomes softer as 
we increase the Keplerian disk rate. We also found that an earlier 
prediction that QPOs occur when the infall time scale roughly matches 
with the cooling time scale, originally obtained using a power-law 
cooling, remains valid even for Compton cooling. Our findings agree 
with the general observations of QPOs.

\subsubsection{Current activity and future plan}

At present, we are working to include the effects of photon bending
into our Monte Carlo simulation code. 
In our simulations, the low energy photons are generated within the
accretion disk (from the surface of the Keplerian disk). They travel
through the space-time around the black hole before escaping from the
accretion disk. The photon trajectory can be calculated by solving 
the geodesic equations in a given space-time. For the present case,
we solve the geodesic equations in the Schwarzschild space-time 
(Chattopadhyay, Garain \& Ghosh 2012). 
% Chattopadhyay I., Garain S. K., Ghosh H., submitted, Proceedings of International
%Conference on Astrophysics and Cosmology (2012), Tribhuvan University, Nepal
The equations and the solution procedure are described in Appendix A.
We are in the process of incorporating the photon transport in 
Schwarzschild space-time into our simulation. Our approach is the 
emitter-to-observer approach i.e., the photons are originated inside the
accretion disk and are traced till it reaches the observer. This 
approach is required particularly when the scattering process is 
included (Schnittman \& Krolik 2013).
% Schnittman J. D., Krolik J. H., 2013, arXiv:astro-ph 1302.3214v2 

We also started to include the effects of viscosity in our
time dependent radiation hydrodynamic simulation code. Very recently
Giri \& Chakrabarti (2013) has shown that the Keplerian disk can be
% Giri K., Chakrabarti S. K., 2013, MNRAS, 430, 2836
formed by incorporating the effects of viscous transport
and radiative cooling. A high viscosity in the equatorial plane produces 
a Keplerian disk while lower viscosity away from the equatorial plane 
fails to convert the sub-Keplerian flow into a Keplerian disk. Till now,
instead of adding viscosity to the flow, we directly included a Keplerian
disk as the supplier of the seed photons in the equatorial plane. 
However, the most self-consistent solution needs to incorporate the 
viscosity and produce the Keplerian disk, and its spectrum {\it ab initio}. 
This is being investigated into.
%This requires generation of photons through bremsstrahlung process and 
%Comptonize them to produce the multicolor blackbody spectrum from the 
%Keplerian disk.

Another work we are planning to do in near future is to fit the 
observed spectrum using our simulation code. For this, we have to simulate
several spectra by varying the flow parameters. From the fitting, we
shall be able to calculate different flow parameters e.g., the accretion 
rates of Keplerian and sub-Keplerian flow etc. 

\clearpage
%~~~~~~~~~~~~~~~~~~~~~~~~~~~~~~~~~~~~~~~~~~~~~~~~~~~~~~~~~~~~~~~~~~~~~~~~~~~~~~~

	\reseteqn
	\resetsec
	\resetfig
	\resettab

%%%%~~~~~~~~~~~~~~~~~~~~~~~~~~~~~~~~~~~~~~~~~~~~~~~~~~~~~~~~~~~~~~~~~~~~~~~~~~~~~~
\alpheqn
\resec
\refig
\retab

%%%%%%%%%%%%%%%%%%%%%%%%%%%%%%%%%%%%%%%%%%%%%%%%%%%%%%%%%%%
% Chapter 6 : APPENDIX
%%%%%%%%%%%%%%%%%%%%%%%%%%%%%%%%%%%%%%%%%%%%%%%%%%%%%%%%%%%
\def\k{{\bf k}}
\def\aug{{\tilde{\cal H}}}
\def\ahat{{\mathaccent "7E {\cal A}}}
\def\bhat{{\mathaccent "7E {\cal B}}}
\def\chat{{\mathaccent "7E {\cal F}}}
\def\dhat{{\mathaccent "7E {\cal D}}}
\def\shat{{\mathaccent "7E {\cal S}}}
\def\proj{{\cal P}}
\def\ket{\vert \vert  \{ \emptyset \} \rangle}
\def\ket2{\vert \vert \otimes \{ R \} \rangle}
\def\sqr{$^{2}$}
\def\T{$T$}
\def\P{$P$}
\def\Q{$Q$}
\def\G{$G$}
\def\M{{\bf M}}
\def\H{{\bf H}}
\def\I{{\bf I}}
\def\Pr{{\bf P}}
\def\Tr{{\bf T}}
\def\eq{\enskip =\enskip}
\def\pls{\enskip+\enskip}
\def\mns{\enskip -\enskip}
\def\oper{O^{\bf k}_{R}}
\def\aug{\tilde{\cal H}}
\def\envfn{\chi^{\alpha}_{RL}}
\def\struc{S^{\alpha}_{RL,R^{\prime}L^{\prime}}}
\def\rR{r_{R}}

\newpage
\appendix
\markboth{\it Geodesic equations in Schwarzschild space-time}
{\it Geodesic equations in Schwarzschild space-time}
\chapter{GEODESIC EQUATIONS IN SCHWARZSCHILD SPACE-TIME}

The Geodesic equations (Weinberg 1972) in Schwarzschild space-time are the following:
%Weinberg S., {\it Gravitation and Cosmology}, 1972, John Willey \& Sons, UK
$$
\frac{d^2r}{dp^2}+\frac{A^\prime(r)}{2A(r)}\left(\frac{dr}{dp}
\right)^2 - \frac{r}{A(r)}\left(\frac{d\theta}{dp}\right)^2 - 
\frac{r \sin^2\theta}{A(r)}\left(\frac{d\phi}{dp}\right)^2+\frac
{B^\prime(r)}{2A(r)}\left(\frac{dt}{dp}\right)^2 = 0, \nonumber\\ 
$$
\begin{equation}\label{eqno6.1}
\frac{d^2\theta}{dp^2}+\frac{2}{r}\frac{d\theta}{dp}\frac{dr}{dp} 
- \sin\theta \cos\theta \left(\frac{d\phi}{dp}\right)^2 = 0, \\
\end{equation} 
$$
\frac{d^2\phi}{dp^2}+\frac{2}{r}\frac{d\phi}{dp}\frac{dr}{dp} + 
2\cot\theta \frac{d\phi}{dp} \frac{d\theta}{dp}= 0, \nonumber \\
$$
$$
\frac{d^2t}{dp^2}+\frac {B^\prime(r)}{B(r)}\frac{dt}{dp}\frac{dr}{dp}=0, \nonumber
$$
where, $A(r)=\left(1-\frac{1}{r}\right)^{-1}$ and 
$B(r)=\left(1-\frac{1}{r}\right)$.\\
The last equation of the set of Eq. (\ref{eqno6.1}) gives 
$$
\frac{dt}{dp}=\frac{E}{B(r)}=\frac{E}{1-\frac{1}{r}},
$$
where, $E$ is a constant of motion (energy-at-infinity) (ST83). 
Using this equation, we have replaced the derivative w.r.t affine parameter 
$p$ to $t$ from the above equations to get the following equations: 
$$
\frac{d^2r}{dt^2}-\frac{3}{2}\frac{1}{r(r-1)}\left(\frac{dr}{dt}\right)^2 
- (r-1)\left(\frac{d\theta}{dt}\right)^2 - (r-1)\sin^2\theta
\left(\frac{d\phi}{dt}\right)^2 + \frac{r-1}{2r^3}=0, \nonumber \\
$$
\begin{equation}\label{eqno6.2}
\frac{d^2\theta}{dt^2}+\frac{2r-3}{r(r-1)}\frac{dr}{dt}\frac{d\theta}{dt} 
- \sin\theta \cos\theta \left(\frac{d\phi}{dt}\right)^2=0, \mathrm{~and~} \\
\end{equation}
$$
\frac{d^2\phi}{dt^2}+\frac{2r-3}{r(r-1)}
\frac{dr}{dt}\frac{d\phi}{dt} + 2\cot\theta \frac{d\theta}{dt}\frac{d\phi}{dt}=0. \nonumber
$$

The set of Eq. (\ref{eqno6.2}) is solved using the fourth order
Runge Kutta method. To start the integration, we need to know the 
initial location of the photon ($r, \theta, \phi$) and its initial
direction of movement ($v^{\hat{r}}, v^{\hat{\theta}}, v^{\hat{\phi}}$),
where, 
$$
v^{\hat{r}} = {r \over (r-1)}{dr\over dt}; \;
v^{\hat{\theta}} = {r\sqrt{r} \over \sqrt{(r-1)}}{d\theta\over dt}; \;
v^{\hat{\phi}} = {r\sqrt{r} \sin \theta\over \sqrt{(r-1)}}{d\phi\over dt}.
$$
In the Monte Carlo simulation, the initial location of a photon and its
propagation direction are randomized using the given distribution functions.
With these initial values, the photon path is computed by solving the above
equations till it suffers a scattering or leaves the electron cloud. 
In case the photon get scattered by an electron, it is tracked subsequently 
from the scattering location with the new direction of propagation.

	\reseteqn
	\resetsec
	\resetfig
	\resettab
%\newpage
%\vskip 3cm
%\vskip 2cm
%\input{Sircorrect/appen.tex}  %imp   
%~~~~~~~~~~~~~~~~~~~~~~~~~~~~~~~~~~~~~~~~~~~~~~~~~~~~~~~~~~~~~~~~~~~~~~~~~~~~~~
\vfill\eject
\pagestyle{newheadings}
{\baselineskip 15pt
\def\etal{{\sl et al.}}
\def\aug{\vert R,L,\{\emptyset\}\rangle }
  \def\ket{\vert \vert	\{ \emptyset \} \rangle}
  \def\ket2{\vert \vert \otimes \{ R \} \rangle}
  \def\sqr{$^{2}$}
\def\dpr{\prime\prime}
\def\tpr{\prime\prime\prime}
\def\pr#1{ Phys.Rev. {\bf B #1}}
\def\pj#1{\proj_{{\cal #1}}}
\def\barr#1{{\overline{#1}}}
\def\jpc#1{J.Phys. Condensed Matter {\bf #1}}
\def\prl#1{ Phys. Rev. Lett. {\bf #1}}
\def\.#1{\mathaccent 95#1}
\def\^#1{\mathaccent 94 #1}
\def\~#1{\mathaccent "7E #1}
\def\Ir{{\mbox{I}}}
\def\Mr{{\mbox{M}}}
\def\Hr{{\mbox{H}}}
\def\sund{\mathaccent 22{\sigma}}
\def\equal{\enskip =\enskip}
\def\plus{\enskip +\enskip}
\def\minus{\enskip -\enskip}
\def\eq{\enskip =\enskip}
\def\pls{\enskip +\enskip}
\def\mns{\enskip -\enskip}
\def\Gund{\mathaccent 22 {G}}
\def\ul#1{\underline{#1}}
\def\ac#1{\mathaccent 95#1}
\def\td#1{\mathaccent "7E#1}
\def\un#1{\underline{#1}}
\def\nbox{\raisebox{.6ex}{\fbox{{\scriptsize{\phantom{$\sqrt{}$}}}}}$\:$}
\def\ybox{\raisebox{.6ex}{\fbox{{\scriptsize{$\sqrt{}$}}}}$\:$}
\def\pbox#1{\raisebox{.6ex}{\fbox{{#1}}}$\:$}
\def\c#1{\mbox{\bf #1}}
\def\und#1{$\underline{\mbox{\bf #1}}\:$}
\def\unit{{\cal I}}
\def\trans{{\cal T}}
\def\proj{{\cal P}}
\def\T{$T$}
\def\P{$P$}
\def\Q{$Q$}
\def\G{$G$}
\def\M{{\bf M}}
\def\H{{\bf H}}
\def\I{{\bf I}}
\def\Pr{{\bf P}}
\def\Tra{{\bf T}}
\def\diag{\varepsilon_{i}}
  \def\proj{{\cal P}}
  \def\trans{{\cal T}}
  \def\ket{\vert \vert	\{ \emptyset \} \rangle}
  \def\ket2{\vert \vert \otimes \{ R \} \rangle}
  \def\sqr{$^{2}$}
\def\k{{ ( k}}
  \def\ahat{{\mathaccent "7E  A}}
  \def\bhat{{\mathaccent "7E  B}}
  \def\chat{{\mathaccent "7E  C}}
  \def\fhat{{\mathaccent "7E  F}}
  \def\dhat{{\mathaccent "7E  D}}
  \def\shat{{\mathaccent "7E  S}}
  \def\phat{{\mathaccent "7E  P}}
  \def\jhat{{\mathaccent "7E  J}}
  \def\khat{{\mathaccent "7E  K}}
  \def\ohat{{\mathaccent "7E o}}
\def\ve{\varepsilon}
\def\car{\{{\cal C}\}}
\def\gt{\; > \;}
\def\lt{\: < \:}

}

%~~~~~~~~~~~~~~~~~~~~~~~~~~~~~~~~~~~~~~~~~~~~~~~~~~~~~~~~~~~~~~~~~~~~~~~~~~~~~~
\end{document}